\documentclass[pre,aps,superscriptaddress,twocolumn,amsmath,amssymb,floatfix,10pt]{revtex4-2} 

\usepackage[breaklinks,colorlinks = true,linkcolor = magenta,urlcolor=magenta,citecolor=red,hypertexnames=false]{hyperref}
\usepackage{graphicx}
\usepackage{multirow}
\usepackage{epsfig}
\usepackage{epstopdf}

\usepackage{natbib}

\usepackage{setspace}
\usepackage{graphicx}% Include figure files
\usepackage{dcolumn}% Align table columns on decimal point
\usepackage{bm}% bold math
\usepackage{color}
\usepackage{latexsym}
\usepackage{textcomp}
\usepackage[section]{placeins}
\usepackage{subfigure}

\usepackage{tabularx}
\usepackage[dvipsnames]{xcolor}
\usepackage{lipsum}
\usepackage{dsfont}
\usepackage{braket}
\usepackage{float}
\usepackage{xr}
\usepackage{wrapfig}
\usepackage[format=plain,labelformat=brace,labelsep=space,justification=RaggedRight,font=default,labelfont=bf,textfont={small,stretch=0.95},margin=0pt,indention=0pt,parindent=0pt,hangindent=0pt,singlelinecheck=true]{caption}
\DeclareCaptionLabelFormat{adja-page}{\hrulefill\\#1 #2 \emph{(previous page)}}

\usepackage{silence}
\allowdisplaybreaks

\def\be{\begin{equation}}
\def\ee{\end{equation}}

\makeatletter
\newsavebox{\@brx}
\newcommand{\llangle}[1][]{\savebox{\@brx}{\(\m@th{#1\langle}\)}%
  \mathopen{\copy\@brx\mkern2mu\kern-0.9\wd\@brx\usebox{\@brx}}}
\newcommand{\rrangle}[1][]{\savebox{\@brx}{\(\m@th{#1\rangle}\)}%
  \mathclose{\copy\@brx\mkern2mu\kern-0.9\wd\@brx\usebox{\@brx}}}
\makeatother

\begin{document}
\title{Tensor Electromagnetism and Emergent Elasticity in Jammed Solids}
\author{Jishnu N. Nampoothiri}
\affiliation{Martin Fisher School of Physics, Brandeis University, Waltham, MA 02454 USA}
\affiliation{Centre for Interdisciplinary Sciences, Tata Institute of Fundamental Research, Hyderabad 500107, India}
\email{jishnu@brandeis.edu}
\author{Michael D'Eon}
\affiliation{Martin Fisher School of Physics, Brandeis University, Waltham, MA 02454 USA}
\email{mdeon@brandeis.edu}
\author{Kabir Ramola}
\affiliation{Centre for Interdisciplinary Sciences, Tata Institute of Fundamental Research, Hyderabad 500107, India}
\email{kabirramola@gmail.com}
\author{Bulbul Chakraborty}
\affiliation{Martin Fisher School of Physics, Brandeis University, Waltham, MA 02454 USA}
\email{bulbul@brandeis.edu}
\author{Subhro Bhattacharjee}
\affiliation{International Centre for Theoretical Sciences, Tata Institute of Fundamental Research, Bengaluru 560089, India}
\email{subhro@icts.res.in}

\date{\today}

%%%%%%%
\begin{abstract}
The theory of mechanical response and stress transmission in disordered, jammed solids poses several open questions of how non-periodic networks -- apparently indistinguishable from a snapshot of a fluid -- sustain shear. We present a stress-only theory of {\it emergent elasticity} for a non-thermal amorphous assembly of grains in a jammed solid, where each grain is subjected to mechanical constraints of force and torque balance. These grain-level constraints lead to the Gauss's law of an emergent $U(1)$ tensor electromagnetism, which then accounts for the mechanical response of such solids. This formulation of amorphous elasticity has several immediate consequences. The mechanical response maps exactly to the static, dielectric response of this tensorial electromagnetism with the polarizability of the medium mapping to emergent elastic moduli. External forces act as vector electric charges whereas the tensorial magnetic fields are sourced by momentum density. The dynamics in the electric and magnetic sectors, naturally translate into the dynamics of the rigid jammed network and ballistic particle motion respectively. The theoretical predictions for both stress-stress correlations and responses are borne out by the results of numerical simulations of frictionless granular packings {in} the static limit of the theory in both 2D and 3D.  
 
\end{abstract}

\maketitle
%\tableofcontents

\section{Introduction}

{ A collection of non-Brownian particles often form jammed solids if the imposed pressure or shear stress exceeds a threshold. Examples abound in nature and industry: sand piles~\cite{Behringer_2018,Geng2001}, shear-jammed grains~\cite{bi2011jamming} or dense suspensions~\cite{Brown_2014}, and gels~\cite{delgado_mao}.  The {\it solidity} of these systems emerges from the imposed stress itself: the rigid structure is created in response to stress and therefore, there is no {unique} zero-stress {\it reference} solid network~\cite{Cates1998,Bouchaud2002}.} It is the mechanical response to {\it additional} stress that determines whether the system is rigid or not. Under these circumstances, one may conclude that both of the fundamental ideas of {the theory of crystalline} elasticity-- the existence of a strain tensor defined with reference to a unique stress-free {spontaneously broken-symmetry} configuration, and a free energy relating stress to strain-- {need to be re-visited and may have to be abandoned~\cite{Otto2003,Bouchaud2002}.}

Curiously however, the mechanical properties of jammed solids bear many similarities to the crystalline elasticity~\cite{Lemaitre2018,Gelin2016,Geng2001,Otto2003,jamming_epitome_2003}. For example, the response of a granular pile~\cite{Geng2001,Otto2003} or that of frictionless jammed packings~\cite{Ellenbroek_thesis,jamming_epitome_2003} {to point forces}, can be described in terms of effective elastic moduli and elastic Green's functions. However, these {\it elastic} moduli do not necessarily satisfy the usual symmetry requirements~\cite{Otto2003} and depend on preparation protocols~\cite{jamming_epitome_2003,Otto2003}. Stress-stress correlations also exhibit power-law decays as expected for elastic media~\cite{Lemaitre2018,Gelin2016}. This poses a puzzle since, {\it e.g.} the theory of crystalline elasticity  is based on the existence of a periodic reference structure -- emerging from spontaneously broken translation symmetry in crystals -- which then defines a strain field via systematic coarse-graining~\cite{chaikin1995principles,landau2012theory}. While the stress field is well defined in jammed solids, the lack of a unique reference configuration makes the definition of the strain field and the associated free energy much less apparent. This puzzle and associated issues gave birth to the quest for a stress-only continuum theory of the elasticity of jammed solids~\cite{Bouchaud2002}. In frictionless systems, these are the minima in a complex energy landscape, the exploration of which  has led to a concentrated effort to understand plasticity in amorphous solids~\cite{Falk2011,McNamara2016,Karmakar2010,itamar_elasticty} based on notions of non-affine displacements.

\begin{figure}[!tbp]
\includegraphics[width=0.37\textwidth]{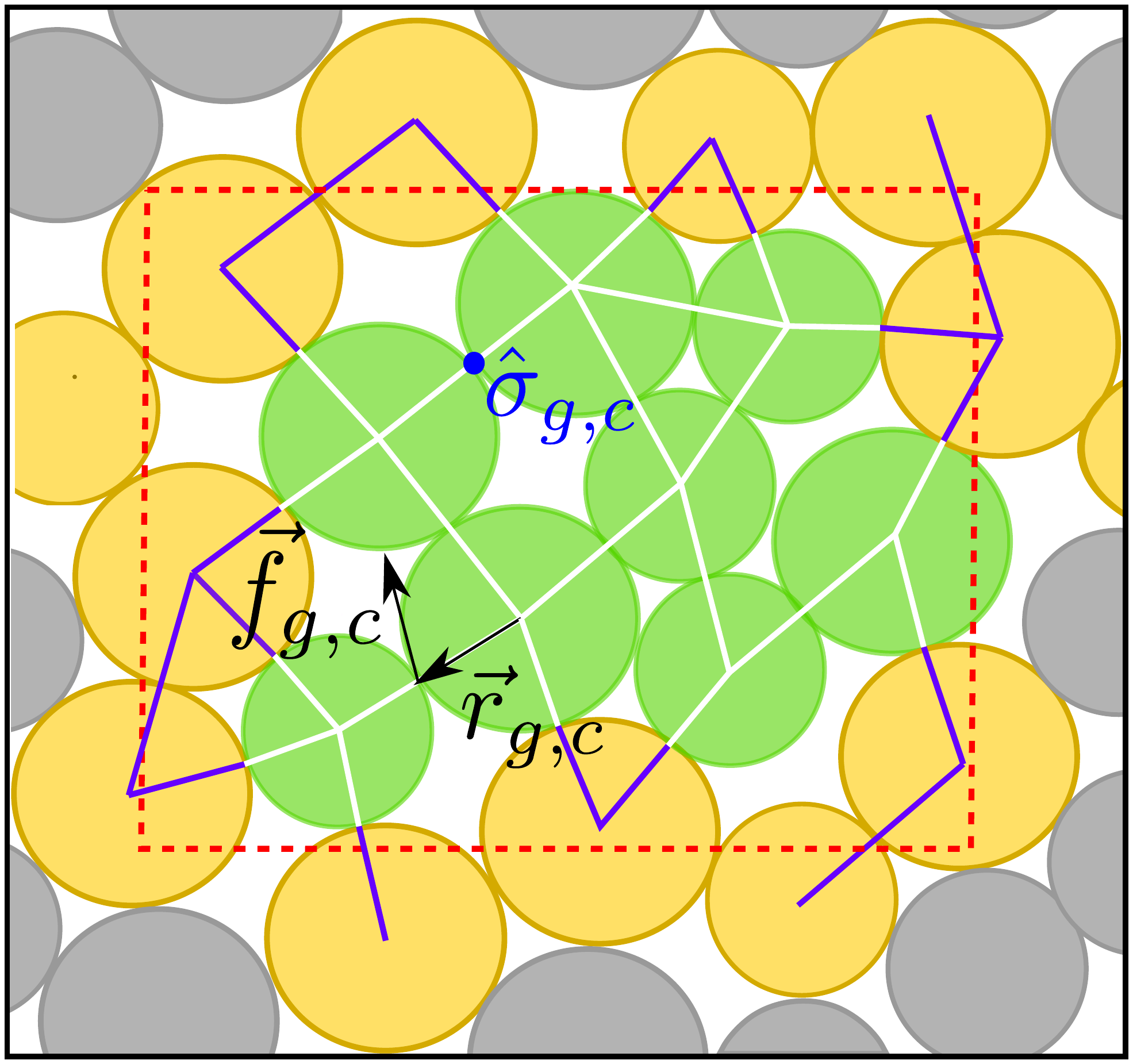}
\caption{{\bf A schematic depiction of a packing of grains (discs) in two dimensions (2D): }
{Each grain is in mechanical equilibrium and satisfies the microscopic constraints of force and torque balance that leads to a non-trivial contact network. A large packing consists of many local structures, which can be coarse grained (say over the red box with volume  $\Omega_r$)  to produce a continuous stress field.
Grains with all contacts in $\Omega_r$ are identified as bulk grains (colored green), whereas grains with a partial overlap with $\Omega_r$ are identified as boundary grains (colored yellow). Each contact has two contributions to the coarse grained stress tensor $\sigma(\boldsymbol{r})$ (detailed in Eq.~\eqref{eq:stress_coarse}). The white (violet) contact links contribute to the (anti)-symmetric part of the total stress tensor.
}
\label{fig:network}}
\vspace{-2mm}
\end{figure}

At the heart of the mechanical response of jammed solids is the athermal, non-Brownian, nature of these assemblies consisting of configurations at mechanical equilibrium implemented locally: each grain is in a state of force and torque balance as shown in Fig.~\ref{fig:network}. These local constraints lead to a non-trivial contact network that is in force and torque balance~\cite{liu2011jamming}. The emergence of such disordered structures from local constraints of mechanical equilibrium is challenging to incorporate in any continuum theory. In the naive continuum limit, force and torque balance do not provide enough equations to uniquely determine the stress distribution~\cite{Bouchaud2002}. Stated differently, since it is not possible to define a strain field with respect to a unique stress-free state, the well-known  compatibility relations of linear elasticity theory are missing, and  the linear response coefficients of crystalline elasticity stemming from  proportionality between stress and strain-- the so called elastic moduli, are not well defined~\cite{landau2012theory}. {Although we focus on jammed solids in this paper, similar stress-bearing structures in other non-thermal solids such as  gels  also emerge from a complicated interplay of external and internal stresses~\cite{delgado_mao}. }

A new framework has been put forward~\cite{PhysRevLett.125.118002} to supply the {\it missing equations} to obtain the much  cherished {\it stress-only} description of granular elasticity. Central to this framework is a gauge theoretic structure that arises from-- (1) the lack of the well defined and unique zero-stress reference configuration and (2) the local mechanical equilibrium of each grain in an athermal solid that serves as a Gauss's law for a {\it tensor electric field}. In this mapping, the forces act as {\it electric} charges of the so called {\it vector charge theory} (VCT)~\cite{Pretko2017} of the  {\it tensor electromagnetism} that naturally incorporates the conditions of force and torque balance~\cite{PhysRevLett.125.118002}. Determining the stress distribution in an amorphous solid then maps to the problem of solving electrostatics in the presence of a {\it dielectric} in the VCT. This stress-only framework is completely devoid of any reliance on a reference structure or displacement fields, which should not have any measurable consequences in amorphous solids. 

In this paper, we uncover the complete structure of the gauge-theoretic framework, whose static limit describes {the {\it elasticity} of granular solids. The gauge theory provides a compatibility condition for the stress without referring to displacement fields that form  a strain tensor. The structure of this amorphous elasticity theory is remarkably similar to classical elasticity theory except for the crucial distinction that there is no strain tensor. The elastic moduli that appear in this theory are not material properties, but emerge from the imposed stress. We demonstrate that these elastic moduli can be obtained from measurements of stress-stress correlations and stress response to localized perturbing forces. This provides a powerful alternative to strain-based measurements of elastic moduli in amorphous solids~\cite{Tanguy,dePablo} and predicts natural dynamical extensions.} Further, this may also provide a natural mechanism for generating the non-local mechanics and rheology of granular materials~\cite{Reddy2011,Henann2014}. Given the broad scope of this paper and the diversity of results presented,  we begin with a concise summary of the salient ideas, results and their implications.
%%%%%%%%%%%%%%%%%%

\vspace{-4mm}

\section{The roadmap}
\label{sec:roadmap}

\paragraph*{Emergent Gauge Theories: }

Gauge-theoretic notions such as parallel transport have been widely applied to the study of defects and plasticity in both crystalline~\cite{pretko_radzhihovsky,Kleinert:1989tk} and amorphous solids~\cite{Baggioli:2021vk,Moshe_PNAS}. However, the framework of  amorphous elasticity that  we present in this paper is based on strict local constraints, {with} no assumptions about the underlying geometry of the contact network. To motivate this, we begin with a short summary of well-established models in strongly correlated condensed matter systems where local energetic constrains result in low energy emergent U(1) electromagnetism.

Our understanding of correlated phases of condensed matter is replete with examples, {where due to energetic constraints, the original degrees of freedom cease to be valid fields at low energies and necessitates  a description in terms of new effective degrees of freedom.} This emergence of new degrees of freedom, while applicable to collective modes such as phonons and spin-waves in a broken symmetry system~\cite{anderson2018basic,chaikin1995principles}, leads to, at first glance, a startling yet qualitatively new outcome, when the energetic constraints are local.  This situation often arises in ``frustrated" systems with competing interactions-- both classical and quantum. The most pertinent examples for the present work are the ``water" ice crystals (hexagonal phase of ice)~\cite{giauque1933molecular,giauque1936entropy,pauling1935structure,gingras2009spin} or their spin analogues-- the classical spin ice~\cite{anderson1956ordering,bramwell2001spin,gingras2009spin,harris1997geometrical,ramirez1999zero}.

\begin{figure}[!tbp]
\includegraphics[width=0.475\textwidth]{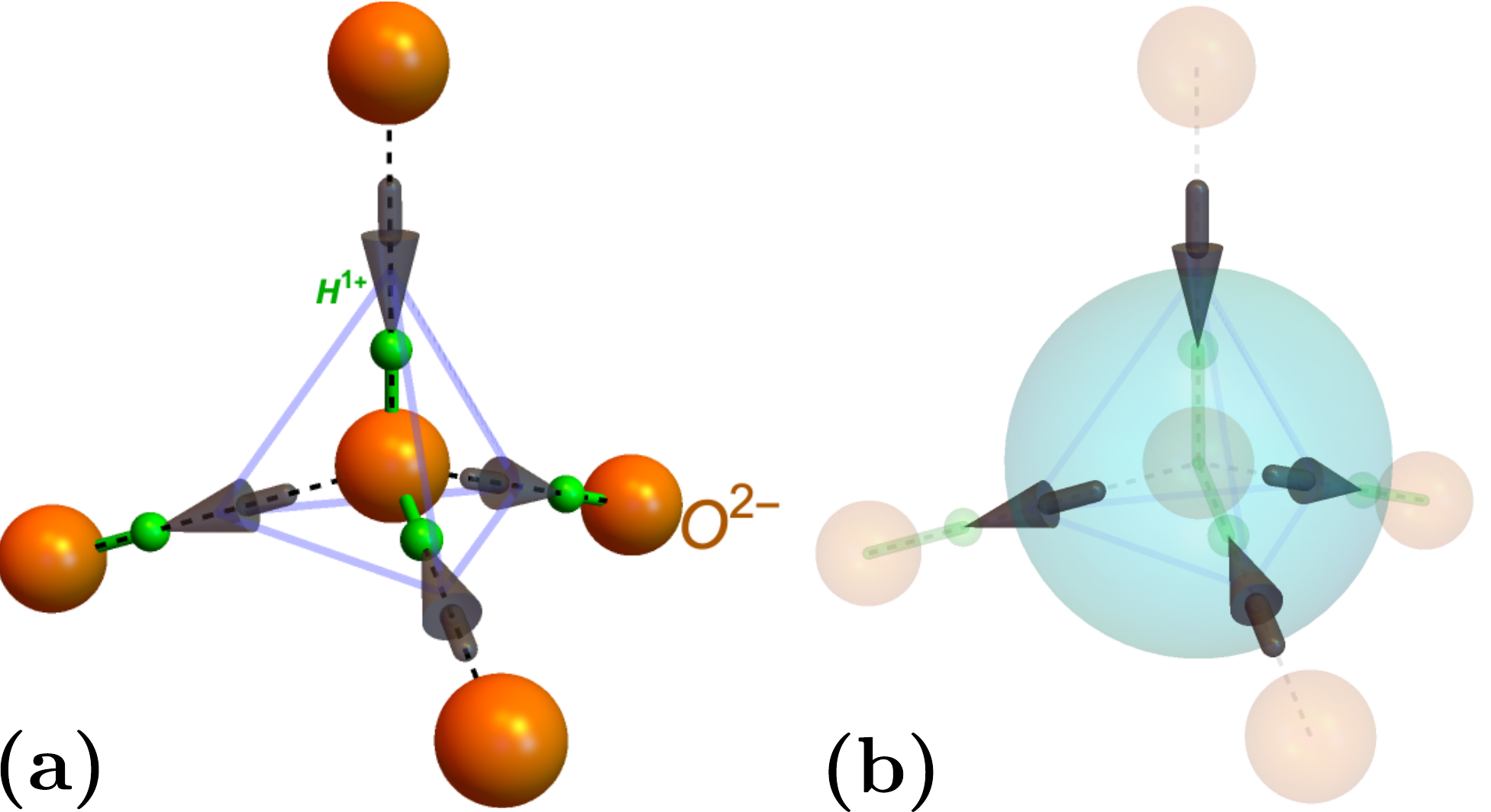}
\caption{{\bf Emergent electromagnetism in hexagonal phase of water-ice and spin ice :} (a) In the hexagonal phase of ice-crystals each O$^{2-}$ ion has four H$^{1+}$ ion (proton) along the O-O bonds with two being near and two being far that gives rise to the six possible configuration one of which is shown, in accordance with Bernal-Fowler {\it ice rules}~\cite{bernal1933theory} or 2-in-2-out rules in terms of the displacement of the proton from the mid-point of the O-O bond shown in arrows. (b) The arrows are mapped to an  emergent magnetic field $\mathfrak{b}$, such that the ice-rules implement its zero divergence, $\nabla\cdot\mathfrak{b}=0$. The low energy description is in terms of an emergent magnetostatics which does not depend on any specific configuration of the protons (see text for details).}
\vspace{-3mm}
\label{fig_waterice}
\end{figure}

In particular, for water ice as shown in Fig.~\ref{fig_waterice}, the O$^{2-}$ ions inside the water ice crystal form a diamond lattice, with each of them bonded to four H$^{1+}$ ions (protons)-- two covalently and two via hydrogen bonding-- leading to ${}^4C_2=6$ possible hydrogen configuration around each oxygen obeying the {\it 2-near-2-far} Bernal-Fowler ``ice-rules"~\cite{bernal1933theory}. In terms of the displacement vectors of the hydrogen atom about the O-O bond mid-points, the above ice rules translate into the {\it 2-in-2-out} configurations which is exactly realised in a class of rare-earth pyrochlore magnets aptly named spin-ice, where these displacement vectors are magnetic moments~\cite{gingras2009spin,anderson1956ordering,villain1979insulating,isakov2005spin}. For both these systems, as was shown by L. Pauling (in context of water ice) the total number of states that satisfy the ice-rules grows exponentially in system size leading to an extensive residual entropy~\cite{pauling1935structure,villain1979insulating}.  

The low energy long-wavelength theory of such proton disorder in water ice or spin configuration in spin ice is quite different from phonons or spin-waves as there is no single reference configuration about which the low energy {excitations can be defined}. Neither does each proton/spin independently survive as a valid low energy degree of freedom since inverting the position of one proton/flipping one spin immediately violates the ice-rules and hence is energetically penalised. However, the total displacement of hydrogen (calculated from any particular oxygen) in case of water ice or the total magnetic moment per tetrahedron for spin ice can be set equal to zero. Considering an imaginary closed surface enclosing each oxygen/tetrahedra the ice rules on the four displacement vectors/spins naturally translate into a {\it Gauss's law}, $\nabla\cdot\mathfrak{b}=0$, for an {\it emergent magnetic field}, $\mathfrak{b}$ which is proportional to the displacement vector/spin moment~\cite{isakov2004dipolar}. The long-wavelength theory at low energy then takes the form of magnetostatics and can successfully account for the dipolar spin correlations~\cite{isakov2004dipolar}, characterised by pinch-points as observed in experiments~\cite{bramwell1998frustration,fennell2009magnetic}.

Crucial to the discussion in this paper, are the following two aspects of the emergent magnetostatics described above. (i) The mapping does not depend on any particular reference spin/proton configuration among all those that are allowed by ice-rules. Instead, any one configuration can be taken as the reference and the observables are averaged over all such {\it ice configurations}. It is in this sense that the reference configuration does not have any measurable consequence in terms of spin correlations. In the language of the {\it emergent magnetostatics}, the above structure of non-observability translates into two ice-configurations being related by a {\it gauge transformation} of the emergent electromagnetism  which is quite different from the case of spin-waves or phonons. (ii)  The mapping to magnetostatics immediately allows us to write a coarse-grained long-wavelength free energy in terms of the emergent magnetic field $\mathfrak{b}$, that automatically accounts for the local constraints, the ice rules,  and hence serves as a valid long-wavelength theory.
\vspace{2mm}
\paragraph*{Emergent Elasticity:} The above considerations can be generalized to the problem of elasticity of jammed solids, which is the focus of this paper. The difference with crystalline elasticity, as we briefly summarise in Section~\ref{sec:problem}, is rooted in the lack of a unique reference configuration ~\cite{Bouchaud2002} defined by broken symmetry. Instead, the force and torque balance constraints on each grain (Eq.~\eqref{eq:mecheq}) lead to a {\it Gauss's law}. In the rest of this paper, elaborating on the ideas presented in reference~\cite{PhysRevLett.125.118002},  we demonstrate that an elegant framework for elasticity of jammed solids can be obtained in terms of this Gauss's law, which corresponds to that of a class of $U(1)$ gauge theories involving symmetric, rank-2 tensors-- the VCT~\cite{Pretko2017} (also see Appendix~\ref{appendix:A}). The VCT Maxwell's equations provide the requisite {\it missing} constitutive relations, giving rise to {\it emergent} elasticity that fully captures the mechanical response of jammed  solids. Similar to the case of water ice or spin ice, VCT does not depend on any reference configuration owing to the gauge redundancy, whose implications we explore in detail. Within this emergent electromagnetism, the mechanical response leads to emergent elastic moduli, that {remarkably} have a completely different origin-- the  underlying disordered network-- than, for example, spontaneous translation symmetry breaking in crystalline elastic solids. We refer to this full gauge-theoretic framework {for granular solids}, as the Vector Charge Theory of Granular mechanics and dynamics (VCTG). We note that while there are various kinds of tensor electromagnetism known in literature~\cite{Pretko2017},  unless specified, we use it here interchangeably with VCT.

\vspace{2mm}

\paragraph*{Short summary of our results: }
The rest of this paper is organised as follows. In Section~\ref{sec:problem}, we provide a short summary of the classical theory of crystalline elasticity followed by the challenges encountered in applying this classical framework to athermal, disordered solids which we refer to as ``jammed solids''.  In Section~\ref{sec:reformulation_athermal}, we provide a step-by-step construction of the central equations of VCTG, starting from the application of Newton's laws to an assembly of particles in mechanical equilibrium. The force and torque balance equations are then naturally interpreted~\cite{PhysRevLett.125.118002} in terms of the two conservation laws of the VCT-- the charge (Eq.~\eqref{eq:chargeconserve}) and charge-angular momentum (Eq.~\eqref{eq:amcon}), respectively. In Sections \ref{sec:reformulation_athermal} A and B,  we present the mapping of the electrostatic limit of a VCT dielectric to the VCTG theory of the elastic response of jammed solids. The details of the mapping, summarised in Table \ref{tbl:dictionary}, provide the requisite  dictionary.  Crucially, we show that this emergent elasticity theory has the same structure as classical elasticity but with the physical displacement field replaced by a gauge potential, {\it i.e.} a stress-only formulation that does not rely on the concept of strain. The contact forces, created by external forces such as gravity in a sandpile,  play the role of ``bound charges'', and therefore, naturally, the elastic moduli entering the emergent elasticity theory map to the rank-4 polarizability tensor of the VCT dielectric.
 
{In Section \ref{sec:correlations}, we derive the stress-stress correlations, $\llangle\sigma_{ij}\sigma_{kl}\rrangle$,  {the average over an ensemble of jammed configurations}, starting with the Landau-Ginzburg action Eq.~\eqref{eq:lagrangian_density} for the static limit of the VCTG . The final results are summarised in Eq.~\eqref{eq:2d_stress_correlations} and \eqref{eq:3d_stress_correlations} for two (2D) and 3D (3D), respectively (explicit forms in the language of the VCT dielectric are given in Appendix \ref{appendix:A}). A characteristic feature of the stress correlations is their singular behaviour in momentum space in the zero momentum limit, {\it i.e.}, $|{\bf q}|\rightarrow 0$. The singularity shows up in the characteristic ``pinch point" in the angular dependence of the correlation functions, as shown in Figs. \ref{fig:2d_correlations},~\ref{fig:2d_full} for 2D and Figs.~\ref{fig:3d_correlations_fig2},~\ref{fig:3d_correlations_fig3} for 3D. {In real-space, these translate to positive, long-ranged power law correlations in the longitudinal direction and rapidly decaying correlations in the transverse direction, as shown in Figs. \ref{fig:2d_correlations} and \ref{fig:2d_full} for 2D. It is this longitudinal correlation that appears dramatically as ``force-chains'' in granular media~\cite{PhysRevLett.125.118002,Wang:2020aa}.} Further, we analyze the stress response to weak external forces (with net sum zero) on a jammed granular assembly (such as shown in Fig. \ref{fig:response_schematic}).} In Section~\ref{sec:numerical}, we present detailed comparisons between theoretical predictions and numerical results of stress correlations and response in both 2D and 3D jammed, frictionless solids. This section also demonstrates how to compute the emergent elastic moduli from these measurements.

Finally, we take up the issue of the dynamics of the granular solid and delve deeper into the gauge structure of the theory and its meaning in the context of amorphous solids in Section~\ref{sec:tensorEM_emergent}.  We make predictions about dynamics, assuming that the analog of Ampere's law exists in VCTG and mapping the magnetic charge to momentum density. We also discuss inertial vs non-inertial dynamics and the roles played by the  electric and magnetic sectors of the theory for these two different classes. Following the dynamics discussion, we address the mechanical interpretation of the gauge structure underlying VCTG, and the connection to strain tensors and displacement fields. Various technical details and additional information are provided in different appendices.

\vspace{-4mm}

%%%%%%%%%%%%%%%%%%%

\section{Elasticity of solids and the granular problem}
\label{sec:problem}
{We begin with a brief summary of the classical theory of elasticity~\cite{landau2012theory,chaikin1995principles}-- one of the most well established field theories in physics, since the challenges faced in constructing a theory of granular mechanics are best understood by contrasting such solids with crystals.} The theory of elasticity of crystalline solids emerges from the spontaneous breaking of translation symmetry {and associated  rigidity}~\cite{chaikin1995principles,landau2012theory,anderson2018basic} that establishes relationships between the stress ($\hat{\sigma}$) and strain ($\hat{\gamma}$) tensors:
\begin{subequations}\label{eq:continuum_elasticity}
\begin{align}
   \partial_i \sigma_{ij} &= f_j^{~\rm{external}},\label{eq:mecheq}\\
     \gamma_{ij}=\frac{1}{2}(\partial_i u_j+\partial_j u_i) &\implies \epsilon_{iab} \epsilon_{jcd} \partial_a \partial_c \gamma_{bd}=0,\label{eq:compatibility}\\
     \sigma_{ij} =& K_{ijkl} \gamma_{kl}, \label{eq:stress_strain}
\end{align}
\end{subequations}
Eq.~\eqref{eq:mecheq} follows from momentum conservation and relates the stress of a volume element at rest to external forces, $f_i$, via Newton's law. The response of the particles in a crystal are assumed to be affine, which leads to Eq.~\eqref{eq:compatibility} relating the macroscopic strain tensor to the displacement field, $\boldsymbol{u}$, measured from the unique crystalline reference structure. The fact that the strain is derivable from a displacement field then leads to the elastic compatibility relation~\cite{landau2012theory} implied in Eq.~\eqref{eq:compatibility}. The constitutive relation of linear elasticity is then derived from a free energy, $\mathcal{F}$, through the relation $\sigma_{ij} = \frac{\partial \mathcal{F}}{\partial u_{ij}}$, leading to Eq.~\eqref{eq:stress_strain} with $K_{ijkl}$ being  the elastic moduli. 

Eqs.~\eqref{eq:continuum_elasticity} provide more than enough equations to solve for the $d(d+1)/2$ independent components of the $d \times d$, symmetric stress tensor for any imposed external force.  The consequences are well known: (a) the elastic moduli are purely  material properties (with the total number of independent moduli being fixed by the space-group of the crystal~\cite{ashcroft1976solid}), (b) stress-stress correlations exhibit power-law decay, (c) the response to any external force is completely determined by the elastic Green's function obtained from Eqs.~\eqref{eq:mecheq} and \eqref{eq:compatibility} and (d) there is a non vanishing sound speed, related to the elastic moduli.

The above discussion reinforces the well-established fact that it is the emergence of  long-range order and the associated rigidity~\cite{anderson2018basic} of crystals that distinguishes its mechanical response from fluids, which are disordered.  Unlike elastic solids, fluids cannot sustain a shear stress and flow in response to it. 
 
\vspace{2mm}

\paragraph*{Challenges of athermal elasticity and various approaches: \label{subsec:amorphous_elasticity}}

Applying the above paradigm of crystalline elasticity  to athermal disordered solids (see Fig.~\ref{fig:network} for an example of a two dimensional (2D)  jammed packing of soft disks) is fraught with  difficulties~\cite{Behringer_2018,Bouchaud2002}. To begin with, there is no obvious broken symmetry and hence  no unique reference structure about which displacements can be defined. Alternatively stated, each mechanically balanced jammed configuration is equally eligible as a  reference configuration. {This allows for} a redundancy, in the definition of displacements that the coarse-grained {\it observables} are expected to be insensitive to. Further, there is no free-energy functional since these solids are not in thermal equilibrium.  Therefore, a rigorous basis for a stress-strain constitutive relation is lacking. For granular materials, there are additional difficulties because of the frictional and/or purely compressive nature of inter-particle interactions-- the configurations are only well defined in presence of external boundary forces~\cite{Otto2003}.

Intriguingly, however, observations of stress response in jammed solids, including frictional, granular solids indicate that the response is elastic {over some range of imposed shear although plasticity inevitably sets in at some critical {\it yield} stress. In the {\it elastic} regime, the elastic moduli appear to be dependent on preparation protocols used to create the jammed states, and they do not necessarily obey the symmetries that emerge from a free-energy function in thermal solids.  A well-know example is the response of a sandpile to a point force, where it is well established that the preparation conditions can qualitatively change the response~\cite{Behringer_2018,Bouchaud2002,Geng2001,Otto2003}. Theoretical ~\cite{Henkes2009,deGiuli_PRL} and numerical studies~\cite{Lemaitre2018} have shown that the stress-stress correlations of amorphous solids exhibit the same power-law decay as in elasticity theory while the stress correlations in granular media  are characterized by a pinch-point singularity in Fourier space~\cite{Henkes2009,Lois2009}.}

{The puzzle of  elasticity of jammed solids is rooted in the question of how non-periodic networks, which are apparently indistinguishable from a snapshot of a fluid can sustain shear?  The resolution to this puzzle lies in special properties of the rigid network that are uncovered through the use of tools of rigidity percolation~\cite{Moukarzel:1995aa,vanhecke_rigidity} and Maxwell counting~\cite{lubensky_kane}. Disordered networks require special properties to possess states of self-stress that are necessary to support external loads. The only loads that a disordered network can support, are the ones that have some overlap with these states of self-stress~\cite{lubensky_kane}.}
It is thus plausible to ask if coarse grained field theoretic descriptions exist that can account for the non-trivial properties of jammed networks and also account for their mechanical responses~\cite{Behringer_2018}? Many questions  naturally follow.  If such a description exists, then what is the nature and universal features of such a field theory and what are the {\it appropriate variables} that one should use to derive such field theories that account for the underlying kinetic constraints placed by the network? It appears, that any attempt to construct such a field theoretic framework for the mechanical properties of jammed solids must essentially answer: (a) how to obtain the stress state within a continuum formulation and (b) how to incorporate information about the structural disorder at the microscopic scale into a continuum formulation, {correctly accounting} for the kinetic constraints. { It is due to this imposition of kinetic constraints that the theory of elasticity of jammed solids is related to that of spin-ice and other frustrated systems~\cite{chalker2011geometrically,moessner2011quantum}, as discussed in Section~\ref{sec:roadmap}. }

%%%%%%%%%%%%%%
\vspace{-3mm}
{{\section{Reformulation of the Theory of Mechanical Response of Jammed Solids}~\label{sec:reformulation_athermal}}
We now construct a  theory of the stress response of jammed solids in terms of a mapping to a generalised theory of  electromagnetism involving vector charges. This section provides details of the structure introduced in reference ~\cite{PhysRevLett.125.118002}, and extends it to dynamics}. We use granular solids as the paradigmatic example but, as will be clear from our discussions, the considerations are much more general--{\it e.g.}, the arguments can be easily generalized to gels or other amorphous solids at zero temperature.
 
In a jammed {solid}, every grain is in mechanical equilibrium and thus satisfies the following constraints of force and torque balance (see Fig.~\ref{fig:network}). In the presence of a body force, $\boldsymbol{f}$, such as gravity, grain $g$, satisfies:
\begin{eqnarray}
\sum_{c \in g} \boldsymbol{f}_{g,c}  =\boldsymbol{f},~~~~~\sum_{c \in g} \boldsymbol{r}_{g,c} \times \boldsymbol{f}_{g,c}  = 0,
\label{eq:forcebal}
\end{eqnarray}
respectively. Here, $\boldsymbol{f}_{g,c}$ is the contact force on the grain, $\boldsymbol{r}_{g,c}$ is the vector joining the center of grain $g$ to the contact $c$ (Fig.~\ref{fig:network}), and the sum is over all contacts belonging to $g$. The force-moment tensor for a contact is:
\begin{equation}
\hat {\sigma}_{g,c} =  \boldsymbol{r}_{g,c} \otimes \boldsymbol{f}_{g,c}.
\label{eq:stress_micro}
\end{equation}
Note that because $\boldsymbol{f}_{g,c}$ is defined as the force on the grain, the above definition of stress is negative to the conventional definition~\cite{landau2012theory} and follow the conventions used in granular mechanics~\cite{Behringer_2018}. A coarse-grained stress tensor field, $\hat {\sigma} (\boldsymbol{r})$ is obtained by summing $\hat {\sigma}_{g,c}$ over all {the contact points, $g,c$, included in a coarse-graining volume $\Omega_r$,  centered at $\boldsymbol{r}$:
\begin{equation}
\hat {\sigma} (\boldsymbol{r}) = \frac{1}{\Omega_r} \sum_{g,c \in \Omega_r} \boldsymbol{r}_{g,c} \otimes \boldsymbol{f}_{g,c}.
\label{eq:stress_coarse}
\end{equation} 
{This stress tensor is not symmetric for grains with frictional interactions~\cite{Goldenberg2002}. However, since grains are in force and torque balance, the grain-level stress tensor $\hat \sigma_g \propto \sum_{c \in g} \boldsymbol{r}_{g,c} \otimes \boldsymbol{f}_{g,c}$ is symmetric. Therefore, the antisymmetric contribution to $\hat {\sigma} (\boldsymbol{r})$ arises only from contacts at the boundary (see Fig.~\ref{fig:network}). The implications of this will become clearer as we investigate conservation principles of the emergent gauge theory.}}

{The local constraints, upon coarse-graining,  lead to Eq.~\eqref{eq:mecheq} for the stress tensor.} 
The torque balance condition ensures that the stress defined {in the bulk of $\Omega(r)$} {(for example counting only the green ``bulk" grains of Fig.~\ref{fig:network})} is a symmetric tensor:  $\sigma_{ij}=\sigma_{ji}$. {The symmetric property and Eq.~\eqref{eq:mecheq} leads to a mapping of the force and torque balance conditions to charge and charge-angular momentum conservation in VCT with  Eq.~\eqref{eq:mecheq} mapping to the electric Gauss's law (see Appendix~\ref{appendix:A}).
 
The question we now address is whether the dynamical generalization of Eq.~\eqref{eq:mecheq}, Newton's second law, 
\begin{align}
    \partial_i\sigma_{ij}=f_j-\partial_t\pi_j,
    \label{eq:newton}
\end{align}
can me mapped to the Maxwell's equations of VCT. In Eq. \eqref{eq:newton}, $\pi$ denotes the momentum density. Note that the relative signs are different from usual definition~\cite{pretko_radzhihovsky} due to the opposite convention for the stress defined in Eq.~\eqref{eq:stress_micro}. Also, Eq.~\eqref{eq:mecheq} does not uniquely fix the stress field, since:
\begin{align}
    \sigma_{ij}\rightarrow \sigma_{ij}+\xi_{ij},
    \label{eq_xidef}
\end{align}
is also a valid solution where the field $\xi_{ij}$ has zero divergence, {\it i.e.}, $\partial_i\xi_{ij}=0$. The space of the solution for the zero divergence field $\xi_{ij}$ is large and in case of crystalline elasticity this can be related to a general function of strain~\cite{landau2012theory,PhysRevA.6.2401}.  We exploit this freedom (eq. \eqref{eq_xidef}) for the dynamical case of Eq. \ref{eq:newton} to define: 
\begin{align}
    \partial_{i}\xi_{ij}+\partial_t\pi_j=0,
    \label{eq:faraday1}
\end{align}
which reduces to a zero divergence condition for the static case while the stress continues to obey Eq.~\eqref{eq:mecheq}. Any mapping of the dynamics  to VCT Maxwell's equations, needs the analog of a magnetic field, in addition to an electric field. As we show below, a mathematically consistent set of equations that reproduce three of the VCT Maxwell's equations can be constructed by using the $\xi$ field, and a symmetric tensor field
}
% Centrally, Eq.~\eqref{eq:mecheq} does not uniquely fix the stress field as is well known from the theory of elasticity~\cite{landau2012theory} since:
% \begin{align}
%     \sigma_{ij}\rightarrow \sigma_{ij}+\xi_{ij},
%     \label{eq_xidef}
% \end{align}
% is also a valid solution where the field $\xi_{ij}$ has zero divergence, {\it i.e.}, $\partial_i\xi_{ij}=0$. The space of the solution for the zero divergence field $\xi_{ij}$ is large and in case of crystalline elasticity this can be related to a general function of strain~\cite{PhysRevA.6.2401}. In fact this can be fruitfully applied even to the dynamical version of Eq.~\eqref{eq:mecheq} which is given by
% \begin{align}
%     \partial_i\sigma_{ij}=f_j-\partial_t\pi_j,
%     \label{eq:newton}
% \end{align}
% where $\pi_j$ is the momentum density. Note that the relative signs are different from usual definition~\cite{pretko_radzhihovsky} due to the opposite convention for the stress defined in Eq.~\eqref{eq:stress_micro}. $\xi_{ij}$, defined via Eq.~\eqref{eq_xidef}, now obeys:
% Eq.~\eqref{eq:newton}, 
% \begin{align}
%     \partial_{i}\xi_{ij}+\partial_t\pi_j=0,
%     \label{eq:faraday1}
% \end{align}
% which reduces to a zero divergence condition for the static case while the stress continues to obey Eq.~\eqref{eq:mecheq}.
$B_{ij}=B_{ji}$ such that 
\begin{align}
   \partial_iB_{ij}=\pi_j. 
   \label{eq:mag}
\end{align} 
We can then rewrite Eq.~\eqref{eq:faraday1} as 
\begin{align}
    \partial_{i}(\xi_{ij}+\partial_tB_{ij})=0.
    \label{eq:faraday2}
\end{align}
One solution can be obtained by setting:
\begin{align}
    \xi_{ij}+\partial_tB_{ij}=0.
    \label{eq:faraday3}
\end{align}
Any function $\xi_{ij}$ that satisfies the above equation is valid. By drawing a parallel with crystalline elasticity~\cite{landau2012theory}, we can choose $\xi_{ij}$  to be of the form:
\begin{align}
    \xi_{ij}=\partial_a\partial_c\left(\epsilon_{iab}\epsilon_{jcd}\sigma_{bd}\right),
    \label{eq:choice}
\end{align}
such that Eq.~\eqref{eq:faraday3} now can be written as
\begin{align}
    \epsilon_{iab}\epsilon_{jcd}\partial_a\partial_c\sigma_{bd}+\partial_tB_{ij}=0.
    \label{eq:faradayfinal}
\end{align}
At first glance, the choice of $\xi_{ij}$ in Eq.~\eqref{eq:choice} seems to be in contradiction with Eq.~\eqref{eq:faraday1} since  the divergence of $\xi_{ij}$  appears to be identically zero. From our extensive knowledge of topological defects in effective field theories with a UV cut-off~\cite{RevModPhys.51.591} scale, however,  we know that fields such as $\sigma_{ij}$ can have singular contributions. {Taking these singular contributions into account,} Eq.~\eqref{eq:faraday1} has the connotation of the conservation of the magnetic charge, whose density is given by the momentum density and whose current is related to the singular contributions of the stress field:
\begin{align}
    \tilde{J}_{ij}=\epsilon_{iab}\epsilon_{jcd}\partial_a\partial_c\sigma^{\rm singular}_{bd} ~.
    \label{current_magnetic}
\end{align}
Eqs.~\eqref{eq:mecheq},~\eqref{eq:mag} and~\eqref{eq:faradayfinal} directly map to three of the  Maxwell's equations of VCT: Eqs.~\eqref{eq:mecheq},~\eqref{eq:mag}, are the electric and magnetic Gauss's laws, respectively, and~\eqref{eq:faradayfinal} is the Faraday's law (see Appendix~\ref{appendix:A}).  {The analog of Ampere's law is missing and its origin is not clear in our current  understanding. Yet, appealing to the complete gauge structure, we postulate the existence of such a fourth equation-- the Ampere's law
\begin{align}
    \epsilon_{iab}\epsilon_{jcd}\partial_a\partial_c B_{bd}&=\partial_t \sigma_{ij}+J_{ij}
    \label{eq_ampere_vctg}
\end{align}
where $J_{ij}$ is the total electric current that ensures the conservation of electric charge (force-balance), we have the full set of ``Maxwell's equations''. 
In Section \ref{sec:tensorEM_emergent} below, we revisit the entire question of dynamics in granular solids in context of VCTG and discuss a consequence of the full Maxwell structure-- the emergent photons-- for the dynamics of jammed granular solids.}

{It is important to note that the well known fact, that the jammed solid has uniform density and conserves particle number leads to the zero divergence of momentum density, $\partial_i\pi_i=0$ via the usual continuity equation. This, along with the magnetic Gauss's law (Eq.~\eqref{eq:mag}) leads to $\partial_i\partial_jB_{ij}=0$ whence the magnetic field can be written as a double curl of a locally defined tensor gauge potential (see Appendix \ref{appendix:A}), except at the magnetic charge, similar to usual electromagnetism \cite{jackson}.
}

The above mapping of mechanical properties to VCT is what we refer to as VCTG. {In passing we note that in 2D, Eq.~\eqref{eq:mecheq} {\it with $f_j =0$ }can be mapped to the {\it Faraday's equation} for a version of tensor gauge theories with scalar charges~\cite{pretko_radzhihovsky}. This mapping has been used to study defects in crystals with the scalar charges representing disclinations and differs from our VCTG fundamentally because their formulation explicitly uses the strain field, well-defined in crystalline structures unlike the present case of jammed amorphous solids.}

In this paper we focus primarily on the study of the static limit of VCTG in two and three dimensions  to show that it describes the elastic response of jammed solids. We briefly return to the issue of Ampere's law and its possible implications for the dynamical response of jammed solids in Section~\ref{sec:tensorEM_emergent}.

%%%%%%%%%%%%%%%%%%
\vspace{-4mm}
\subsection{The electrostatic limit}
\label{sec:electrostatic}
  
{From the discussion above, elasticity of jammed solids appears to be directly related to the electrostatic limit of the Maxwell's equation of VCT, if we follow the tentative mapping in which the force in Eq.~\eqref{eq:mecheq} maps to the vector electric charge and the coarse-grained stress maps to the electric field due to this charge:}
\begin{align}
     f_i \xrightarrow{\mathrm{?}} \rho_i,~~~~\sigma_{ij} \xrightarrow{\mathrm{?}}  E_{ij}.
     \label{eq:mapping}
 \end{align}
The static limit of Eqs.~\eqref{eq:newton} and~\eqref{eq:faradayfinal} then are in exact analogy with Eqs.~\eqref{eq:vectorelectrostatics} of the vector charge theory. With this mapping to electrostatics, the stress tensor satisfies  $\partial_i\sigma_{ij}=f_j$ and $\epsilon_{iab}\epsilon_{jcd}\partial_a\partial_c\sigma_{bd}=0$. 
As already summarised in our earlier paper~\cite{PhysRevLett.125.118002}, the two conservation laws of the VCT (Eqs.~\eqref{eq:chargeconserve} and \eqref{eq:amcon}) directly correspond to the conditions of force and torque balance in the bulk of the granular material {\it i.e.}
\begin{equation}
    \int d^3\boldsymbol{r}~\rho_i(\boldsymbol{r})=0,~~~~\int d^3\boldsymbol{r}~\epsilon_{ijk}r_j\rho_k(\boldsymbol{r})=0.
\end{equation}
The issue of conservation at the boundary, however, is a bit more subtle, and is related to how granular networks develop in response to external stresses-- a topic that we now turn to.
%%%%%%%%%%%%%%%%%%%%%%%

\subsection{The VCT Dielectric}

\begin{table*}[htbp!]
\begin{tabularx}{\textwidth}{|X | X|}
\hline
&\\
\multicolumn{1}{|>{\large}c|}{\textbf{Vector Charge Theory (VCT)}}  & \multicolumn{1}{>{\large}c|}{\textbf{Emergent Elasticity (VCTG)}} \\
&\\
\hline
\multicolumn{2}{|c|}{}\\
\multicolumn{2}{|c|}{\textbf{Vacuum Formulation (only free charges present)}}\\
\hline
Electric Field tensor: $E_{ij}$ & Stress Tensor  $\sigma_{ij}$ \\
\hline
Electric charge: $\rho_i$ & Force on grain  $f_{i}$ \\
\hline
Gauss's Law: $\partial_i E_{ij} = \rho_j$ &  $\partial_i \sigma_{ij} = f_j$ \\
\hline
Charge conservation & Force balance \\
\hline
Charge angular momentum conservation & Torque balance \\
\hline
Magnetic charge: $\tilde{\rho}_i$ & Momentum density  $\pi_{i}$ \\
\hline
Magnetic Gauss's law: $\partial_i B_{ij}=\tilde{\rho}_j$ & $\partial_i B_{ij}= \pi_j$\\
\hline
Faraday's Law: $\partial_a \partial_c ( \epsilon^{i a b} \epsilon^{j c d} E_{b d}) = -\tilde{J}_{i j}-\partial_t B_{ij}$ & $\partial_a \partial_c ( \epsilon^{i a b} \epsilon^{j c d} \sigma_{b d}) = -j_{i j}-\partial_t B_{ij}$\\
\hline
Ampere's law: $ \partial_a \partial_c ( \epsilon^{i a b} \epsilon^{j c d} B_{b d}) = {J}_{i j}+\partial_t E_{ij}$ & $ \partial_a \partial_c ( \epsilon^{i a b} \epsilon^{j c d} B_{b d}) = {J}_{i j}+\partial_t \sigma_{ij}$\\
\hline
Magnetic charge conservation: $\partial_t \tilde{\rho}_j +\partial_i \tilde{J}_{ij}=0$ & Momentum Conservation: $\partial_t \pi_j +\partial_i j_{ij}=0$ \\
\hline
Magnetic current: $\tilde{J}_{ij}$ & $j_{ij} = \epsilon^{i a b} \epsilon^{j c d} \partial_a \partial_c \sigma_{bd}^{\rm{singular}}$\\
\hline
\multicolumn{2}{|c|}{}\\
\multicolumn{2}{|c|}{\textbf{Dielectric Formulation (both free and bound charges are present)}}\\
\hline
Bound charges are those created in response to external charges & Contact forces are created in response to imposed external forces \\
\hline
$\boldsymbol{\rho} = \boldsymbol {\rho}_{\rm free} +  \boldsymbol {\rho}_{\rm bound}$ & $\boldsymbol {f} = \boldsymbol{f}_{\rm external} + \boldsymbol {f}_{\rm contact}$ \\
\hline
Dipole Moment:  a tensor $P_{ik} = \rho_i  d_k$ &  At every contact $(\sigma_c)_{ik} = (f_c)_i (r_c)_k $ \\
\hline
$\partial_i P_{ij} =- (\rho_j)_{\rm bound}$  &  $\partial_i (\sigma_c)_{ij} = -(f_{\rm contact})_j$  \\
\hline
Polarizable elements: molecules etc.  &  Contacts between grains: unpolarized contacts are ones where the grain shapes are not distorted (no force at that contact) \\
\hline
Electric Displacement Field tensor: $D_{ij}=E_{ij}+P_{ij}$ & Stress Tensor  $\sigma_{ij}=E_{ij}+(\sigma_c)_{ij}$; $E_{ij}= \frac{1}{2}(\partial_i \varphi_j + \partial_j \varphi_i)$. \\
&$\varphi_i$ is a gauge potential that plays the role of $u_i$ the displacement vector in crystals.\\
\hline
Gauss's Law: $\partial_i D_{ij} = \rho_j^{\rm free}$ &  $\partial_i \sigma_{ij} = f_j^{\rm external}$ \\
\hline
\end{tabularx}
\caption{The dictionary\label{tbl:dictionary} of the Vector Charge Theory (VCT) - Vector Charge Theory of Granular Mechanics (VCTG) mapping. Eq. \eqref{eq_correspondance} and the discussion following it emphasises the difference between VCTG and classical elasticity theory.}
\vspace{-2mm}
\end{table*}

%%%%%%%%%%%%%%%%%%%
A crucial difference, which we already alluded to in the introduction as well as  in reference~\cite{PhysRevLett.125.118002}, is that a granular solid is  well defined {\it only} in presence of rigid boundary conditions where external force ({\it external charge}) is applied. Further, in the presence of friction between the grains, the boundary also supports a non-zero external torque. The mechanical response in the bulk therefore, depends on the details of these boundary forces and torques. {In other words, it is the boundary forces/charges that give rise to non-zero stress/electric field inside the bulk.} It is therefore desirable to formalise the above electromagnetic mapping  by including the external boundary charges-- a situation similar to that of electromagnetism in the presence of a dielectric medium.  {Therefore we now turn to the  dielectric formulation of the VCT.}

Following standard electrostatics~\cite{jackson}, we divide charges in a dielectric into
\begin{align}
    \rho_i=\rho_i^{\rm free}+\rho_i^{\rm bound},
\end{align}
such that the Gauss's law is now written as
\begin{align}
    \partial_i E_{ij}=\rho_i^{\rm free}+\rho_i^{\rm bound}.
    \label{eq:displacement_1}
\end{align}

The tensor dipole moment of such a charge distribution (see Appendix~\ref{appendix:A}) is given by,
\begin{align}
    P_{jk}=\int d^3~{\bf r'}~r'_k~\rho_j({\bf r'}).
    \label{eq:dipole_exp}
\end{align}
It is immediately clear that the anti-symmetric part of the net dipole moment is nothing but the charge-angular momentum -- the total torque in granular elasticity {\it i.e.},
{\small
\begin{align}
    P^A_{jk}=\frac{P_{jk}-P_{kj}}{2}=\frac{1}{2}\int d^3\boldsymbol{r}~(\rho_j r_k-\rho_k r_j)=\frac{1}{2}\epsilon_{ijk}\mathcal{T}_i,
    \label{eq_pator}
\end{align}
}
where $\mathcal{T}_i$ is the torque. 
The torque balance, in addition to the force,  in mechanical equilibrium thus has a natural interpretation within VCT. Notably however, the anti-symmetric part is not only conserved, but, is a purely a boundary term as is clear from Eq.~\eqref{eq:surfacedipole}. {This result provides a natural explanation for the hyper-uniformity of torque fluctuations }observed in reference~\cite{PhysRevLett.126.075501}. For frictionless grains, however, the torque is identically zero.

The multipole expansion within VCT, discussed in Appendix~\ref{appendix:A}, leads to:
\begin{align}
    \rho_i^{\rm bound}=-\partial_j\mathcal{P}_{ij}(\boldsymbol{r}),
    \label{eq:bound_dipole}
\end{align}
where $\mathcal{P}_{ij}(\boldsymbol{r})$ is the dipole moment density. Eq.~\eqref{eq:displacement_1} can now be re-written as
\begin{align}
    \partial_i D_{ij} =\rho^{\rm free}_j,~~~{\rm with}~~~
    D_{ij}=E_{ij}+\mathcal{P}_{ij}
     \label{eq_displacementgauss}
\end{align}
where $D_{ij}$ is the tensor electric displacement field.
For a linear dielectric, similar to reference~\cite{PhysRevLett.125.118002}, the polarizability is proportional to the electric field,
\begin{align}
    \mathcal{P}_{i j} =\chi_{i j k l} E_{k l},
    \label{eq:suscept}
\end{align}
where $\chi$ is the polarizability tensor. Using Eq.~\eqref{eq:suscept}, the electric displacement tensor can be re-written as 
 \begin{align}
     D_{ij} = (\delta_{ijkl} + \chi_{ijkl}) & E_{kl} \equiv (\Lambda^{-1})_{ijkl}E_{kl}.
     \label{eq:E_D_relation}
 \end{align}   
{with $\Lambda^{-1}$ being a rank-4 dielectric tensor.} Since $E_{i j}$ is symmetric, { $\chi_{i j k l}$ is given by a linear combination of symmetric and anti-symmetric parts in the indices $i$ and $j$ as}
\begin{align}
    \chi^{A}_{i j k l} = \chi_{i j k l} - \chi_{j i k l},~~~\chi^{S}_{i j k l} = \chi_{i j k l} + \chi_{j i k l}.
\end{align}
This leads to symmetric and anti-symmetric contributions to $\mathcal{P}_{i j}$: 
\begin{align}
    \mathcal{P}^{A}_{i j} = \chi^{A}_{i j k l} E_{k l},~~~~\mathcal{P}^{S}_{i j} = \chi^{S}_{i j k l} E_{k l}.
\end{align}
These antisymmetric contributions imply that, for frictional grains,  a boundary force can lead to a finite antisymmetric polarisation, {\it i.e.} the boundary torque, which may lead to shearing of the boundary layer~\cite{Artoni,Shojaaee}.

Therefore, the field equations of the VCT for a linear dielectric in the static limit are given by Eqs.~\eqref{eq_displacementgauss} and \eqref{eq:E_D_relation} along with,
\begin{align}
     %\partial_i D_{ij} &= \rho_j^{\rm{free}},\nonumber\\
     E_{ij}=\frac{1}{2}(\partial_i\varphi_j+\partial_j\varphi_i) &\implies \epsilon_{iab} \epsilon_{jcd} \partial_a \partial_c E_{bd}=0.%,\nonumber\\
    % D_{ij} = (\delta_{ijkl} + \chi_{ijkl})& E_{kl} \equiv {\Lambda^{-1}_{ijkl}}E_{kl}.
    \label{eq:linear_dielectric_field_eqns}
\end{align}
Comparing the structure of the above {\it gauge theory} with that of  crystalline elasticity presented in Eq.~\eqref{eq:continuum_elasticity}~\cite{landau2012theory}, 
{shows a clear correspondence\label{mapping} between VCT and the  theory {of elasticity:}
% {we can\label{mapping}, therefore,  construct a mapping between VCT and the  theory {of elasticity as 
\begin{align}
    \hat{D} \leftrightarrow \hat{\sigma},~~\hat{E} \leftrightarrow \hat{\gamma},~~\Lambda^{-1}_{ijkl} \leftrightarrow K_{ijkl}.
\label{eq_correspondance}
\end{align}
}
The important distinction with classical elasticity is the replacement of the strain tensor by $\hat{E}$ {along with the fact that the displacement fields are replaced by gauge potentials ($\boldsymbol{\varphi}$) and hence do not have any observable consequence}. This formulation, therefore, addresses the lack of a canonical definition of a strain tensor $\hat{\gamma}$, arising from the fact that there is no unique reference state about which we can define displacement fields $\boldsymbol{u}$. {Further, we note that because of the antisymmetric contribution to $\mathcal{P}_{ij}$, the { electric displacement field, and hence the stress tensor} is not necessarily symmetric but can have an antisymmetric boundary contribution. {This stress-only elasticity formulations is what we refer to as VCTG.} Indeed, in frictional granular materials, the possibility of an anti-symmetric contribution  to the stress tensor has been widely recognized and is often addressed via the theoretical framework of Cosserat elasticity~\cite{forest2005mechanics}. The VCTG formulation clearly identifies the anti-symmetric boundary contribution to $\hat{\sigma}$ as arising from the dipole moment tensor $\hat{P}$.} {To summarize the above discussion, we have shown that mapping to a U(1) gauge theory with vector charges, provides a fully consistent `` stress-only elasticity'' framework for jammed solids} captured by the equations:
\begin{align}
     \partial_i \sigma_{ij} &= f_j^{\rm~ external},\nonumber\\
     E_{ij}=\frac{1}{2}(\partial_i\varphi_j+\partial_j\varphi_i) &\implies \epsilon_{iab} \epsilon_{jcd} \partial_a \partial_c E_{bd}=0,\nonumber\\
     \sigma_{ij} = (\delta_{ijkl} + \chi_{ijkl})& E_{kl} \equiv {\Lambda^{-1}_{ijkl}E_{kl}.}
    \label{eq:VCTG_field_eqns}
\end{align}
 A detailed glossary of the mapping of various quantities of interest in VCTG, a stress-only formulation of granular elasticity, to its  counterparts in VCT is given in Table~\ref{tbl:dictionary}.

The VCTG framework unifies existing theories that use stress-based, gauge potentials~\cite{Ball2002,Henkes2009,deGiuli_PRL,DeGiuli2020} under one common umbrella. The structure of the theory is completely parallel to that of elasticity theory with two crucial differences:  (i) the analog of the strain tensor is constructed from gauge potentials and {\it not} from physical displacement fields that refer to a preferred stress-free structure, and (ii) the elastic modulus tensor emerges from the properties of the networks that are created by boundary stresses and thus depend both on preparation protocols and material properties (force laws). Experimental protocols that involve boundary strains can be handled within our theory by translating these to boundary forces that are created by these strains. In this context, it is intriguing to note that the phenomenon of shear thickening is suspensions is much better characterized by imposed stresses than by imposed shear rates~\cite{Brown_2014}. In Section~\ref{sec:numerical}, we discuss numerical studies of jammed frictionless granular solids that test the predictions of the theory presented above.
%%%%%%%%%%%%%%%%%%%%%%%%%%

\vspace{-5mm}

{\section{Stress correlations, pinch points and ``emergent" elastic moduli}~\label{sec:correlations}}
Having outlined the mapping of the mechanical response of granular solids to the VCT, we now calculate observables that can be directly compared with experiments and numerical simulations of jammed solids. To obtain the stress-stress correlations for an arbitrary $\hat{\Lambda}$, we begin with the Landau-Ginzburg action that gives rise to the field equations of the VCTG given in Eq.~\eqref{eq:VCTG_field_eqns}:
\begin{equation}
\vspace{-1mm}
\mathcal{S} = \int d^{d} \boldsymbol{r} ~\frac{1}{2g}\sigma_{ij}(\boldsymbol{r}) E_{ij}(\boldsymbol{r})
\label{eq:lagrangian_density}
\end{equation}
supplemented with the Gauss's law in Eq.~\eqref{eq_displacementgauss}. In the above equation we have explicitly used $\sigma_{ij}$ instead of $D_{ij}$ for familiarity, keeping in mind the mapping in Eq.~\eqref{eq_correspondance}.

This action formulation is based on the assumption that all configurations that have the same stress state, and satisfy the kinetic constraints of force and torque balance are equally likely. This equiprobability ansatz is the basis of the Edwards stress ensemble~\cite{Blumenfeld:2009aa,Bi2015,DeGiuli2020}. In a more general context, this ansatz is an extension of the theories of frustrated magnets and hard-core dimer models, where, in the absence of any imposed bias,  all states in the degenerate low energy manifold are taken to be equally likely~\cite{chalker2011geometrically,moessner2011quantum}.  Jammed solids with no body forces such as gravity correspond to the charge free sector of VCTG. In this limit Eq.~\eqref{eq_displacementgauss} reduces to $\partial_{i} \sigma_{ij} = 0$, 
subject to the  boundary conditions, which can produce a non-zero stress field in the system.

Similar to conventional  electromagnetism, the electrostatic potentials can be constructed in two ways, either by imposing the zero divergence condition or the zero (double) curl condition. The latter leads to the $\boldsymbol{\varphi}$ potentials (Eq.~\eqref{eq:VCTG_field_eqns}), and is more suited to the calculation of the stress response. However, the stress-stress correlations are more conveniently computed by using potentials that solve the zero divergence condition, by expressing the stress tensor in terms of the following potentials~\cite{xu_dual}:
{
\begin{equation}
\Delta{\sigma_{ij}}=\left\{\begin{array}{l}
\epsilon_{ia}\epsilon_{jb} \partial_{a} \partial_{b} \psi~~~~~~~~~~~d=2\\
\epsilon_{iab}\epsilon_{jcd}\partial_a\partial_c\psi_{bd}~~~~~~d=3\\
\end{array}\right.
\label{eq:potential_form}
\end{equation}
Here $\Delta \sigma_{ij} \equiv \sigma_{ij}-\llangle \sigma_{ij} \rrangle$
represents the fluctuation of the stress tensor with respect to the mean value of each packing.}
In the context of elasticity theory, this potential formulation is akin to the well-known Airy stress function~\cite{chaikin1995principles} in $d=2$ and Beltrami stress function~\cite{Gurtin1973} in $d=3$. The potential formulation using $\psi_{ab}$ as defined above is dual to the potential formulation using $\varphi_a$, defined in Eq.~\eqref{eq:VCTG_field_eqns}. 
In Appendix~\ref{appendix:A}, we present the stress-stress correlations derived using the $\varphi_a$-potential which are identical to those derived using the $\psi$ formulation, as expected.

In performing specific computations, it is convenient to represent the ${d(d+1)}/{2}$ components of the tensors $\sigma_{ij}$ and $E_{ij}$ as vectors, $\ket{\Sigma}$, and $\ket{\mathcal{E}}$, respectively, using the standard Voigt representation using the bulk symmetry:  $\sigma_{ij} = \sigma_{ji}$, $E_{ij} = E_{ji}$. We have used calligraphic symbols for the objects in Voigt notation (see Eqs.~\eqref{eq:voigt_notation_2d} and~\eqref{eq:voigt_notation} for explicit forms in $d=2$ and $3$ respectively) to indicate that the components of these vectors transform differently-- {\it e.g.}, the components $\Ket{\Sigma}$ transform as components of a rank-2 Cartesian tensor. Also, for notational simplification, we express the  polarization tensor $\Lambda_{ijkl}$, defined in Eq.~\eqref{eq:E_D_relation}, in the Voigt notation as $\hat{\Lambda}$:
\begin{equation}
\Ket{\mathcal{E}}=\hat{\Lambda}\Ket{\Sigma}.
\label{eq:D_E_new_relation_2d}
\end{equation}
Using this notation, and setting $g=1$, the action in ~\eqref{eq:lagrangian_density} can be written as:
\begin{eqnarray}
    \mathcal{S} &=& \int d^{d} \boldsymbol{r}~\Braket{\Sigma| \hat{R} |\mathcal{E}}.
    \label{eq:lagrangian_matrix_2d}
\end{eqnarray}
where $\hat{R}$ is a $\frac{d(d+1)}{2}\times \frac{d(d+1)}{2}$ diagonal matrix given by,
\begin{equation}
    R_{ii} = 
    \begin{cases}
    1~~~~ \textrm{for}~~~i =1,\dots,d\\
    2~~~~ \textrm{for}~~~i =(d+1),\dots,\frac{d(d+1)}{2}.
    \end{cases}
\end{equation}

Due to the translation invariance of the system, the computation of the correlations is simplified in Fourier space. Representing the action (Eq.~\eqref{eq:lagrangian_matrix_2d}) in terms of the Fourier transformed fields $\Ket{\tilde{\Sigma}(\boldsymbol{q})}$, we get 
{
\begin{eqnarray}
\mathcal{S} &=& \int d^{2} \boldsymbol{q} \Braket{ \tilde{\Sigma}(\boldsymbol{q}) | \hat{R} \hat{\Lambda} | \tilde{\Sigma}(-\boldsymbol{q})}.
\label{eq:lagrangian_matrix_fourier}
\end{eqnarray}
}%

Below, we derive the correlations using the above Landau-Ginzburg action in both 2D and 3D and then compare them with our numerical calculations (see Section~\ref{sec:numerical}). The correlations between the components of the stress tensor $\llangle \sigma_{ij}\sigma_{kl} \rrangle$ have six symmetry related combinations in 2D, and $21$ different combinations in 3D. We consider the two cases separately.

\vspace{4mm}
%%%%%%%%%%%%

\paragraph*{Two Dimensions}

In 2D, the components of the stress tensor can be expressed in Voigt notation as
{
\begin{equation}
    \Ket{\Delta \Sigma} =
    \begin{bmatrix}
    \Delta{\sigma_{xx}}\\
    \Delta{\sigma_{yy}}\\
    \Delta{\sigma_{xy}}
    \end{bmatrix}.
    % ~
    % \Ket{\Delta\mathcal{E}} =
    % \begin{bmatrix}
    % \Delta E_{xx}\\
    % \Delta E_{yy}\\
    % \Delta E_{xy}
    % \end{bmatrix}.
    \label{eq:voigt_notation_2d}
\end{equation}
Therefore, in Fourier space, in terms of the scalar potential, $\psi$, via Eq.~\eqref{eq:potential_form}, we have 
\begin{equation}
\Ket{{\Delta{\tilde\Sigma}}(\boldsymbol{q})} =\Ket{\mathcal{A}(\boldsymbol{q})} \tilde{\psi}(\boldsymbol{q})
~,~~
\Ket{\mathcal{A}(\boldsymbol{q})} = 
\begin{bmatrix}
q_y^2\\
q_x^2\\
-q_x q_y\\
\end{bmatrix}.
\label{eq:D_psi_relation_2d}
\end{equation}
}
where the $\tilde{\psi}$ is the scalar potential $\psi$ in Fourier space. The  partition function is given by
\begin{eqnarray}
Z = \int [\mathcal{D}\tilde{\psi}]~ e^{-\mathcal{S}[\tilde{\psi}]},
\label{eq_partition_2d}
\end{eqnarray}
where from Eq.~\eqref{eq:lagrangian_matrix_fourier}, we have
\begin{eqnarray}
    \mathcal{S} = \int d^{2} \boldsymbol{q} ~ \Braket{\mathcal{A}(\boldsymbol{q})|\hat{R}\hat{\Lambda}|\mathcal{A}(-\boldsymbol{q})} \tilde{\psi}(\boldsymbol{q})\tilde{\psi}(-\boldsymbol{q}).
\end{eqnarray}

{This immediately leads to
\begin{equation}
    \llangle[\Big]\tilde{\psi}\left(\boldsymbol{q}\right)\tilde{\psi}\left(-\boldsymbol{q}\right)\rrangle[\Big] = \Braket{\mathcal{A}(\boldsymbol{q})|\hat{R}\hat{\Lambda}|\mathcal{A}(-\boldsymbol{q})}^{-1}.
    \label{eq:psi_psi_correlation_2d}
\end{equation}
Using Eq.~\eqref{eq:D_psi_relation_2d}, we obtain the gauge-invariant stress correlators:
{\small
\begin{align}
 \llangle[\Big] \Delta\tilde{\Sigma}(\boldsymbol{q})_{i} \Delta\tilde{\Sigma}(-\boldsymbol{q})_{j} \rrangle[\Big] &= \mathcal{A}(\boldsymbol{q})_{i} \mathcal{A}(-\boldsymbol{q})_{j} \llangle[\Big]\tilde{\psi}\left(\boldsymbol{q}\right)\tilde{\psi}\left(-\boldsymbol{q}\right)\rrangle[\Big].
\label{eq:2d_stress_correlations}   
\end{align}}
}%
The correlations between the fluctuations of the components of the stress tensor $\llangle \Delta\sigma_{ij}\Delta\sigma_{kl} \rrangle$ is obtained using Eq.~\eqref{eq:voigt_notation_2d}. 
 The explicit forms for these correlations for a specific form of $\hat\Lambda^{-1}$ given in Eq.~\eqref{eq:lambda_2d_iso} (see the discussion below and in the next section about this form), are given in  Eq.~\eqref{eq:2d_correlations_explicit}. Notably, these correlations depend only on a particular ratio, $K_{2D}$,  of the elements of $\hat\Lambda^{-1}$. This theoretical prediction that all correlations scale in the same way provides a robust test of the theory, which we test via numerical simulations in the next section.

\begin{figure*}[!htbp]
\includegraphics[width=0.9\textwidth]{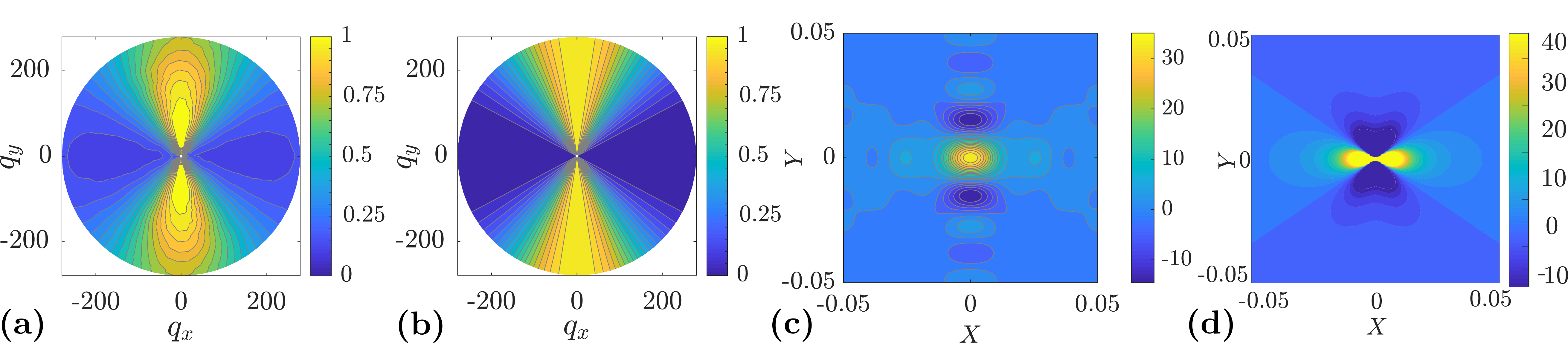}
\caption{{Comparison of the correlation function, $C_{xxxx}$ (Eq.~\eqref{eq_cxxxx}), obtained from numerical simulations with the theoretical predictions in 2D. Panels \textbf{(a)}  and \textbf{(b)} show the Fourier space results from numerics and theory respectively. Panels \textbf{(c)} and \textbf{(d)} show the real space correlations, obtained from numerical simulations and VCTG predictions respectively, close to the origin. There are approximately $9 \times 9$ grains in the region shown in panel \textbf{(c)}. The modulations seen are at the grain scale, reflecting the particulate nature of the jammed solid, and the large $q$ (ultra-violet) cutoff seen in panel (a). From panel \textbf{(d)}, it is clear that the VCTG describes the envelope of the power-law decays and the strongly anisotropic nature of the correlations, but fails to capture the grain-scale structure. The comparison given, is for a system of $8192$ grains with packing fraction $\phi = 0.88$, averaged over $239$ configurations under isotropic compression.} The full comparison for all the remaining $5$ correlations are given in Fig.~\ref{fig:2d_full}.
~\label{fig:2d_correlations}}
\vspace{-3mm}
\end{figure*}

An important feature of these correlations is their independence of $q \equiv |\boldsymbol{q}|$. They are only dependent on the angular variable $\theta$. This is clearly seen from Eqs.~\eqref{eq:D_psi_relation_2d}, \eqref{eq:psi_psi_correlation_2d}, \eqref{eq:2d_stress_correlations} and the explicit forms given in Eq.~\eqref{eq:2d_correlations_explicit}. For example, $\llangle  \Delta\sigma_{xx} \Delta\sigma_{xx} \rrangle \propto \sin^4\theta$ depends only on the polar angle $\theta$. Hence, the correlations display singular behavior as one approaches $q \rightarrow 0$ producing a pinch-point at $q=0$~\cite{Prem2018} (see Figs.~\ref{fig:2d_correlations} and~\ref{fig:2d_full} where this ``pinch-point" structure is clearly visible). Such pinch-point behaviour has previously been identified in the literature~\cite{Henkes2009,DeGiuli2020,Lois2009,McNamara2016,Wang:2020aa} as a salient feature of the stress correlations of granular systems. The current formulation shows that these features emerge purely from the requirement of gauge invariance, and even more importantly, VCTG allows us to compute the explicit angular distributions for {\it} any $\hat {\Lambda}^{-1}$, as shown in Appendix \ref{appendix:A}.
%%%%%%%%%%%%%%%%%%%%%

\vspace{4mm}

\paragraph*{Three Dimensions:}

In 3D, the components of the stress tensor can be expressed in Voigt notation as

{\small
\begin{equation}
    \Ket{\Delta \Sigma} =
    \begin{bmatrix}
    \Delta \sigma_{xx}\\
    \Delta \sigma_{yy}\\
    \Delta \sigma_{zz}\\
    \Delta \sigma_{xy}\\
    \Delta \sigma_{xz}\\
    \Delta \sigma_{yz}
    \end{bmatrix}.
    % ~~~~~~
    % \Ket{\Delta \mathcal{E}}=
    % \begin{bmatrix}
    % \Delta E_{xx}\\
    % \Delta E_{yy}\\
    % \Delta E_{zz}\\
    % \Delta E_{xy}\\
    % \Delta E_{xz}\\
    % \Delta E_{yz}
    % \end{bmatrix}.
    \label{eq:voigt_notation}
\end{equation}}
Using  Eq.~\eqref{eq:potential_form}, the action can be expressed in terms of the potential $\psi_{ab}$. The gauge redundancy of the theory implies that
the stress tensor is invariant under:  
\begin{align}
%    \psi_{aa}&\rightarrow \psi_{aa}+2\partial_a\lambda_a~~~~~~~~~~\forall~a=x,y,z\nonumber\\
    \psi_{ab}&\rightarrow \psi_{ab}+\partial_a\lambda_b+\partial_b\lambda_a,~~~\forall~a,b=x,y,z,
\end{align}
%\vspace{1mm}
where $\lambda_a$ is a 3-component vector field that generates the gauge transformation. We can use this freedom to choose a gauge to set either the three diagonal components or the three off-diagonal components of $\hat{\psi}$ to zero. In classical elasticity theory, these two choices correspond to the Maxwell stress function~\cite{sadd2009elasticity} (off-diagonal), and Morera stress function~\cite{sadd2009elasticity})(diagonal). Here we choose the ``Morera" form:
\begin{equation}
\vspace{3mm}
 \hat{\psi} = 
\begin{bmatrix}
0 & \psi_{xy} & \psi_{xz}\\
\psi_{xy} & 0 & \psi_{yz}\\
\psi_{xz} & \psi_{yz}& 0~
\end{bmatrix},
\vspace{-6mm}
\end{equation}\\
which in the Voigt notation is:
\begin{equation}
\Ket{\psi'} =
\begin{bmatrix}
\psi_{xy}\\
\psi_{xz}\\
\psi_{yz}\\
\end{bmatrix}.
\end{equation}

In Fourier space, analogous to Eq.~\eqref{eq:D_psi_relation_2d}, we have:
{\small
{
\begin{equation}
\Ket{\Delta\tilde{\Sigma}(\boldsymbol{q})} =\mathcal{A}(\boldsymbol{q}) \Ket{\tilde{\psi}'(\boldsymbol{q})},
\label{eq:D_psi_relation}
\end{equation}
}

and the transformation matrix in Fourier space is given by
\begin{equation}
\mathcal{A}(\boldsymbol{q}) = 
\begin{bmatrix}
0 & 0 & -2 q_y q_z\\
0 & -2 q_x q_z & 0\\
-2 q_x q_y & 0 & 0\\
-q_z^2 & q_y q_z & q_x q_z\\
q_y q_z & -q_y^2 & q_x q_y\\
q_x q_z & q_x q_y & -q_x^2\\
\end{bmatrix}.
\vspace{2mm}
\label{eq:matrixA_definition}
\end{equation}
}%
The corresponding 3D  Landau-Ginzburg action (Eq.~\eqref{eq:lagrangian_matrix_fourier}) is:
\vspace{-7mm}
{\small
\begin{eqnarray}
\mathcal{S} &=& \int d^{3} \boldsymbol{q} ~\Braket{\tilde{\psi'}(\boldsymbol{q}) | \underbrace{ \mathcal{A}(\boldsymbol{q})^{T} ~\hat{R}~ \hat{\Lambda}~ \mathcal{A}(-\boldsymbol{q}) }_{\hat{\mathcal{B}}(\boldsymbol{q})}| \tilde{\psi'}(-\boldsymbol{q})}.
\label{eq:lagrangian_matrix_fourier_3d}
\end{eqnarray}
}

We note that the matrix $\hat{\mathcal{B}}(\boldsymbol{q})$ in Eq.~\eqref{eq:lagrangian_matrix_fourier_3d} is symmetric. Using a partition function similar to Eq.~\eqref{eq_partition_2d}, we obtain, in the Morera gauge
\begin{eqnarray}
\llangle[\Big] \tilde{\psi}'_{i}(\boldsymbol{q}) \tilde{\psi}'_{j}(-\boldsymbol{q}) \rrangle[\Big] = (\hat{\mathcal{B}}(\boldsymbol{q})^{-1})_{ij},
\label{eq:psi_psi_correlation_3d}
\end{eqnarray}

{
which leads to the gauge-invariant stress-stress correlators:
\begin{eqnarray}
\llangle[\Big] \Delta\tilde{\Sigma}_{i}(\boldsymbol{q}) \Delta\tilde{\Sigma}_{j}(-\boldsymbol{q}) \rrangle[\Big] =  \mathcal{A}(\boldsymbol{q})^{T}_{ik} \mathcal{A}(-\boldsymbol{q})_{jl} (\hat{\mathcal{B}}(\boldsymbol{q})^{-1})_{kl}.
\label{eq:3d_stress_correlations}
\end{eqnarray}
}
For the simplest form of the polarizability tensor $\hat{\Lambda}^{-1}=K\mathds{1}$, these simplify to:
{\small
\begin{widetext}
\begin{equation}
C_{ijkl}^{~\mathrm{vacuum}}(\boldsymbol{q}) \equiv \llangle[\Big]\Delta\sigma_{ij}(\boldsymbol{q})\Delta\sigma_{kl}(-\boldsymbol{q}) \rrangle[\Big] =  K\left [ \frac{1}{2}\left(\delta^{ik} \delta^{jl}+\delta^{il} \delta^{jk}\right)+
\frac{q^i q^j q^k q^l}{q^4}
-\frac{1}{2}\left(\frac{\delta^{ik} q^j q^l}{q^2}+\frac{\delta^{jk} q^i q^l}{q^2}+\frac{\delta^{il} q^j q^k}{q^2}+\frac{\delta^{jl} q^i q^k}{q^2}\right)\right ].
\label{eq:theory_original_corr_fn}
\end{equation}
\end{widetext}
}
As expected,  this structure is identical to the VCT E-field correlators~\cite{Prem2018}. Analogous to 2D, these correlators are independent of $q$ and dependent only on the angular variables $\theta$ and $\phi$, producing a pinch-point at $q=0$. This structure survives any modifications to any ${\bf q}$ independent $\hat{\Lambda}^{-1}$. The forms, given in Eq.~\eqref{eq:3d_correlations_explicit} for the $\Lambda^{-1}$ defined in  Eq.~\eqref{eq:lambda_3d_iso} (see discussion below about this form), provides an explicit verification.  We have presented numerical evidence of pinch-point singularities in three-dimensional jammed solids
in Figs.~\ref{fig:3d_correlations_fig2}  and Fig.~\ref{fig:3d_correlations_fig3}.

The detailed fitting of the numerical data to the above expressions presented in the next section  provides strong support for the predictions of  VCTG. The detailed angular variations of $C_{ijkl}$ are strongly affected by the form of $\hat{\Lambda}^{-1}$, and these provide the signatures needed for determining the emergent elastic moduli tensor from the stress-stress correlations, a topic that we discuss next. 

\vspace{2mm}

\paragraph*{Emergent Elastic Moduli:}
In the above procedure for computing the stress-stress correlations of jammed solids in 2D and 3D, the only unknown in this formulation is the matrix $\hat{\Lambda}^{-1}$. This plays a role of the elastic modulus tensor in continuum elasticity. However, within VCTG, the moduli correspond to the dielectric constants via Eq.~\eqref{eq:E_D_relation} or~\eqref{eq:D_E_new_relation_2d}. Since these are determined by the polarizability of the medium they are constrained by the spatial symmetries of the contact networks in jammed solids, akin to the spatial symmetries of a crystalline solid. The symmetry property of the elastic moduli of thermal solids is imposed by the presence of a free-energy, which does not exist for jammed solids~\cite{Otto2003}. 
{Microscopic details of the contact forces, such as the positivity condition in dry granular materials,  or the Coulomb condition of static friction on the tangential forces enter the field theory through $\hat{\Lambda}^{-1}$, which we refer to
% Any other constraints, such as the positivity of the contact forces or the Coulomb condition of static friction on the tangential forces only renormalise $\hat{\Lambda}^{-1}$. We dub this 
as ``emergent elastic moduli''. They are the coupling constants of the field theory, which we determine from numerical and experimental measurements, as is the practice in  any field theory.  Translating to the language of elasticity theory, computing the elastic moduli is not within the scope of the theory; they are material inputs.}

We can obtain $\hat{\Lambda}^{-1}$ by fitting the theoretical forms of the VCTG correlations to the observed stress-stress correlations through Eq.~\eqref{eq:2d_stress_correlations} or~\eqref{eq:3d_stress_correlations}. Performing these fits for an arbitrary form of $\hat{\Lambda}^{-1}$ is complicated due to the large number of parameters involved. It is therefore simpler to posit specific forms for $\hat{\Lambda}^{-1}$ using the analogy between spatial symmetries of jammed contact networks and crystals. This reduces the number of parameters with which to perform the fits to our configuration averaged numerical data (next section) comprehensively.  Note that this procedure for determining $\hat{\Lambda}^{-1}$ relies on the {\it fluctuations} of the stress and not its response to a boundary strain.  The procedure for determining  $\hat{\Lambda}^{-1}$ in experimentally or numerically generated jammed solids can then be summarized as: (i) measure $C_{ijkl}(\boldsymbol{q}) \equiv \llangle[\Big] \Delta\tilde{\sigma}_{i}(\boldsymbol{q}) \Delta\tilde{\sigma}_{j}(-\boldsymbol{q}) \rrangle[\Big]$ for a given ensemble of jammed networks, (ii) choose forms of $\hat{\Lambda}^{-1}$ consistent with the symmetries of the configuration averages of these networks (e. g., isotropy, uniaxial anisotropy), and (iii) fit the  numerical data to the corresponding VCTG predictions.   

In the next section, we apply  this approach to  numerically generated ensembles of jammed frictionless soft discs and spheres at a give $A_G$, the average grain area (2D) or volume (3D) respectively. The packings we study in both 2D and 3D are under isotropic compression and therefore, we perform the fit assuming an isotropic form of $\hat{\Lambda}^{-1}$~\cite{archer2012introduction}.  This form is parametrized by two Lam\'e constants~\cite{landau2012theory}. As we will show in the next section, this form works extremely well. As was shown in our earlier paper~\cite{PhysRevLett.125.118002}, in a sheared granular system, $\hat{\Lambda}^{-1}$ has less symmetry, in which case one additional parameter is required to describe the stress-stress correlations in 2D.
%%%%%%%%%%%%%%%%%%%%%%%%%%%
\vspace{2mm}
\paragraph*{Stress Response to a point force:}
The disorder averaged response of jammed solids to a perturbing force ${\boldsymbol{f}}^{\rm p}$,   $\llangle\sigma_{ij}({\bf r})\rrangle_{f^{\rm p}}$,
can be calculated using the dielectric theory of VCTG developed above. The $\llangle ... \rrangle$ above denotes an average over an ensemble of configurations with the perturbing  force ${\boldsymbol{f}}^{\rm p}$ held fixed. Such experiments have been performed on sandpiles~\cite{Geng2001}, and motivated the quest for a stress-only framework for granular elasticity~\cite{Bouchaud2002} as well as theories of anisotropic elasticity~\cite{Otto2003}. In Fourier space,  
\begin{equation}
     \llangle \sigma_{ij} ({\bf q}) \rrangle_{f^{\rm p}} =G_{ijk}({\bf q})  f^{\rm p}_k ({\bf q}) ~,
     \label{eq:genresponse_stress}
\end{equation}
where the Green's function can be computed using standard methods based on the Landau-Ginzburg action (Eq.~\eqref{eq:lagrangian_density}) (see Appendix \ref{appendix:A} for details):
\begin{align}
    G_{ijk}({\bf q})=&\left[\frac{i g (\Lambda^{-1}_{abij}+\Lambda_{ijab}^{-1})}{4}\right] \left[ q_a \mathcal{C}^{-1}_{kb} + q_b \mathcal{C}^{-1}_{ka} \right] \nonumber \\
     \mathcal{C}_{ij} =& -g q_a q_b \Lambda^{-1}_{iajb}.
    \label{eq_greensexp}
\end{align}

The above expression allows us to compare the VCTG predictions to numerical and experimental observations in jammed solids. 
In real space,
\begin{equation}
    \llangle \sigma_{ij}(\boldsymbol{r}) \rrangle = \int d^2r~ G_{ijk} (\boldsymbol{r}-\boldsymbol{r}\,') f^{\mathrm{p}}_k(\boldsymbol{r}\,').
    \label{eq_stress_resp}
\end{equation}
In particular, for $\Lambda^{-1}=\mathbb{I}$, the above equation reduces to Eq.~\eqref{eq_electgreens} with $G_{ijk}=\mathcal{G}_{ijk}^d({\bf r})$, the vacuum Green's function given by Eq.~\eqref{eq:2d_vacuum_real_Green's}.

\section{Numerical Investigation of Jammed Soft Spheres~\label{sec:numerical}}
In this section we use numerical simulations of jammed frictionless spheres in 2D and 3D to test the predictions of VCTG which applies to stress fields averaged over all contact networks for an ensemble of jammed packings. Details of the numerical methods and additional numerical results are provided in the Supplemental Material~\cite{SI}.

The ensembles of jammed packings studied here are characterized by a  fixed value of the area (volume in 3D) per grain, $A_G$. In such an ensemble, the pressure and packing fraction both fluctuate, however the packing fraction, $\phi$, fluctuations are purely due to the presence of rattlers. Therefore, we use fixed $A_G$ and fixed $\phi$, interchangeably to characterize our ensemble. Several key consistencies between VCTG and both numerical and experimental  measurements of stress distributions in 2D systems were already discussed in reference~\cite{PhysRevLett.125.118002}. Here, in addition to elaborating on these earlier results, we verify the VCTG predictions (i) of the  correlations of stress fluctuations in both 2D and 3D jammed solids, (ii) the response of jammed solids to localized external forces in 2D, and (iii) compute the emergent elastic moduli of 2D and 3D jammed solids as a function of pressure (see Table \ref{tbl:stiffness_consts}).

A jammed contact network is generated by starting with a random arrangement of soft frictionless bidisperse grains interacting via a one-sided harmonic potential~\cite{corey_packing_protocol}, in a simulation box that fixes $A_G$, and minimizing the energy. The stress field $\hat{\sigma}(\boldsymbol{r})$, inside a jammed solid is obtained from the force moment tensor defined in Eq.~\eqref{eq:stress_micro}. Disorder averaging over multiple such contact networks, we compute $\llangle \sigma_{ij} \rrangle$ and their correlations, $\llangle \Delta\sigma_{ij} \Delta\sigma_{kl} \rrangle$. A  coarse-graining length scale is chosen either through a box  size in real space (linear dimension of $\approx 5$ grain diameters in $2D$) or by implementing a large $q \equiv |\boldsymbol{q}|$ cutoff $q_{max}$ in Fourier space (corresponding to linear dimensions of $\approx 3$ grain diameters). 

We first present results comparing the numerically obtained correlations, $\llangle \Delta\sigma_{ij} \Delta\sigma_{kl} \rrangle$, to VCTG predictions in both 2D and 3D. Then, we present results of comparisons between the numerical results of stress response to theoretical predictions in 2D.  

\vspace{-3mm}

\subsection{Numerical results for Stress-stress Correlation functions in jammed packings}

\paragraph*{Two dimensions:}
We begin by examining the stress-stress correlations in packings of $N=8192$ bidisperse soft frictionless disks in 2D, under isotropic compression. Our objective is two fold: (i) test the broad features of the VCTG predictions, and (ii) determine the emergent elastic modulus tensor, $\hat\Lambda^{-1}$ by fitting the numerical results to the theoretical predictions. Following the earlier discussion on expected symmetries of the emergent elastic moduli (Section \ref{sec:correlations}), we choose an isotropic form for $\hat{\Lambda}^{-1}$ for isotropically jammed solids:
{\small
\begin{equation}
\hat{\Lambda}^{-1}_{\mathrm{iso}}
=\begin{bmatrix}
\lambda+2\mu & \lambda & 0\\
\lambda & \lambda+2\mu & 0\\
0 & 0& 2\mu
\end{bmatrix}.
\label{eq:lambda_2d_iso}
\end{equation}
}%

\begin{figure}[!htbp]
\vspace{-4mm}
\includegraphics[width=0.35\textwidth]{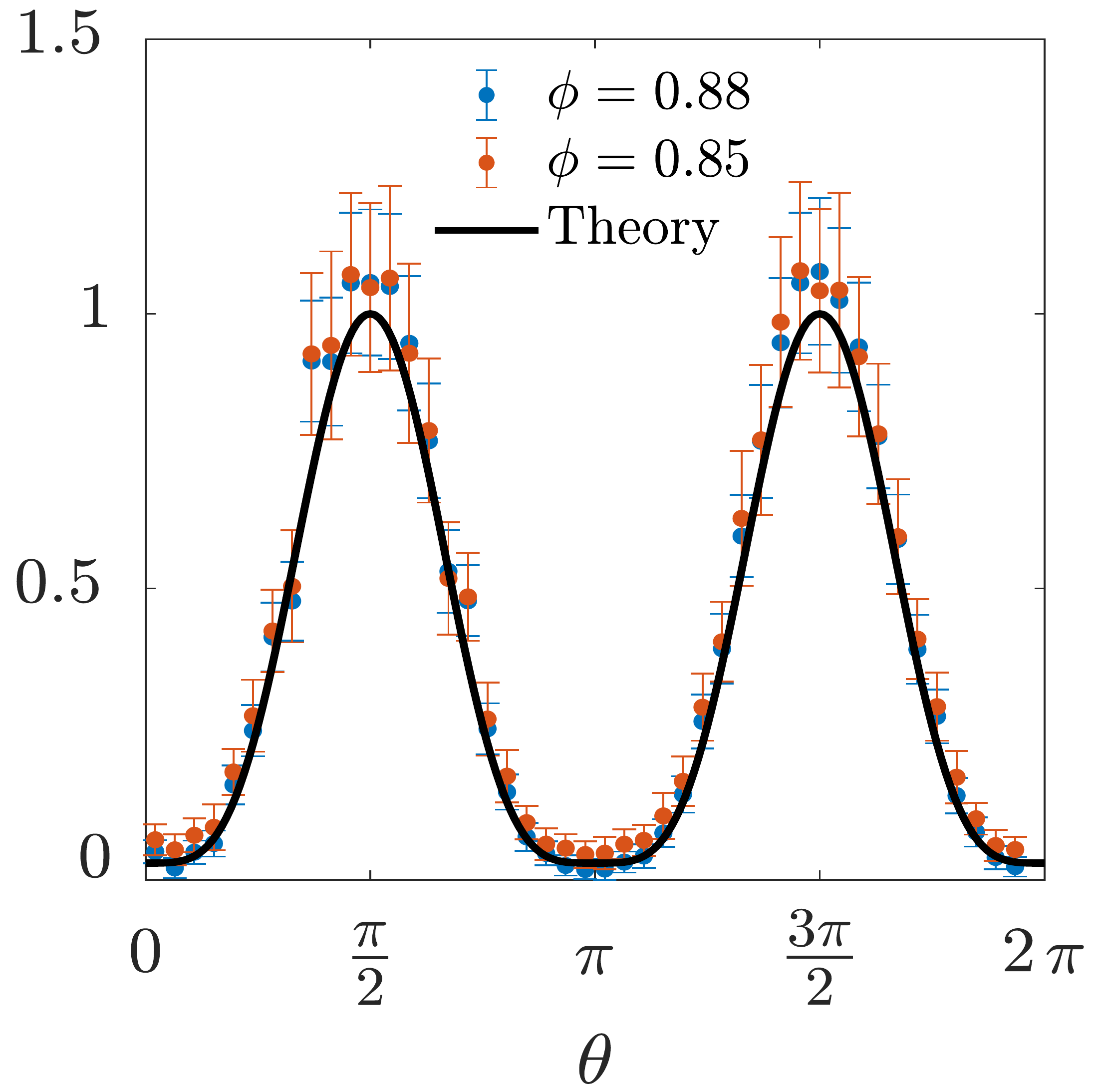}
\vspace{-2mm}
\caption{\label{fig:2d_corr_Cxxxx_angular} Quantitative comparison of the radially averaged $C_{xxxx}$ correlation function obtained from numerical simulations, along with the theoretical predictions, in 2D. The comparisons given, are for a system of 8192 grains at two packing fractions $\phi = 0.88$ and $\phi = 0.85$. Each curve has been scaled by the maximum value of the corresponding angular $C_{xxxx}$ correlation.}
\end{figure}

We compute the stress-stress correlations predicted by VCTG (Eq.~\eqref{eq:2d_stress_correlations}) using this form of $\hat\Lambda^{-1}$ (Section~\ref{sec:correlations}), and determine  $\lambda$ and $\mu$ by fitting to numerical results. We emphasize that this is a two parameter fit to six different correlation functions in Eq.~\eqref{eq:2d_stress_correlations} (explicit forms are given in Eq.~\eqref{eq:2d_correlations_explicit}). In 2D, these correlations turn out to depend {\it only} on the following combination of $\lambda$ and $\mu$, denoted by $K_{2D}$ via the constant pressure, $P = \frac{1}{2}(\sigma_{xx}+\sigma_{yy})$, correlations, {\it i.e.}:

\begin{figure*}[!tbp]
\includegraphics[width=0.85\textwidth]{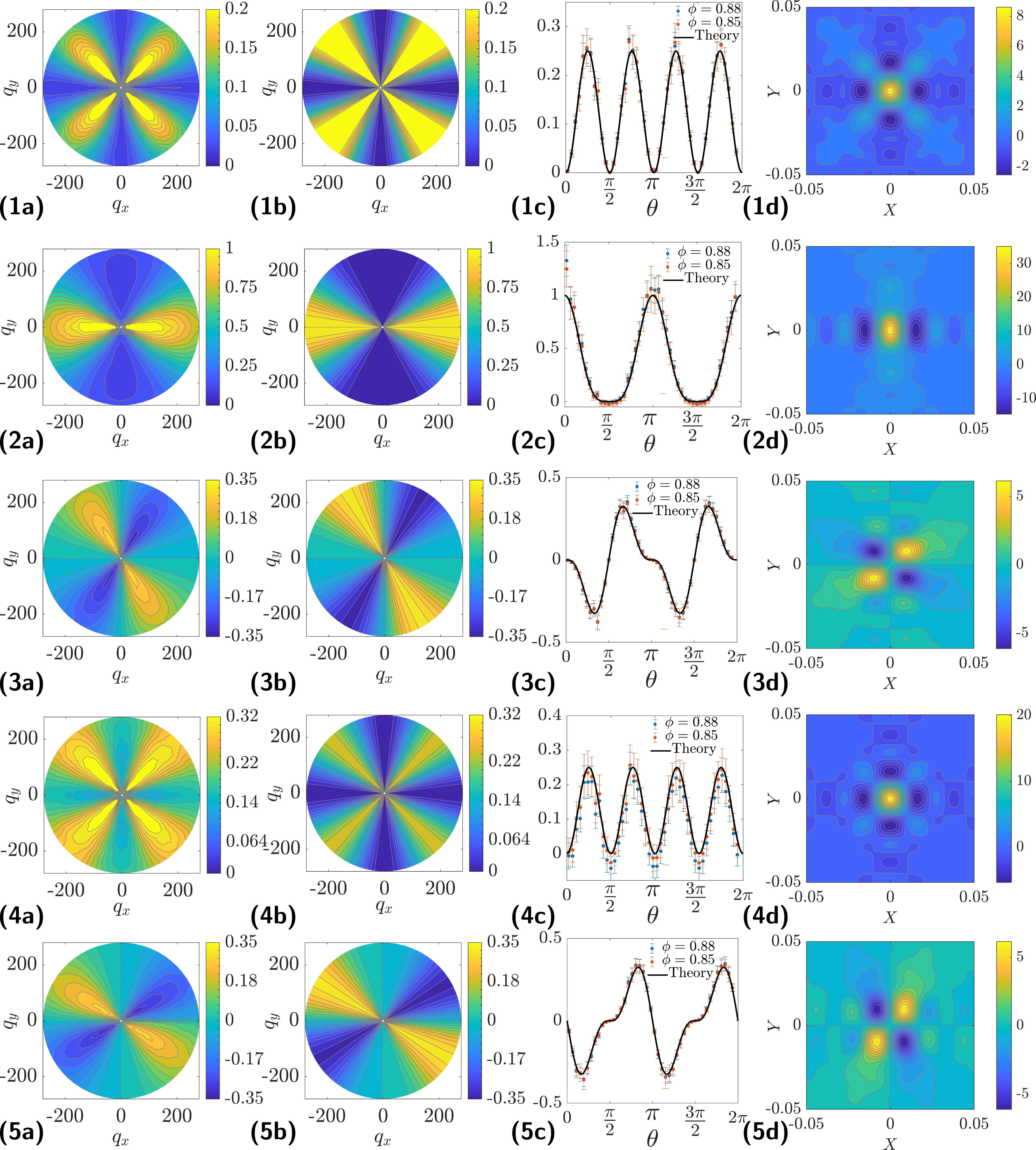}
\caption{ Comparison of the correlation functions obtained from numerical simulations with the theoretical predictions, in 2D. The comparisons given, are for a system of $8192$ grains at two packing fractions $\phi=0.88$ and $\phi=0.85$. The four columns show respectively: (a) the numerical results in $q$-space, (b) the theoretical predictions in $q$-space, (c) quantitative comparison between the radially averaged (integrated over $|\boldsymbol{q}|$) correlations, plotted as a function of the angle $\theta$ and (d) the numerical results in real space obtained by an inverse Fourier transform of the correlations in Fourier space. The modulations seen in (d) are at the grain scale. The contour plots are for $\phi = 0.88$ and the quantitative comparison in the third column shows the results for both $\phi = 0.85$ and $\phi = 0.88$. In the quantitative comparison plots, the data has been scaled by the maximum value of the angular $C_{xxxx}$ correlation function. Each row presents the results for a specific correlation function. The five rows present respectively: (1) $C_{xyxy}$, (2) $C_{yyyy}$, (3) $C_{xxxy}$, (4) $C_{xxyy}$ and (5) $C_{xyyy}$ whose explicit forms are given in the Appendix (see Eq.~\eqref{eq:2d_correlations_explicit}).~\label{fig:2d_full}}
\vspace{-7mm}
\end{figure*}

{\small
\begin{align}
    \vspace{-2mm}
    \llangle \Delta P(\boldsymbol{q})\Delta P(-\boldsymbol{q}) \rrangle &= \mu\left(\frac{\lambda+\mu}{\lambda+2\mu}\right) \equiv  K_{\mathrm{2D}} = \frac{\mu}{2(1-\nu)},\label{eq:2d_K_definition}
\end{align}
}%

where, $\nu = \frac{\lambda}{2(\lambda + \mu)}$ is the {\it Poisson's ratio}. All other stress correlations are given by:

{\small
\begin{align}
    C_{ijkl}^{\mathrm{~isotropic}}(\boldsymbol{q}) &= 4K_{\mathrm{2D}}~C_{ijkl}^{\mathrm{~vacuum}}(\boldsymbol{q}).
\end{align}
}%

{\small
\begin{table*}[!htbp]
\begin{center}
\begin{tabular*}{0.79\textwidth}{|c|c|c|c|c|c|c|}
\hline
&  &  &  &  &  &   \\
\textbf{Grain Area} & \textbf{Dimension}  & $\llangle z -z_{iso}\rrangle$ &
\textbf{Grain}  &
\textbf{Packing}  &
\textbf{Poisson's} &
$\llangle \Delta\mathbf{P}(\boldsymbol{\mathbf{q}})\Delta\mathbf{P}(-\boldsymbol{\mathbf{q}}) \rrangle $  \\
$A_G$ &  &  & \textbf{Pressure}  & \textbf{Fraction}  & \textbf{Ratio}  &  \\
 &  &   & $\llangle P_G \rrangle$  & $\phi$  & $\nu$  & $K$ \\
\hline
$1.04 \times 10^{-4}$ & 2D & $0.294$ & $2.24 \times 10^1$ & $0.835$ &  $-$  & $0.0537$ \\
\hline
$1.05 \times 10^{-4}$  & 2D & $0.463$ & $5.25 \times 10^1$ & $0.852$ &  $\nu \in [0.4,0.5]$  & $0.257$ \\
\hline
$1.06 \times 10^{-4}$  & 2D & $0.595$ & $8.38 \times 10^1$ & $0.865$ & $-$  & $0.583$\\
 \hline
$1.07 \times 10^{-4}$ &  2D & $0.705$ & $1.16 \times 10^2$ & $0.877$ & $\nu \in [0.4,0.5]$  & $0.992$\\
 \hline
$2.44 \times 10^{-5}$  & 3D  & $0.839$ & $1.26\times 10^2$ & $0.655$ & $\nu  \approx 0.40$  & $0.17$\\
\hline
$2.56 \times 10^{-5}$ & 3D  & $1.68$ & $5.75 \times 10^2$ & $0.689$ & $\nu  \approx 0.40$ & $0.18$\\
\hline
\end{tabular*}
\caption{Emergent elastic moduli, obtained using a combination of stress-correlation and stress-response measurements. The ensembles in 2D and 3D are characterized by a fixed grain area/volume ($A_G$). $\llangle z -z_{iso}\rrangle$ and $\llangle P_G \rrangle$ denote averages over the ensemble of the number of contacts per grain, and the pressure per grain, respectively. The two and three dimensional results are from systems of size $8192$ and $27000$ respectively.
\label{tbl:stiffness_consts}}
\vspace{-8mm}
\end{center}
\end{table*}
}%

As an example (from Eq.~\eqref{eq:2d_correlations_explicit}), 
{\small
\begin{align}
    C_{xxxx}(q,\theta) &=4K_{2D} \sin^4\theta
    \label{eq_cxxxx}
\end{align}
}%
is plotted in Fig. \ref{fig:2d_correlations}(b) and the agreement with numerical results in \ref{fig:2d_correlations}(a) is evident. { Deviations from theoretical predictions are seen at large $q \approx 2\pi/{(\rm grain ~ diameter)}$, as expected from the particulate nature of the jammed state.}  $K_{2D}$ is obtained by fitting the numerical  pressure correlations but we are unable to determine $\lambda$ and $\mu$ independently from the stress-stress correlations alone. We will show in the next section that combining measurements of stress response and stress-stress correlations, allows us to determine both. 

The pinch-point singularities in the stress-stress correlations in ${\bf q}$-space (Fig.~\ref{fig:2d_correlations}(a) and (b)) lead to  power-law decays of the correlations  in real space (Fig.~\ref{fig:2d_correlations}(c)), $\llangle \Delta\sigma(\boldsymbol{r}) \Delta\sigma(0) \rrangle \propto 1/|r|^d$.  An equally important feature is the strong anisotropy of these correlations which display sharp variations with the angle in real-space. For example, the correlation of $\sigma_{xx}$ is largest along the $x$ direction, vanishes along the $(1,1)$ direction and becomes negative along the $y$ direction, {crossing zero at a distance $\approx {\rm grain~diameter}$, as is evident from the numerical data in Fig.~\ref{fig:2d_correlations}(c)).  This grain-scale modulation, which is  consistent with the large $q$ cutoff seen in Fig.~\ref{fig:2d_correlations}(a), cannot be captured by the continuum, VCTG description. }Further, we can deduce from the rotational invariance properties of the stress tensor, that a projection along any direction $\hat{\boldsymbol{\alpha}}$ will exhibit the slowest decay along $\hat{\boldsymbol{\alpha}}$, while decaying rapidly in the transverse direction~\cite{Wang:2020aa}. In the continuum theory, this strong anisotropy is the manifestation of force-chains, and is a direct consequence of the Gauss's law constraint on the stress tensor: transverse correlations are strongly suppressed in relation to longitudinal correlations.

We have further tested the quantitative agreement between the VCTG and the configuration averaged numerical data (for two different packing fractions $\phi = 0.85~\mathrm{and}~0.88$), in Fig.~\ref{fig:2d_corr_Cxxxx_angular} for $C_{xxxx}$ with the best-fit values of the $K_{\mathrm{2D}}$ given in Table~\ref{tbl:stiffness_consts}. The contour plots presented in Figs.~\ref{fig:2d_correlations},~\ref{fig:2d_full} are obtained from $239$ distinct packings with the packing fraction $\phi = 0.88$. Similar comparisons for all the five remaining  correlations are given in Fig.~\ref{fig:2d_full}, all of which show excellent agreement between the numerical results and the prediction from VCTG. The excellent agreement between theory and numerical data supports our assertion regarding the symmetry properties of  emergent elastic moduli.

Table \ref{tbl:stiffness_consts} shows that  $K_{\mathrm{2D}}$ decreases with the mean background grain pressure $P_G$. This is expected since the solid ceases to exist at $P_G=0$, the unjamming point. Viewed from the perspective of the types of jammed networks that can be created with purely repulsive interactions, stress fluctuations become completely frozen as the imposed pressure goes to zero since the stress tensor is positive definite. 

\begin{widetext}
    \begin{minipage}{0.97\linewidth}
        \begin{figure}[H]
            \centering
            \includegraphics[width=0.84\textwidth]{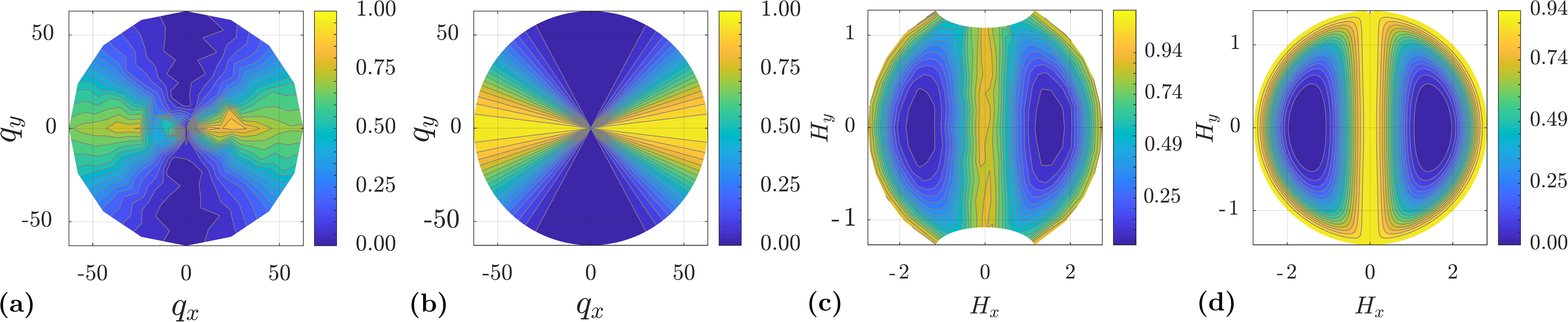}
            \caption{Comparison of the correlation functions obtained from numerical simulations with the theoretical predictions, in 3D. The comparisons are done on a system of $27000$ grains at packing fraction $\phi = 0.69$.  Panels  (a) and (b) show respectively,  the numerical and theoretical forms for a slice of the correlation function $C_{yyyy}(\boldsymbol{q})$  on the XY-plane ($\theta=\pi / 2$). The pinch-point structure at $|\boldsymbol{q}|=0$ is clearly visible. Panels (c) and (d) show the numerical and theoretical results for $C_{yyyy}(\boldsymbol{q})$ respectively. The results are presented in the Hammer projection~\cite{Hammer} coordinates $H_x$ and $H_y$. The missing regions in the numerics is due to difficulties in sampling around $\theta=0$ and $\theta=\pi$.
            ~\label{fig:3d_correlations_fig2}} 
        \end{figure}    
    \end{minipage}
\end{widetext}

Interpreted in the VCTG framework in a linear dielectric, since no dipoles (contact forces) can be created without external pressure, the polarization vanishes. Earlier studies of jammed frictionless spheres in 2D and 3D have investigated how the ratio of the shear to bulk modulus varies with the distance to the unjamming point $(\phi-\phi_J)$, and identified a non-trivial power law~\cite{jamming_epitome_2003}. While we have not undertaken a systematic study of this variation close to the unjamming point, our simulations, restricted to be deep within the jammed region, clearly indicate that the Poisson ratio, $\nu\approx 1/2$ (see Table \ref{tbl:stiffness_consts}). This implies that the bulk modulus, $\approx \lambda$ is much larger than the shear modulus, $\mu$. The variation of $K_{\mathrm{2D}}$ with pressure and the deviation of the contact number from isostaticity, $\llangle z -z_{iso}\rrangle$, is also consistent with earlier studies~\cite{Henkes2009}. 

\vspace{3mm}

\paragraph*{Three Dimensions:}

We extend the above analysis to 3D by performing numerical simulations of $27000$ soft frictionless bidisperse spherical grains. We generated $374$ packings of grains in mechanical equilibrium under isotropic compression with packing fractions $\phi = 0.69$. In 3D, the form of the isotropic $\hat{\Lambda}^{-1}$ is given by, 
{\small
\begin{equation}
\hat{\Lambda}^{-1}_{\rm iso}
= \begin{bmatrix}
\lambda+2\mu & \lambda & \lambda & 0 & 0 &0\\
\lambda & \lambda+2\mu & \lambda & 0 & 0 &0\\
\lambda & \lambda & \lambda+2\mu & 0 & 0 &0\\
0 & 0 & 0 & 2\mu & 0 & 0\\
0 & 0 & 0 & 0 & 2\mu & 0\\
0 & 0 & 0 & 0 & 0 & 2\mu
\end{bmatrix}.
\label{eq:lambda_3d_iso}
\end{equation}
}%
Unlike the two dimensional case, the three dimensional correlations depend on both $\lambda$ and $\mu$ and are qualitatively different from the vacuum correlations. This allows us to determine both both $\lambda$ and $\mu$ from the stress stress correlations. The best fit estimates for the parameters $\lambda$ and $\mu$  are given in Table~\ref{tbl:stiffness_consts}.

\begin{figure}[!hbp]
\includegraphics[height=17cm]{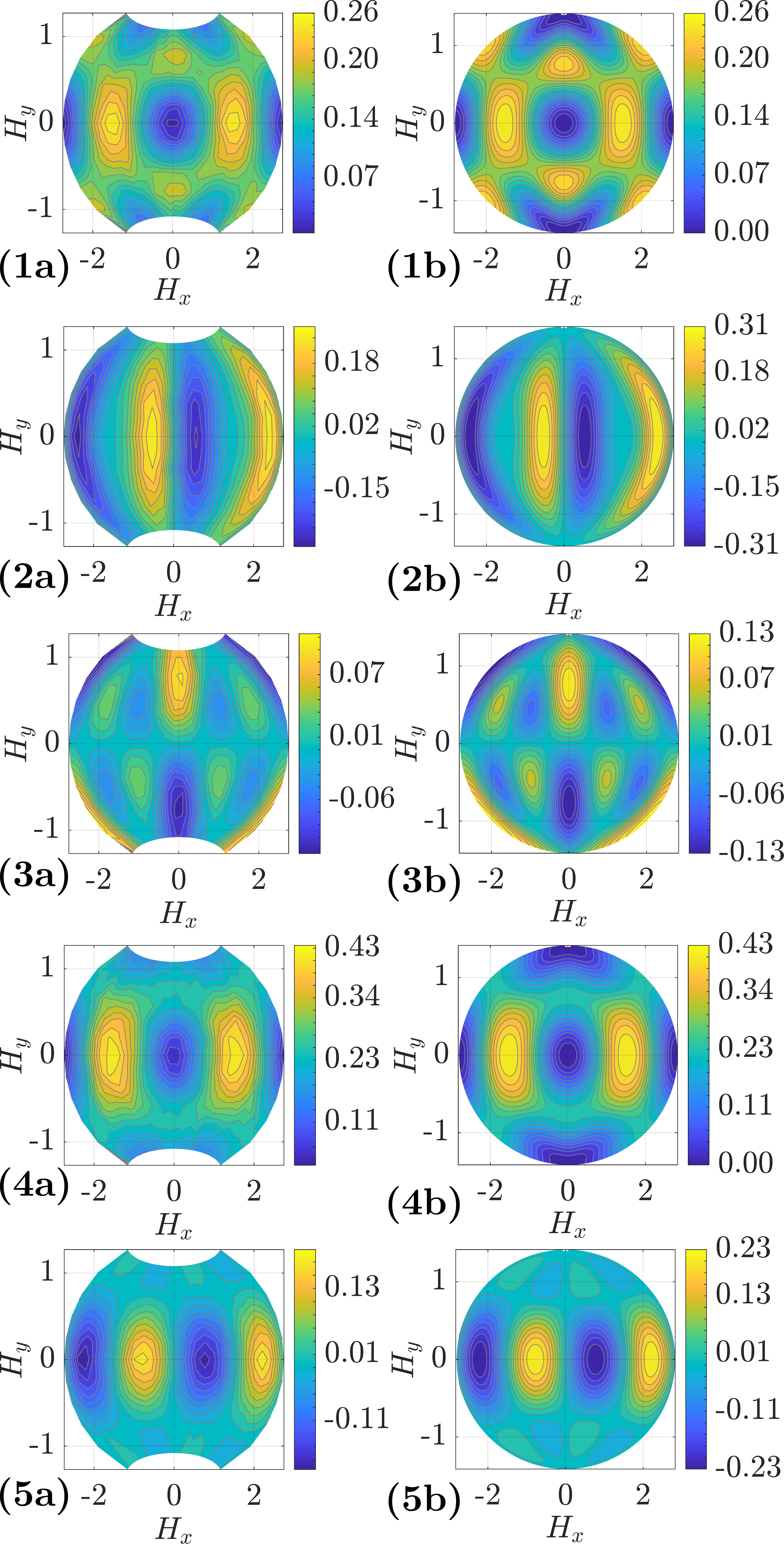}
\caption{Comparison of the correlation functions obtained from numerical simulations with the theoretical predictions, in 3D. The comparisons are done on a system of $27000$  grains at packing fraction $\phi = 0.69$. Columns \textbf{(a)} and \textbf{(b)} show the numerical and theoretical results for the correlations respectively. The rows show respectively: \textbf{(1)} $C_{xzxz}$, \textbf{(2)} $C_{xyyy}$, \textbf{(3)} $C_{xyyz}$, \textbf{(4)} $C_{xxzz}$ and \textbf{(5)} $C_{xyzz}$. The comparisons for the remaining $15$ correlations are given in the Appendix Fig.~\ref{fig:3d_full}.~\label{fig:3d_correlations_fig3}}
\vspace{-4mm}
\end{figure}

The explicit forms of the three dimensional correlation functions are given in the Appendix (see Eqs.~\eqref{eq:3d_correlations_explicit1}-\eqref{eq:3d_correlations_explicit}). For example, 
\begin{align}
    C_{yyyy}(q,\theta,\Phi) &= 4\mu\left(\frac{\lambda+\mu}{\lambda+2\mu}\right) \left[\sin ^2\theta  \cos ^2\Phi +\cos ^2\theta \right]^2,
\end{align}
as predicted by  VCTG, is compared to  numerical results in Fig.~\ref{fig:3d_correlations_fig2} where $\theta$ and $\Phi$ are polar and azimuthal angles respectively.  {In Fig.~\ref{fig:3d_correlations_fig3}, we have presented representative samples for the $21$ possible distinct stress-stress correlations in 3D, grouping them by symmetry}. The agreement with theoretical predictions is remarkable, and the pinch-point singularities are evident. Notably, for these isotropically jammed solids,  all 21 correlation functions are controlled by only two parameters.

We have also presented additional supporting evidence for the pinch-point singularities in 3D in the Supplemental Material~\cite{SI}, where we have depicted the behavior of the correlation function $C_{yyyy}$ integrated over the angular coordinates. It is clear that the integrated correlation function scales as $|\boldsymbol{q}|^{2}$, which is consistent with a function that is independent of $|\boldsymbol{q}|$ and therefore purely a function of the angular coordinates. 
Similar to 2D, the pressure ($P = \frac{1}{3}(\sigma_{xx}+\sigma_{yy}+\sigma_{zz})$) correlation in 3D is independent of $\boldsymbol{q}$, and given by: 
{\small
\begin{equation}
    \llangle \Delta P(\boldsymbol{q})\Delta P(-\boldsymbol{q}) \rrangle = \frac{4\mu}{9}\left(\frac{3\lambda+2\mu}{\lambda+2\mu}\right) \equiv  K_{\mathrm{3D}} = \frac{4\mu}{9}\left(\frac{1+\nu}{1-\nu}\right).
    \label{eq:3d_K_definition}
\end{equation}
}%

\begin{figure*}[!htbp]
\includegraphics[width=0.85\textwidth]{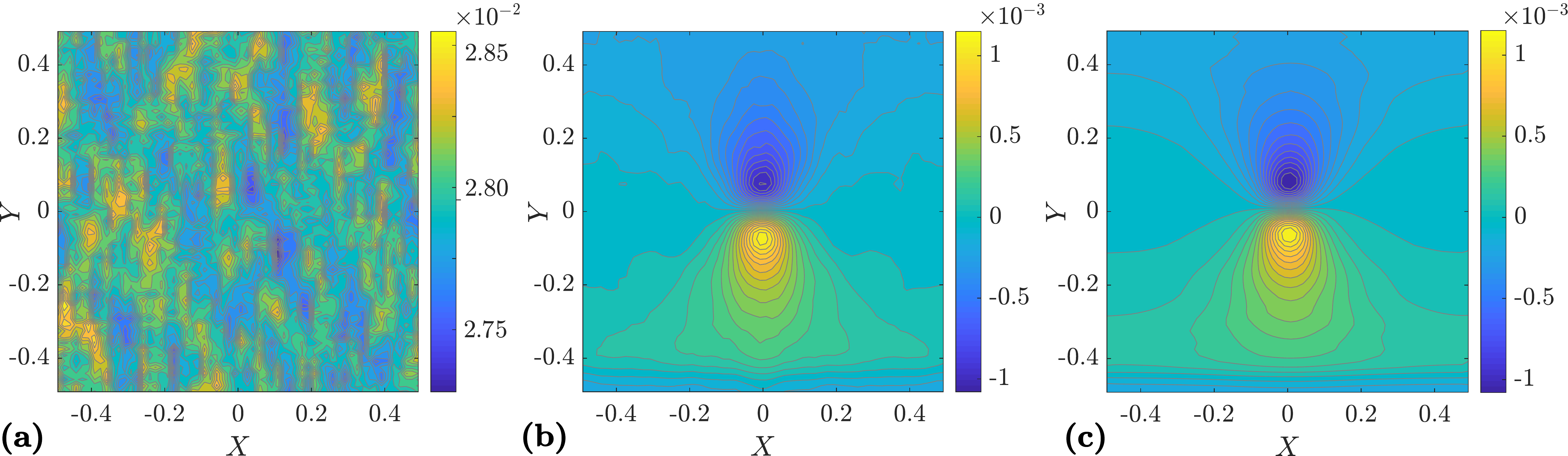}
\caption{Comparisons between the theoretical predictions and numerical response to a point force in 2D for $\llangle \sigma_{yy} \rrangle$ component of the stress tensor. Panel \textbf{(a)} displays the results for $yy$ component of the background stress fluctuations. Panel \textbf{(b)} depicts the numerical response, computed as a difference of $\llangle \sigma_{yy} \rrangle$ before and after the point force is applied. Panel \textbf{(c)} shows the theoretical predictions for the response, computed for the simulation geometry from the Green's function using Eqs.~\eqref{eq:2d_response_vacuum_general},~\eqref{eq:Greens_fn_dielectric_vac} and~\eqref{eq:charge_distbn_response}, using the parameters for $\phi =0.88$ given in Table~\ref{tbl:stiffness_consts}. The numerical result for $\llangle \sigma_{yy} \rrangle$ has been symmetrized about the Y-axis for noise reduction: $\sigma_{yy}(x,y)=\frac{1}{2}(\sigma_{yy}(x,y)+\sigma_{yy}(-x,y))$. The comparison is done for packings of $N=8192$ grains with $P_G \in [1.15,1.20]\times 10^2$.
\label{fig:resp_phi088_sigmayy}
\vspace{-3mm}
}
\end{figure*}

\subsection{Numerical results for Stress Response in 2D~\label{subsec:numerical_response}}
The net difference between the stress fields with and without a perturbing force is given by Eq.~\eqref{eq:genresponse_stress} in momentum space or \eqref{eq_stress_resp} in real space. With the isotropic form of $\Lambda^{-1}$ in 2D,  Eq.~\eqref{eq:lambda_2d_iso}, the Green's functions $G_{ijk}({\bf q})$ (Eq.~\eqref{eq_greensexp}) is (with $q^2 \equiv (q_x^2+q_y^2)$):
{\small
 \begin{align}
     G_{ijk}(\boldsymbol{q}) &= \mathcal{G}^{2D}_{ijk}(\boldsymbol{q}) + \\
     &\left(\frac{\nu}{1-\nu}\right)
     \left(\mathcal{G}^{2D}_{ijk}(\boldsymbol{q}) + \frac{i \delta_{ij} q_k - i \delta_{ik} q_j - i\delta_{jk} q_i}{q^2} \right)\nonumber
 \end{align}
 }%

where 
\begin{equation}
    {\mathcal{G}}_{ijk}^{\,2D}(\boldsymbol{q})= \frac{i}{q^4} \left[ q_i q^2 \delta_{jk} + q_j q^2\delta_{ik} - q_i q_j q_k  \right] \label{eq:2d_response_vacuum_general}
\end{equation}
is the VCT Green's function in 2D  obtained by Fourier transforming Eq.~\eqref{eq:2d_vacuum_real_Green's}. The explicit forms of various components are given in Eq.~\eqref{eq:Greens_fn_dielectric_vac}.

\begin{figure}[htbp]
\vspace{-2mm}
\includegraphics[width=0.35\textwidth]{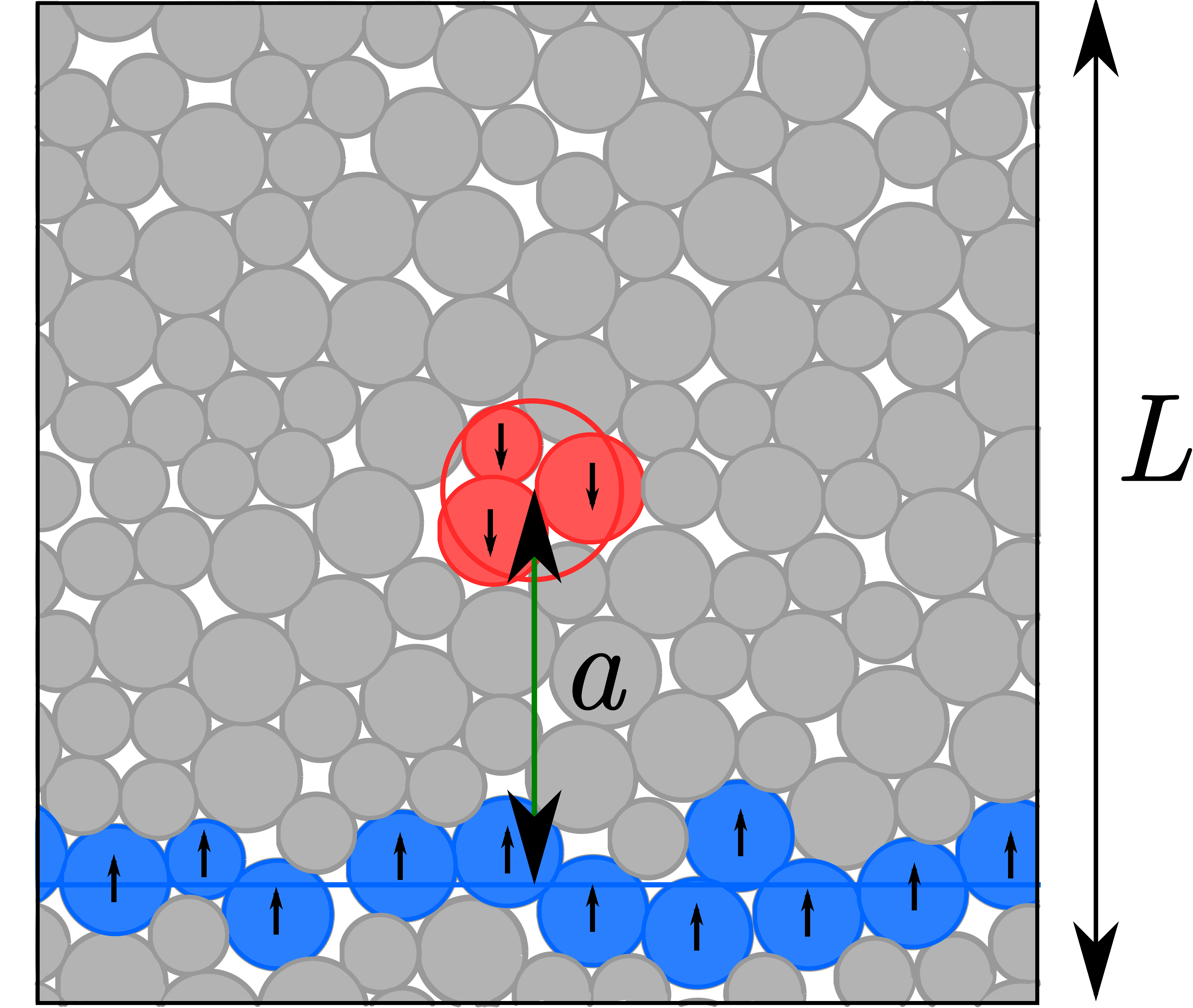}
\caption{\label{fig:response_schematic}Geometry used in the computation of the response to point forces. The grains are in a square box of length $L=1$. The point charge is distributed on grains (colored in red) within an circular area with radius $r_0=0.05$. The compensating forces are distributed on grains (colored in blue) that intersect the $x$-axis (blue line). The spacing between the center of the circle and the compensating line charge is $a=0.45$.}
\vspace{-3mm}
\end{figure}

We test these predictions  in the ensemble of 2D jammed solids used to analyze stress-stress correlation in the previous section. 
Eqs.~\eqref{eq:2d_response_vacuum_general} and \eqref{eq:Greens_fn_dielectric_vac} predict that the Green's functions for an isotropically jammed solid can be written as a linear combination of the VCT Green's functions. An important fact to note is that this mixing depends on the Poisson's ratio, $\nu$. 
Therefore, combining the response measurements with correlations measurements allows us to determine both $K_{2D}$ and $\nu$, and consequently both $\lambda$ and $\mu$. 

\vspace{4mm}

We compute the response by adopting a method similar to the one used in~\cite{Ellenbroek_thesis}. Fig.~\ref{fig:response_schematic} shows the geometry used. The starting configurations are drawn from the same ensemble used in the computation of the stress correlations, corresponding to $\phi = 0.88$. 
% We apply forces over a small region of the system, with the total magnitudes of these forces $F \approx 10$ (with average contact force $\approx 0.1$). 
We distribute localized forces with a total magnitude $F \approx 10$ (compared with the average contact force $\approx 0.1$)
% These localized forces are distributed 
over all grains within a circle of radius $r_0=0.05$ centered at the origin. This  constitutes our implementation of the `point' force (see~\cite{SI} for more details). In order to maintain force balance, we add compensating forces along a line (which we choose to be particles with an overlap with the $x$-axis) as shown in Fig.~\ref{fig:response_schematic}. The external force distribution is therefore given by

{\small
\begin{equation}
\boldsymbol{f}^{\mathrm{p}}(x,y)=\left(\frac{1}{L}~\delta(y)-\frac{1}{\pi r_0^2}~\Theta(r_0^2-x^2-(y-a)^2)\right) F \hat{y} ~ , 
\label{eq:charge_distbn_response}
\end{equation}
}%

where $\delta$ is the Dirac delta function, $\Theta$ is the Heaviside step function and the other quantities are depicted in Fig.~\ref{fig:response_schematic}. This procedure is repeated for each packing in the ensemble, and the average response of the system is computed using these configurations.

\begin{figure}[htbp!]
\includegraphics[width=0.42\textwidth]{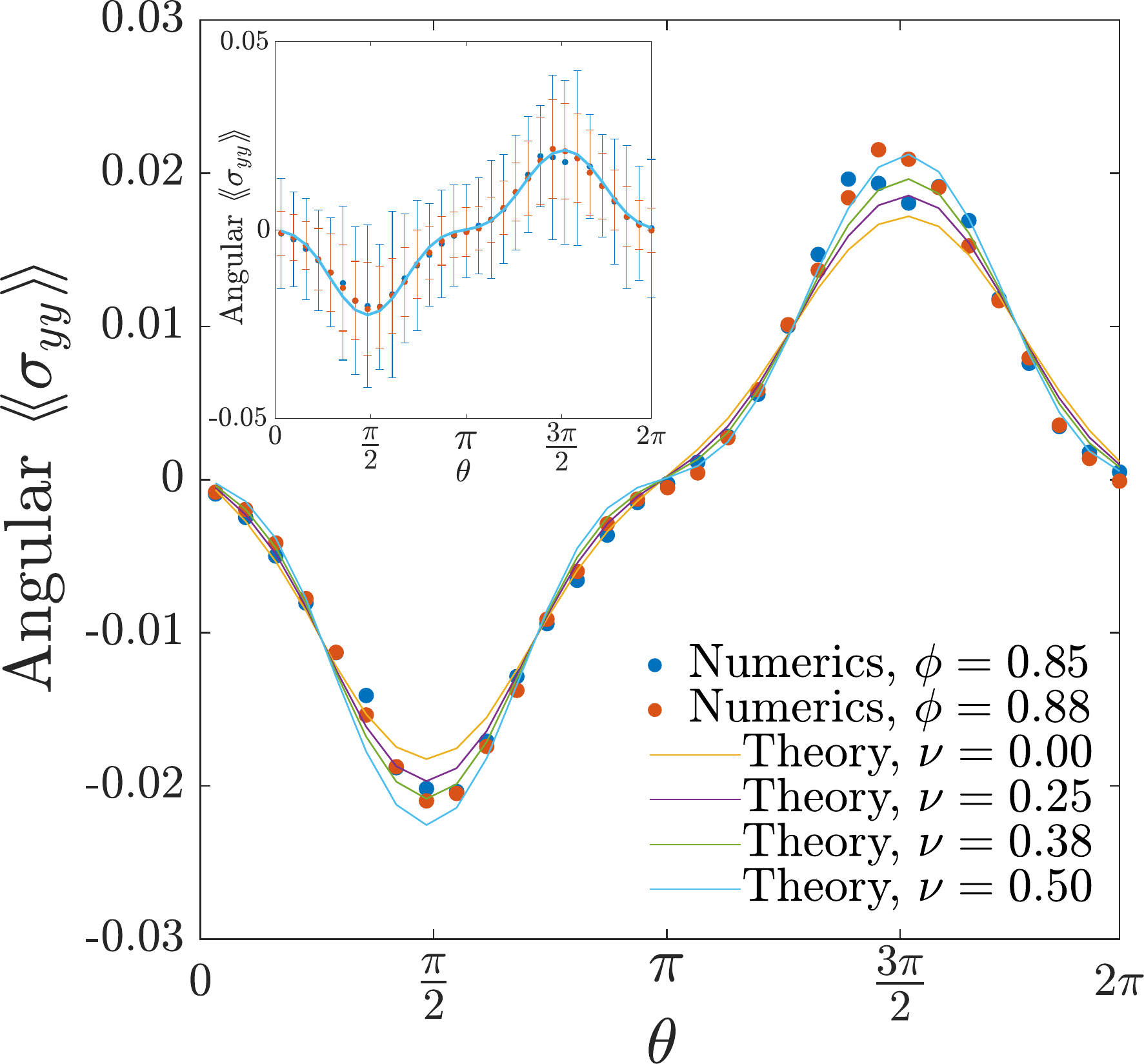}
\caption{Comparisons between theory and numerics for the radially integrated response of the $\llangle \sigma_{yy} \rrangle$ component of the stress tensor. The angular results are obtained by integrating the results given in Fig.~\ref{fig:resp_phi088_sigmayy} on an annulus of~\label{fig:resp_angular_sigmayy}}
\vspace{-4mm}
\end{figure}
\begin{figure}[!htp]
    \captionsetup{labelformat=empty}
    \ContinuedFloat
    \caption{  radius $r \in [0.075,0.3]$ centered on the point charge. The numerical results are presented for two packing fractions $\phi =0.88$ and $0.85$ and are scaled by the magnitude of the point force. The theoretical results are determined by the Poisson's ratio $\nu$ and we plot these results for $\nu=0.00,~0.25,~0.38$ and $0.50$. \textbf{(Inset)} The same results with error bars on the numerical response. The large error bars can be attributed to noise in the background stress, both before and after the application of the external forces.}
    \vspace{-8mm}
\end{figure}

Fig.~\ref{fig:resp_phi088_sigmayy} compares the predictions of VCTG with our numerical results for the $\llangle \sigma_{yy} \rrangle$ component of the stress response to the body force configuration in Eq.~\eqref{eq:charge_distbn_response}. We found that the $\llangle \sigma_{yy} \rrangle$ and $\llangle \sigma_{xx} \rrangle$ response are roughly symmetric about the $Y$-axis,
% \rimm{[X or Y axis ? I thought that the symmetry about the X-axis is broken by the presence of the line charge. The figures are symmetry about the Y (vertical) axis right ? Do you mean X=0 ? ]} 
as expected, and this symmetry becomes more and more evident as we increase the number of configurations in the averaging procedure. Since increasing the number of configurations used in the averaging is computationally expensive, we have symmetrized these responses about the $Y$-axis in order to reduce noise. 

In the first column of Fig.~\ref{fig:resp_phi088_sigmayy}, we display the $\llangle \sigma_{yy} \rrangle$ before the point force is applied. In this case, the system is under isotropic compression and the solution for the stress field from the field equations of VCTG is a uniform stress field. The numerical results are consistent with
\begin{widetext}
    \begin{minipage}{0.97\linewidth}
        \begin{figure}[H]
            \centering
            \includegraphics[width=0.97\textwidth]{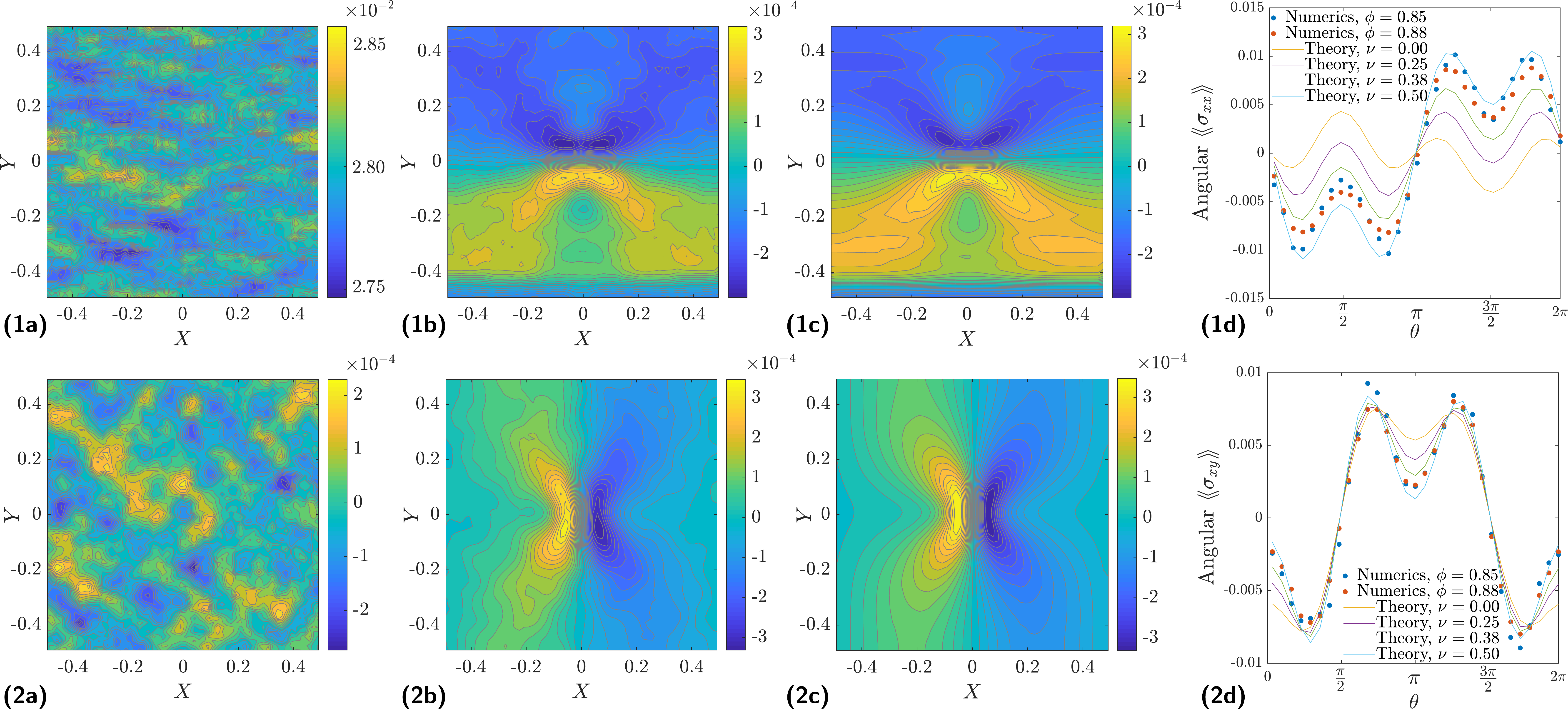}
            \caption{Comparisons between the theoretical predictions and numerical response to a point charge in 2D. The external force distributions used are provided in Eq.~\eqref{eq:charge_distbn_response}. The rows (1) and (2) display the results for $\llangle \sigma_{xx} \rrangle$ and $ \llangle \sigma_{xy} \rrangle$ respectively. The first column (a)  displays the respective components of the background stress fluctuations. The second column (b) displays the response for each component of the stress tensor computed as a difference before and after the external forces are applied. The third column (c) displays the theoretical predictions for the response, computed for the simulation geometry using the Green's function provided in Eqs.~\eqref{eq:Greens_fn_dielectric_vac},~\eqref{eq:2d_response_vacuum_general} and~\eqref{eq:charge_distbn_response}. These theoretical results have been computed using the stiffness constants for $\phi=0.88$ provided in Table~\ref{tbl:stiffness_consts}. The fourth column (d) displays the quantitative comparison for the angular response from theory and numerics, obtained by integrating the response on an annulus of radius $r \in [0.075,0.3]$ centered on the point charge. The numerical results are given for two packing fractions $\phi =0.85~\mathrm{and}~0.88$ and the theoretical results are provided for three different Poisson's ratio $\nu=0.00,~0.25,~0.38~\mathrm{and}~0.50$. The numerical results for the response of $\sigma_{xx}$ have been symmetrized about the $Y$-axis as $\sigma_{xx}(x,y)=\frac{1}{2}(\sigma_{xx}(x,y)+\sigma_{xx}(-x,y))$ in order to reduce the noise in the data. The error bars in the angular response for $\sigma_{xx},~\sigma_{xy}$ are identical in magnitude to the ones provided for $\sigma_{yy}$ in the inset of Fig.~\ref{fig:resp_angular_sigmayy} as they result from the background stress fluctuations in the packings.\label{fig:resp_phi088} }
        \end{figure}
    \end{minipage}
\end{widetext}
this but, as expected exhibit fluctuations about the uniform state. The second column shows the numerical response computed as the difference between the average stress fields with and without the perturbing force Eq.~\eqref{eq:charge_distbn_response}, and the third column is the VCTG prediction of the response to the {\it same} force perturbation.
% the the response to the force distribution in Eq.~\eqref{eq:charge_distbn_response}, computed as the difference of the average stress fields before and after the forces have been applied. The third column shows the response to the force distribution from the VCTG via Eq.~\eqref{eq_stress_resp}. From these results, it is clear that the overall qualitative features of the stress response is accurately captured by the VCTG. 

Stress response measurements are more sensitive than correlation measurements to the inherent fluctuations in the stress background, as seen for example on the left panel of  Fig. \ref{fig:resp_phi088_sigmayy}.
% The numerical measurement of the stress response in granular systems involves the measurement of the of a signal(stress response), over an inherently noisy background(starting stress distribution). 
To ensure that the signal is larger than the noise, we chose a large enough point force and focused on the near-field response.  Since the response decays as a power law,  the further away we are from the localized perturbation, the more difficult it becomes to identify a signal over the noise.
% Also, since the response decays spatially, care must be taken to ensure that quantitative comparisons between theory and numerics are done only in the region close to the point force. This means that our results test the near field predictions of the tensor gauge theory. 

To compute the Poisson ratio from response measurements, we analyze the  `angular'  response obtained by integrating on an annulus of radius $r \in [0.075,0.3]$ centered on the point charge.  The size of this annulus is chosen such that the inner radius is larger than the point charge and the outer radius is small enough that the response has not decayed to the background stress levels.
% The accuracy of the VCTG response predictions is further confirmed by the successful quantitative comparison of the angular response in figures~\ref{fig:resp_angular_sigmayy} and~\ref{fig:resp_phi088}.
Fig.~\ref{fig:resp_angular_sigmayy} presents the comparison between VCTG predictions and and numerical results for $\llangle \sigma_{yy} \rrangle$ at two different packing fractions. Similar comparisons are shown for the remaining components of the stress tensor in Fig.~\ref{fig:resp_phi088}. %The large error bars for the angular response in the inset of  figure~\ref{fig:resp_angular_sigmayy} can be attributed to the background stress fluctuations (see Appendix~\ref{app:numerical_response}).
{The numerical results confirm that that the angular variations are sensitive only to the Poisson's ratio $\nu = \frac{\lambda}{2(\lambda+\mu)}$. 
%Since $\lambda$ and $\mu$ are both positive for our system, due to the repulsive nature of the interactions, we expect $\nu \in [0,0.5]$. 
Quantitatively, we can deduce that $0.4 \le \nu \le 0.5$.  Since $\nu \rightarrow 0.5$ as $\lambda/\mu \rightarrow \infty$, these jammed solids are close to this asymptotic limit.  
It is computationally difficult to make a more accurate determination of $\nu$ since the sensitivity is  reduced  near this asymptotic limit, and as seen from the inset of Fig.~\ref{fig:resp_angular_sigmayy}, there are large error bars due to the background stress fluctuations of the packings. The component of the stress most sensitive to changes in $\nu$ is $\sigma_{xx}$ and, therefore,  measuring this transverse response is essential for determining the elastic moduli.}
% predictions and our numerical results in figures~\ref{fig:resp_angular_sigmayy} and~\ref{fig:resp_phi088} we have $\lambda \gg \mu$, and therefore $\nu \approx 0.5$. 
% Additionally, from the inset of Fig.~\ref{fig:resp_angular_sigmayy}, it is clear that the errors in our response measurements are very large due to the background stress fluctuations of the packings. These two factors present significant difficulties in making an accurate measurement of $\nu$. 

%=========KR============== 

%===========BC========

\section{Tensorial Electrodynamics and dynamics in Emergent Elasticity~\label{sec:tensorEM_emergent}}
%Following the comprehensive discussion of the connection between numerical calculations of the stress state of jammed granular assemblies and the electrostatic limit of the VCTG, and having provided some motivation for the full structure of the tensorial electrodynamics, we now \rimb{provide theoretical implications of the entire set of Tensor  electrodynamics and the gauge structure, as encoded by VCTG, in context of granular media.}%summarise our theory of the mechanical response of granular media and the underlying gauge structure as encoded by VCTG.
The above analysis provide convincing evidence that VCTG captures the continuum description of mechanical response of jammed solids. We now return to the question of dynamics of jammed solids vis-a-vis VCTG.

% above numerical results show compelling evidence that the electrostatic limit of VCTG provides the framework for analyzing the mechanical response of jammed solids. We therefore now attempt to extend the emergent elasticity framework to understand the dynamical properties of such systems.

In Section \ref{sec:reformulation_athermal}, we provided a partial mapping of Newton's equations to the Maxwell's equations of VCT and demonstrated a clear correspondence between the two Gauss's laws and the Faraday's law of VCT by identifying the momentum density as the magnetic charge, external forces as free electric charges, and the magnetic charge current with the double curl of the stress tensor. However, as we mentioned, we currently cannot provide any systematic derivation of the Ampere's law in context of the granular solid from the underlying dynamics.  Instead, we have postulated its existence for completeness of a fully dynamical VCTG theory along with charge conservation (force balance for jammed solids).

% On the electromagnetic side such extensions are quite straight forward and takes the form of Ampere's law given by Eq.~\eqref{eq_ampere} along with the necessary continuity equations for both the electric and magnetic charges. However, unlike the two Gauss's laws and the Faraday's law, we currently cannot provide a systematic derivation of the Ampere's law in context of the granular solid from the underlying dynamics. Instead, as noted above, we postulate it for the moment and indicate its consequence in terms of the dynamics of the jammed assembly. 

We now take a step back to  examine in a bit more detail the nature of the dynamics of an assembly of grains to motivate the Maxwell's equations.  For an assembly of particles with purely contact interactions, the dynamics, when coarse-grained over a UV length-scale {$l_c$},  manifests in two different ways: (1) through rearrangements of the contact network with minimal displacement of the particles themselves (more particularly the displacement of any particle in the assembly is less than the length-scale {$l_c$}), and, (2) The bodily motion of the individual particles for length scales {$\gg l_c$}. At the level of individual particles there is of course no distinction of the two types, but as far as the assembly is concerned, the first denotes the dynamics of the network and the force distribution at the contacts (with few contacts breaking), while the second leads to substantial rearrangement of the network itself. Contact breaking necessarily induces non-affine displacements, and close to the unjamming transition even infinitesimal perturbations can lead to the breaking of contacts~\cite{Schreck2011}. Deep within the jammed state, however it is possible to explore network configurations without inducing contact breaking~\cite{Geng2001}. It is, however, difficult to create an affine strain field inside granular assemblies and special protocols have to be designed to achieve these~\cite{Behringer_2018}.
% \rim{\bf [Do we make a comment about affine and non-affine displacements/strains here ?]}

Indeed it is the rearrangements of contact forces, albeit in the static limit, that identified  the emergent elastic behaviour of granular media with VCTG, the dielectric response of VCT with forces being identified with the vector electric charges. It is important to reiterate that the above separation of the dynamics is only meaningful if the coarse-graining length-scale, {$l_c > a$}, where $a$ is the diameter of the particle and that the VCTG assumes the existence of a finite {$l_c$}.

The response of granular media to small time varying boundary forces are then expected to be completely captured by purely the motion of electric charges if we coarse-grain up to the scale {$l_c$}.
In other words, {$l_c$} defines the smallest region within which a given contact network can achieve force and torque balance which then appears in the VCTG in terms of the coarse grained fields.  In the context of jamming, this is known as the isostatic length scale~\cite{tkachenko1999stress,wyart2005geometric}, which can be deduced from constraint counting,  and diverges at the unjamming transition. The isostatic point is special in that the boundary forces completely determine the stress-state in the bulk. Equating {$l_c$} with the isostatic length scale, it is well-established that this length scale diverges as $\left(\llangle z -z_{iso}\rrangle\right)^{-\nu}$, where $\llangle z \rrangle$ is the average number of contacts per grain, and  $z_{iso}$ defines the minimum number of contacts per grain that are needed to achieve force and torque balance. In $d$ dimensions,  for frictionless spheres, $z_{iso}= 2 d$.  In the limit of infinite friction coefficient, $z_{iso} = (d+1)$. In general, $z_{iso}$ depends non-trivially on the friction coefficient, however the exponent $\nu$ seems to be independent of the friction coefficient~\cite{Somfai2007}. For infinitely rigid particles, hard spheres, $\llangle z \rrangle$ is bounded above by $2 d$.  Since granular materials are composed of particles near this hard sphere limit, $\llangle z \rrangle \approx z_{iso}$, and therefore {$l_c \gg a$}.

At the isostatic point ($\llangle z \rrangle = z_{iso}$), the particles are just touching with zero compression of grains.  Within VCTG, this translates to the medium losing its ability to polarize and create bound charges (contact forces) in response to free charges (external forces). The external forces therefore, completely determine the stress state in the bulk. Most earlier field theories of granular stresses have focused on this limit~\cite{Ball2002,DeGiuli2020,deGiuli_PRL}.
% \rim{\bf [What does the last statement intends to convey in not clear to me.] ({\it BCcomment: The point I was trying to make is that in the isostatic limit, the particle are just touching and therefore this is not a polarizable medium anymore. The external forces completely define the properties of the system including the network})}.

The above interpretation of the two classes of dynamical responses is consistent with the identification of the magnetic charge density with the momentum density~(Eq.~\eqref{eq:mag}) as only the second class of dynamics leads to finite momentum density for the grains when coarse-grained over {$l_c$}. This gives clear meaning to the conservation of magnetic charge density and its cross moments as  nothing but the conservation of momentum and and angular momentum respectively (see Appendix~\ref{appendix:A}) in the absence of external forces or torques as is applicable for a particle in ballistic motion. %\rim{{\bf [May be we need to discuss what we want to say regarding this]} The electric and magnetic charges thus interchange roles as we go from jammed to unjammed: the former being relevant when a contact network can be defined and the latter when it cannot.}

The dynamics of the electric charge (force), and magnetic charge (momentum density) act as sources of the corresponding fields in VCTG with each charge being conserved. With this assumption, we postulate Ampere's law as given by Eq.~\eqref{eq_ampere_vctg}. %\rimb{We have definitively identified stress as the electric field, however, the identity of the magnetic field still alludes us.} % Within VCTG, postulating an Ampere's law, 
% \begin{align}
%     \epsilon_{iab}\epsilon_{jcd}\partial_a\partial_c B_{bd}&=\partial_t E_{ij}+J_{ij}
% \end{align}
%where $J_{ij}$ is the total electric current that ensures the conservation of electric charge (force-balance), we have the full set of ``Maxwell's equations''.
The essence of Faraday's and Ampere's is that they feed back on each other. Thus {the full set of Maxwell's equation implies a third class of dynamical modes related to the ``photons" of the VCT that arises from the self-sustaining feedback of the electric and magnetic sectors. It is known~\cite{Pretko2017} that for VCT, the photon dispersion is given by
\begin{align}
\label{eq:dispersion}
    \omega\sim k^2
\end{align}
%\rim{\bf [Check for d=2]} 
which predicts that the density of states of the analog of low-energy photons in VCTG to be $\rho(\omega)\sim \omega^{(d-2)/2}$. This non-Debye form of low energy density of states differs from the predictions for the phonon density of states jammed frictionless solids.  Based on a Hessian analysis, phonons in frictionless jammed solids
% those estimated from the numerical harmonic estimations using 
shows Debye-like behaviour $\rho(\omega)\sim \omega^{d-1}$ for $\omega$ below a characteristic frequency $\omega^*$~\cite{jamming_epitome_2003}.  However, we note that the low energy density of modes estimated from velocity-velocity correlations~\cite{daniels_acoustic} in granular solids indicate departure from the Debye predictions in disordered packings.  These observations along with our prediction based on VCTG suggests that the low energy density of states has contributions from different origins with the softest ones, presumably, arising from the emergent {\it photons}. A more detailed study of the dynamics is a natural direction to elucidate the nature of these different contributions. %\rimm{\bf [ SB : Will go through the references and polish this up]}
}

\subsection{The meaning of the gauge structure in VCTG}
\label{subsec:gauge}

% \rim{Having discussed comprehensive connection between the numerical calculations of granular assembly and the electrostatic limit of the VCTG and provided some motivation for the possible tensor electrodynamics, we now summarise} our theory of the mechanical response of granular media and the underlying gauge structure as encoded by tensor electromagnetism. 

Central to the gauge structure is the position of the individual grains in the assembly of the athermal granular solid that forms the jammed structure. Unlike in crystalline solids there is no reference unique zero stress configuration for granular particles such that the deviation from such an unique configuration can be used to define a strain field. In other words, a strain field defined by choosing a reference configuration should not have an observable consequence for the mechanical properties. This is exactly the content of 
\begin{equation}
E_{ij} = \frac{1}{2} (\partial_i \varphi_j + \partial_j \varphi_i )~.
\label{eq:Ecompatibility}
\end{equation}
where the $\varphi_i$ are the electrostatic potentials, {\it i.e.}, the potentials $\boldsymbol{\varphi}$  play a role analogous to that of the displacement field in the strain tensor of elasticity theory, however unlike the displacement field, $\boldsymbol{\varphi}$ should not have an observable consequence in a self averaged theory. This non-existence of the reference configuration along with the fact that local force and torque balance needs to be treated at equal level for all configurations results in a coarse grained gauge theory.

We note that while for calculations it may be convenient to choose a reference configuration and define a strain field with respect to it, within our formalism that would  be akin to fixing a gauge. It would therefore be useful to understand the structures of existing theories of elasticity of amorphous solids, and especially the appearance of non-affine strain fields,  from this perspective~\cite{Falk2011}.  A widely used measure of plastic response is based on the extent of non-affine displacements and referred to as $D^2_{min}$~\cite{Falk2011,Utter2008}.   This measure is defined by looking at deviations from a locally defined optimum strain tensor.  We suggest that this procedure 
can be viewed as defining the gauge fixing conditions since the displacements measured from these  states are not physical observables as they are based on the {\it locally} optimized strain.  We plan to investigate the detailed relationships between gauge potentials and non-affine displacements in future work.
%%%%%%%%%%%%%%%%%%%%%%%%

\section{Summary and Outlook}

\paragraph*{Jammed Granular media is a symmetric rank-2 U(1) tensor dielectric:} In this paper we have established a rigorous theoretical framework for the mechanical response of non-thermal, disordered solids such as those found in granular media, dense suspensions and gels.
% mapping between the stress fluctuations and response in static granular materials to tensor electrostatics of polarizable media. 
Our framework does not rely on the existence of a well-defined displacement field and the notion of a continuum strain tensor that is associated with crystalline solids, and instead relies on a mapping between the stress fluctuations and response in such jammed solids to the tensor electrostatics of polarizable media, that we have referred to as VCTG.  The force and torque balance conditions that govern such materials at the microscopic level are naturally interpreted as a Gauss's law, leading to an emergent low-energy description in terms of vector charges, which leads to a continuum Maxwell theory of a $U(1)$ tensor field at long wavelengths. The salient idea underlying the mapping of mechanical response to a gauge theory is that gauge-redundancy is a natural consequence of the lack of a {\it unique} stress-free reference structure in non-thermal amorphous solids.
% This stress-only description~\cite{Bouchaud2002} circumvents the lack of a reference structure in amorphous materials. 

The identification of the electric {displacement field, $D_{ij}$,} in the tensor gauge theory with the stress field in amorphous solids allowed us, through the construction of a Landau-Ginzburg action of a polarizable medium, to predict both stress-stress correlations and stress-response of jammed solids to external forces.  Comparisons between theory and numerical or experimental measurements provides a mechanism for computing  the ``emergent elastic moduli'' of such solids: emergent since these are not material properties but depend on preparations protocols.  We demonstrate the remarkable success of this approach by describing in detail its application to  frictionless jammed solids in both 2D and 3D. In an earlier paper, we had presented results of comparison to experiments on frictional packings in 2D~\cite{PhysRevLett.125.118002}. 

% in the VCTG theory of mechanical response makes concrete predictions about both stress-stress correlations and stress-response to external forces. 
% The predictions from this theory can be concretely compared with both numerics as well as experiments.
%The grain-level constraints of mechanical equilibrium naturally appear as a Gauss's law of an , which we identify as the electric field. 
% The identification of the electric field allowed us, through the construction of a Landau-Ginzburg action of a polarizable medium, to predict the stress-stress correlations in such static granular materials, that display a remarkable match with numerically generated configurations of soft frictionless disks. In addition to numerical simulations, we have also corroborated the predictions of the tensor gauge theory with experiments of photoelastic disks in two dimensions \cite{PhysRevLett.125.118002}.
% Additionally, our mapping also leads to predictions for the response of static granular materials to external forces, which are identified as vector charges. Once again, the predicted response from numerical simulations matches the predictions from our theory remarkably well.

A central feature of the vector-charge theory is that it is a linear theory, as is  evident from the field equations, which are the analog of Maxwell's equations (see Appendix \ref{appendix:A}). This may raise a sense of concern regarding applying it to the theory of granular elasticity which is a non-linear problem in general. It is however important to realise that granular solids {\it only} represent a dielectric for the tensor gauge theory and in general the response of such a dielectric is non-linear. However, deep inside the jammed regime, the dielectric behaves linearly in response to small added forces allowing us to use a dielectric theory with a constant polarizability tensor, which then has a structure analogous to the linear elasticity. {In addition to the possibility of such a non-linear response, the VCTG framework can also be used to analyze the analog of Debye screening in the presence of ``mobile charges'' (contact forces rearranging dynamically), as we show in \ref{appendix:A}.  In VCTG, the screening is anisotropic and the analog of screening length is a screening tensor. }

\paragraph*{Dynamic Response} We have also presented a mapping of the  full tensorial electrodynamics to the dynamic response of granular solids.  A crucial feature  of the mapping is the identification of the analog of a magnetic field that is ``sourced'' by the momentum density.  This dynamical theory points to the  existence of a length-scale $l_c$ that separates the electric and magnetic sectors. Identifying $l_c$ with the well-known isostatic length-scale is a definite avenue for calculations in the immediate future to test the validity of the proposed tensor gauge theory in the dynamic regime, extending the comprehensive tests in the static regime that we have provided in this paper and in Ref. \cite{PhysRevLett.125.118002}. A natural question for experimental tests of the theory, in this regard, is: can the dynamics of the network be probed without generating contact breaking? Experiments on stress transmission in assemblies of photoelastic particles~\cite{Geng2001}, indicates that this is indeed possible. A striking prediction of the full tensorial electrodynamics is the appearance of the $\omega\sim k^2$ dispersing emergent photons (three of them) at low energies and long-wavelengths. These low-energy modes represent the dynamic response of such systems as predicted within the tensor gauge theory, however the exact nature of these modes within numerical simulations and experiments remain to be identified.{In this regard, while we have identified stress with the electric displacement field, the microscopic identification of the magnetic field  in terms of the geometry and dynamics of the granular assembly via Eq.~\eqref{eq:mag} remains to be fully explored.}

\paragraph*{Other open questions:} A natural follow up question pertains to existing theories that use a reference configuration to define a displacement and a strain field. While a natural interpretation of these constructions are that they are {\it gauge-fixed} versions of our VCTG, the concrete mapping between these theories and VCTG requires further investigation following the approach briefly discussed in  Section~\ref{sec:tensorEM_emergent} in the context of measures of non-affine strain in amorphous, jammed solids~\cite{Falk2011}. In the same vein, the  framework of VCTG is always well defined in terms of boundary forces, in contrast to boundary strain which may or may not map to a unique set of boundary forces. A special case where the mapping is expected to work is rheological measurements with boundary conditions that ensure that all boundary forces are normal to the boundary (no-slip condition) and provides for further non-trivial applications of the VCTG. Two similar but distinct extensions of the VCTG formalism are to understand the mechanical response across--(1) the transition from amorphous jammed solids to broken symmetry crystals, and, (2) the connection with the glass transition. For the former, it appears that a mechanism that selects out a unique zero-stress reference configuration and hence {\it breaks down} the gauge structure is required, while for the latter we need to understand the dynamics of VCTG at finite temperature to explore the possibility of growing length and time-scales. In regards to this last point, in a recent paper~\cite{Zippelius}, the emergence of shear rigidity at the glass transition has been investigated using the Zwanzig-Mori formalism.  The correlation functions of the shear stress in the ``glass'' bear remarkable similarity to the predictions of VCTG.

\vspace{-2mm}

%%%%%%%%%%%%%%%%%%%%%%%%%%%%%%%%%%%%%%%%%%
\acknowledgments
{\small
The authors thank A. Seth, R. Moessner, S. S. Ray, A. Dhar, C. Dasgupta, A. Kundu, S. Ramaswamy and P. Chaudhuri for fruitful discussions. S. B. acknowledges funding from Max Planck Partner group Grant at ICTS, Swarna jayanti fellowship grant of SERB-DST (India) Grant No. SB/SJF/2021-22/12 and the Department of Atomic Energy, Government of India, under Project No. RTI4001. JN, MD, and BC  were supported by NSF CBET award number 1916877 and NSF DMR award number 2026834. 
The work of JN and KR was funded in part by intramural funds at TIFR Hyderabad from the DAE.
We acknowledge computational support from the Brandeis HPCC which is partially supported by the NSF through DMR-MRSEC 2011846 and OAC-1920147.
}%
\clearpage
%%%%%%%%%%%%%%%%%%%%%%%%%%%%%%%%%%%%%%%%%%
\appendix

\renewcommand\thefigure{\thesection\arabic{figure}}    
\section{Details of the Vector charge theory (VCT)\label{appendix:A}}
\setcounter{figure}{0}
Here we briefly summarise the electromagnetism of the vector charges which is a particular type of tensor electromagnetism for rank-2 symmetric tensor electric and magnetic fields that are sourced by vector charges and tensor currents~\cite{Pretko2017}. 
\vspace{2mm}
\paragraph{The Maxwell's Equations of VCT:}

The Maxwell equations  in three spatial dimension are given by
\begin{align}
    \partial_i E_{ij}&=\rho_{j} \label{append:Gauss} \\
    \partial_iB_{ij}&=\tilde\rho_{j}\\\label{append:faraday}
    \epsilon_{iab}\epsilon_{jcd}\partial_a\partial_c E_{bd}&=-\partial_tB_{ij}-\tilde J_{ij}\\
    \epsilon_{iab}\epsilon_{jcd}\partial_a\partial_c B_{bd}&=\partial_tE_{ij}+J_{ij}
    \label{eq_ampere}
\end{align}
{where $E_{ij} (B_{ij}$, $\rho_i (\tilde \rho_i)$, and $J_{ij} (\tilde J_{ij})$ are the electric (magnetic) field, charge, and currents respectively.
} A solution to these equations can be formulated in terms of gauge potentials $A_{ij}$ and $\varphi_i$ as:
$E_{ij} = -\frac{1}{2} (\partial_i \varphi_j + \partial_j \varphi_i) - \partial_t A_{ij}$, and $B_{ij} = \epsilon_{iab}\epsilon_{jcd}\partial_a\partial_c A_{bd}$.
The gauge transformations, 
$A_{ij} \rightarrow A_{ij} + \frac{1}{2}(\partial_{i} \psi_j + \partial_j \psi_i)$, and $\varphi_{i} \rightarrow \varphi_i - \partial_t \psi_i$, 
leave  $B_{ij}$ and $E_{ij}$ unchanged, respectively.
{In two spatial dimensions the magnetic field is a scalar. Since we shall use this, it is convenient to write this separately:}
{\begin{align}
    \partial_j {E_{ij}^{\rm{2D}}}&=\rho_{i} \label{append:Gauss2D} \\
    \partial_i B^{\rm{2D}} &=\tilde\rho_{i}\\\label{append:faraday2D}
    \epsilon_{ia}\epsilon_{jb}\partial_i \partial_j {E_{ab}^{\rm{2D}}}&=-\partial_tB^{\rm 2D} -{\tilde J}^{\rm 2D}\\
    \epsilon_{ia}\epsilon_{jb}\partial_i\partial_j B &=\partial_t{E^{\rm{2D}}_{ab}}+{J^{\rm {2D}}_{ab}}~.
    \label{eq_maxwells2D}
\end{align}}
The corresponding potential formulation becomes:
${E_{ij}^{\rm{2D}}} = -\frac{1}{2} (\partial_i \varphi_j + \partial_j \varphi_i) - \partial_t A_{ij}$, and $B^{\rm{2D}} = \epsilon_{ia}\epsilon_{jb}\partial_i\partial_j A_{ab}$

\vspace{4mm}
\paragraph{Conserved quantities in the VCT:}
The VCT conserves the total vector charge (both electric and magnetic)
\begin{align}
    \boldsymbol{Q}=\int d^3\boldsymbol{r}~\boldsymbol{\rho}
    \label{eq:chargeconserve}
\end{align}
and corresponding cross moment in a charge neutral system 
\begin{align}
    \boldsymbol{T}=\int d^3\boldsymbol{r} ~\boldsymbol{\rho}\times\boldsymbol{r}=\int d^3\boldsymbol{r}~\boldsymbol{\mathcal{T}}
    \label{eq:amcon}
\end{align}
where $\boldsymbol{\mathcal{T}}$ is the cross moment density. For electric charges we can re-write the above equation to get
\begin{align}
    {T}_i&=\epsilon_{ijk}~\int d^3\boldsymbol{r}~\left[\rho_j r_k-\rho_k  r_j\right]\nonumber\\
    &=\epsilon_{ijk}~\int d^3\boldsymbol{r}~\left[\partial_{m}E_{mj}r_k-\partial_m E_{mk}r_j\right]
\end{align}
where in the second line we have used Gauss's law (Eq.~\eqref{append:Gauss}). Transferring the integral with partial integration, we get
\begin{align}
    {T}_i&=\epsilon_{ijk}~\int d^3\boldsymbol{r}~\partial_m\left[E_{mj}r_k- E_{mk}r_j\right]\nonumber\\
    %&~~-\epsilon_{ijk}\int d^3\boldsymbol{r}~\left[E_{kj}-E_{jk}\right]
    &=\epsilon_{ijk}~\int dS_m~\left[E_{mj}r_k- E_{mk}r_j\right]
    \label{eq:surfacedipole}
\end{align}
{where, in the second line, we have used the divergence theorem to write it as a pure surface term.}
 
 A similar set of relations can be derived for the magnetic charge cross moment
\begin{align}
    \tilde{T}_i&=\epsilon_{ijk}~\int d^3\boldsymbol{r}~\left[\tilde\rho_j r_k-\tilde\rho_k  r_j\right].
\end{align}
%%%%%%%%%%%%%%%%%%%%%

\vspace{2mm}
\paragraph{The electrostatic limit:}
{The electrostatic limit is obtained  by setting the time derivative and the currents to zero in the Maxwell's equations (Eqs.~\eqref{append:Gauss}-\eqref{eq_ampere}), which gives}
\begin{align}
    \partial_i E_{ij}&=\rho_{j} \nonumber \\
    \epsilon_{iab}\epsilon_{jcd}\partial_a\partial_c E_{bd}&=0
    \label{eq:vectorelectrostatics}
\end{align}
For a general charge distribution, $\rho_i(\boldsymbol{r})$, the {d-dimensional} solution for electric field due to Gauss's Law in tensor electromagnetism~\cite{Pretko2017} is given by
{\begin{align}
    E_{ij}(\boldsymbol{r})=\int d^d\boldsymbol{r}'~\mathcal{G}^{d}_{ijk}(\boldsymbol{r}-\boldsymbol{r}')\rho_k(\boldsymbol{r}')
    \label{eq_electgreens}
\end{align}
where
{\small
\begin{align}
\mathcal{G}^{3D}_{i j k} (\boldsymbol{R}) =& \frac{1}{8 \pi}\left(\frac{\delta_{i k} R_{j}}{R^3} + \frac{\delta_{j k} R_{i}}{R^3} - \frac{\delta_{i j} R_{k}}{R^3} + 3\frac{R_i R_j R_k}{R^5}\right), \label{eq:3d_vacuum_real_Green's}\\
\mathcal{G}^{2D}_{ijk}(\boldsymbol{R}) =& \frac{1}{4\pi} \left(\frac{2 R_i R_j R_k}{R^4} -  \frac{\delta_{ij} R_k}{R^2} +  \frac{\delta_{ik} R_j}{R^2} +  \frac{\delta_{jk} R_i}{R^2}\right),
\label{eq:2d_vacuum_real_Green's}
\end{align}}
}%
with $\boldsymbol{R} = \boldsymbol{r}-\boldsymbol{r}'$.

{To construct the dielectric formulation, we express the bound charge density, $\rho_{i}^{\rm bound}$, in terms of the dipole moment of the induced charge distribution, following the path of usual electromagnetism~\cite{jackson} by constructing a multi-pole expansion of the potential for a localised ``vector charge'' distribution (see ~\cite{SI} for details). As shown in ~\cite{SI}, the electrostatic potential created by many dipole moments in 3D is:
\begin{align}
    \varphi_i^{(2)}(\boldsymbol{r})=\int d^3\boldsymbol{r}'~\partial_k \mathcal{G}^{3D}_{ij}(\boldsymbol{r}-\boldsymbol{r}')\mathcal{P}_{jk}(\boldsymbol{r}')
\end{align}
where $\mathcal{G}^{3D}_{ij}(\boldsymbol{r}-\boldsymbol{r}')$ is given by Eq.~\eqref{eq:potgreenexp}, and $\mathcal{P}_{ij}(\boldsymbol{r})$ is the dipole density which is related to the induced volume charge density via Eq.~\eqref{eq:bound_dipole}. In addition, similar to ordinary electromagnetism~\cite{jackson}, there is an induced surface charge-density given by $\rho^{\rm surf}_i=\hat{n}_j\mathcal{P}_{ij}$, where $\hat{\bf n}$ is normal to the surface.  

As in any dielectric, the polarizability tensor relates the electric displacement field, $D_{ij}$ and the electric field, $E_{ij}$. In the Supplemental Material~\cite{SI}, we provide detailed derivations for the correlations and response in a dielectric characterized by a polarizability tensor, $\hat{\Lambda}^{-1}$, in $q$-space. We summarize the results below. The response is given by:
\begin{equation}
     \langle D_{ij} (\boldsymbol{q}) \rangle_{\rho^{\rm ext}} = G_{ijk}({\bf q})~\rho^{\rm ext}_k ({\bf q}) ~,
     \label{eq:genresponse}
\end{equation}
{where $G_{ijk}({\bf q})$ is given by Eq.~\eqref{eq_stress_resp}.}}

{For the response in 2D, with the form of $\Lambda^{-1}$ given by Eq.~\eqref{eq:lambda_2d_iso}
\begin{align}
    {G}_{xxx}(\boldsymbol{q})&={\mathcal{G}}_{xxx}^{\,2D}(\boldsymbol{q})-\left(\frac{\nu}{1-\nu}\right){\mathcal{G}}_{yyx}^{\,2D}(\boldsymbol{q}), \nonumber \\
    {G}_{xxy}(\boldsymbol{q})&={\mathcal{G}}_{xxy}^{\,2D}(\boldsymbol{q})+\left(\frac{\nu}{1-\nu}\right){\mathcal{G}}_{xyx}^{\,2D}(\boldsymbol{q}), \nonumber \\
    {G}_{yyx}(\boldsymbol{q})&={\mathcal{G}}_{yyx}^{\,2D}(\boldsymbol{q})+\left(\frac{\nu}{1-\nu}\right){\mathcal{G}}_{xyy}^{\,2D}(\boldsymbol{q}), \nonumber \\
    {G}_{yyy}(\boldsymbol{q})&={\mathcal{G}}_{yyy}^{\,2D}(\boldsymbol{q})-\left(\frac{\nu}{1-\nu}\right){\mathcal{G}}_{xxy}^{\,2D}(\boldsymbol{q}), \label{eq:Greens_fn_dielectric_vac} \\
    {G}_{xyx}(\boldsymbol{q})&={\mathcal{G}}_{xyx}^{\,2D}(\boldsymbol{q})+\left(\frac{\nu}{1-\nu}\right){\mathcal{G}}_{xxy}^{\,2D}(\boldsymbol{q}), \nonumber \\
    {G}_{xyy}(\boldsymbol{q})&={\mathcal{G}}_{xyy}^{\,2D}(\boldsymbol{q})+\left(\frac{\nu}{1-\nu}\right){\mathcal{G}}_{yyx}^{\,2D}(\boldsymbol{q}). \nonumber 
\end{align}
with $\mathcal{G}^{2D}_{ijk}({\bf q})$ being given by Eq.~\eqref{eq:2d_response_vacuum_general}.
}

\vspace{5mm}
\paragraph{\hspace{-3mm}Green's Functions for the electrostatic potential:} The general solution to the zero curl condition in Eq.~\eqref{eq:vectorelectrostatics} is 
\begin{align}
    E_{ij}=-\frac{1}{2}\left(\partial_i\varphi_j+\partial_j\varphi_i\right).
    \label{eq:electrostatic}
\end{align}

For a generic charge distribution, $\rho_j(\boldsymbol{r})$, the potential is given by
\begin{align}
    \varphi_i^{3D}(\boldsymbol{r})=\int d^3\boldsymbol{r}'~\mathcal{G}^{3D}_{ij}(\boldsymbol{r}-\boldsymbol{r}')~\rho_j(\boldsymbol{r}'),
    \label{eq:potgreen}
\end{align}
where
\begin{align}
    \mathcal{G}^{3D}_{ij}(\boldsymbol{R})=\frac{1}{8\pi}\left[\frac{3\delta_{ij}}{|\boldsymbol{R}|}+\frac{R_i R_j}{|\boldsymbol{R}|^3}\right]
    \label{eq:potgreenexp}
\end{align}
is the Green's function for the potential $\boldsymbol{\varphi}$ in 3D. 
For a point charge $\rho_j=Q_j\delta^3(\boldsymbol{r})$  Eq.~\eqref{eq:potgreen} reduces to~\cite{Pretko2017}
 \begin{align}
     \varphi_i^{3D}(\boldsymbol{r})=\frac{1}{8\pi}\left(\frac{3Q_i}{\left|r\right|}+\frac{(\boldsymbol{Q}\cdot\boldsymbol{r})r_i}{\left|r\right|^3}\right).
     \label{eq:point_potential}
 \end{align}
{ In 2D, the corresponding Green's Function is:
\begin{align}
    \mathcal{G}^{2D}_{ij}(\boldsymbol{R})=\frac{1}{4\pi}\left[-3 \left[\ln \left(\frac{|\boldsymbol{R}|}{C}\right) \right]\delta_{ij}+\frac{R_i R_j}{|\boldsymbol{R}|^2}\right].
    \label{eq:potgreenexp2D}
\end{align}
Here $C$ is a constant that sets a length scale.
The potential in 2D is: 
{\small
\begin{align}
     \varphi_i^{2D}(\boldsymbol{r})=\frac{1}{4\pi}\left (-3\left[\ln \left(\frac{|\boldsymbol{r}|}{C}\right)\right] Q_i +\frac{(\boldsymbol{Q}\cdot\boldsymbol{r})r_i}{\left|r\right|^2}\right ).
     \label{eq:point_potential2D}
 \end{align}
}
}%

%%%%%%%%%%%%%%%%%

\vspace{3mm}
\paragraph{Stress-stress correlation in 2D and 3D:}
Here we present explicit forms for the stress-stress correlations in $q$-space, corresponding to the form of $\Lambda^{-1}$ given by Eq.~\eqref{eq:lambda_2d_iso}.
In 2D, these are:
\begin{align}
    C_{xxxx}(q,\theta) &=4K_{2D} \sin^4\theta \nonumber\\ 
    C_{yyyy}(q,\theta) &=4K_{2D} \cos^4 \theta \nonumber\\
    C_{xyxy}(q,\theta) &=4K_{2D} \sin^2 \theta \cos^2 \theta \label{eq:2d_correlations_explicit}\\
    C_{xxxy}(q,\theta) &=4K_{2D} \left(-\sin^3 \theta \cos \theta\right) \nonumber\\ 
    C_{xxyy}(q,\theta) &=4K_{2D} \sin^2 \theta \cos^2 \theta \nonumber \\
    C_{xyyy}(q,\theta) &=4K_{2D} \left(-\sin \theta \cos^3 \theta\right) \nonumber
\end{align}
with {\small$K_{2D}=\mu\left(\frac{\lambda+\mu}{\lambda+2\mu}\right)= \frac{\mu}{2(1-\nu)}$}, and $\nu = \frac{\lambda}{2(\lambda+\mu)}$ is the Poisson's ratio. 
The explicit forms of the three dimensional stress-stress correlations are given by,
{\small
\begin{align}
    C_{xxxx}(q,\theta,\Phi) &= 4(\mathcal{K}_1+\mathcal{K}_2) \left[\sin ^2\theta  \sin ^2\Phi +\cos ^2\theta \right]^2 \nonumber\\ 
    C_{yyyy}(q,\theta,\Phi) &= 4(\mathcal{K}_1+\mathcal{K}_2) \left[\sin ^2\theta  \cos ^2\Phi +\cos ^2\theta \right]^2 \nonumber\\
    C_{zzzz}(q,\theta,\Phi) &= 4(\mathcal{K}_1+\mathcal{K}_2) \left[\sin ^4\theta\right]
    \label{eq:3d_correlations_explicit1}
\end{align}
\vspace{3mm}
\begin{align}
        C_{xxxy}(q,\theta,\Phi) =-4(\mathcal{K}_1+\mathcal{K}_2) \Big(&\sin ^2\theta \sin \Phi \cos \Phi\nonumber \\
        & \left[\sin ^2\theta \sin ^2\Phi+\cos ^2\theta\right] \Big) \nonumber\\
        C_{xxxz}(q,\theta,\Phi) =-4(\mathcal{K}_1+\mathcal{K}_2) \Big(&\sin \theta  \cos \theta \cos \Phi \nonumber\\
        & \left[\sin ^2\theta \sin ^2\Phi+\cos ^2\theta\right]\Big) \nonumber\\
        C_{yyyx}(q,\theta,\Phi) =-4(\mathcal{K}_1+\mathcal{K}_2) \Big(& \sin ^2\theta \sin \Phi \cos \Phi \nonumber \\
        &\left[\sin ^2\theta \cos ^2\Phi+\cos ^2\theta\right]\Big) \nonumber\\
        C_{yyyz}(q,\theta,\Phi)=-4(\mathcal{K}_1+\mathcal{K}_2) \Big(& \sin \theta \cos \theta  \sin \Phi \nonumber \\
        &\left[\sin ^2\theta \cos ^2\Phi+\cos ^2\theta\right]\Big) \nonumber\\
        C_{zzzx}(q,\theta,\Phi) =-4(\mathcal{K}_1+\mathcal{K}_2) & \sin ^3\theta \cos \theta \cos \Phi \nonumber\\
        C_{zzzy}(q,\theta,\Phi) =-4(\mathcal{K}_1+\mathcal{K}_2) &  \sin ^3\theta \cos \theta \sin \Phi
\end{align}

\begin{align}
    C_{xyxy}(q,\theta,\Phi) =&(\mathcal{K}_1+\mathcal{K}_2)~ \sin ^4\theta  \sin ^2 2\Phi +\nonumber \\
    &(2\mathcal{K}_1+\mathcal{K}_2)~\cos ^2\theta \nonumber\\ 
    C_{xzxz}(q,\theta,\Phi) =&(2\mathcal{K}_1+\mathcal{K}_2) \sin ^4\theta  \sin ^2\Phi+ \sin ^2\theta \cos ^2\theta \Big( \nonumber \\
    &(2\mathcal{K}_1+3\mathcal{K}_2)\cos^2 \Phi+(2\mathcal{K}_1+\mathcal{K}_2)\Big) \nonumber \\
    C_{yzyz}(q,\theta,\Phi) =&(2\mathcal{K}_1+\mathcal{K}_2) \sin ^4\theta  \cos ^2\Phi+ \sin ^2\theta \cos ^2\theta \Big( \nonumber \\
    &(2\mathcal{K}_1+3\mathcal{K}_2)\sin^2 \Phi+(2\mathcal{K}_1+\mathcal{K}_2)\Big)
    \end{align}

\begin{align}
    C_{xxyy}(q,\theta,\Phi) &= \left[(\mathcal{K}_1+\mathcal{K}_2) \sin ^4\theta \sin ^22\Phi+2 \mathcal{K}_2~ \cos ^2\theta\right] \nonumber\\
    C_{xxzz}(q,\theta,\Phi) &=\left[(\mathcal{K}_1+\mathcal{K}_2)\sin ^2 2\theta \cos^2 \Phi +2 \mathcal{K}_2  \sin ^2\theta \sin ^2\Phi\right] \nonumber\\
    C_{yyzz}(q,\theta,\Phi) &= \left[(\mathcal{K}_1+\mathcal{K}_2)\sin ^2 2\theta \sin^2 \Phi +2 \mathcal{K}_2  \sin ^2\theta \cos ^2\Phi\right]
\end{align}
\begin{align}
     C_{xxyz}(q,\theta,\Phi) &=\sin 2\theta \sin \Phi \left[\mathcal{K}_2 -2(\mathcal{K}_1+\mathcal{K}_2)\cos^2\Phi \sin^2\theta \right]\nonumber\\
     C_{yyxz}(q,\theta,\Phi) &=\sin 2\theta \cos \Phi \left[\mathcal{K}_2 -2(\mathcal{K}_1+\mathcal{K}_2)\sin^2\Phi \sin^2\theta \right] \nonumber\\
      C_{zzxy}(q,\theta,\Phi) &= \sin ^2\theta \sin 2\Phi \left[(\mathcal{K}_1+\mathcal{K}_2)\cos 2\theta +\mathcal{K}_1 \right]
\end{align}

\begin{align}
        C_{xyxz}(q,\theta,\Phi) =-\sin \theta &\cos \theta \sin \Phi \Bigg[(\mathcal{K}_1+\mathcal{K}_2) \Big( \nonumber\\
        &\cos 2\theta-2 \sin ^2\theta \cos 2\Phi\Big)+\mathcal{K}_1\Bigg] \nonumber\\
        C_{xyyz}(q,\theta,\Phi) =-\sin \theta &\cos \theta \sin \Phi \Bigg[(\mathcal{K}_1+\mathcal{K}_2) \Big( \nonumber\\
        &\cos 2\theta+2 \sin ^2\theta \cos 2\Phi\Big)+\mathcal{K}_1\Bigg] \nonumber\\
        C_{xzyz}(q,\theta,\Phi) = \sin ^2\theta& \sin 2\Phi \left[(\mathcal{K}_1+\mathcal{K}_2)\cos 2\theta+\frac{K_2}{2} \right]
        \label{eq:3d_correlations_explicit}
\end{align}
}%
where {\small$\mathcal{K}_1=\mu\left(\frac{\mu}{\lambda+2\mu}\right)$, $\mathcal{K}_2=\mu\left(\frac{\lambda}{\lambda+2\mu}\right)$}. From these, we can obtain $\lambda$ and $\mu$ as {\small$\mu=2\mathcal{K}_1+\mathcal{K}_2,~\lambda=\frac{\mathcal{K}_2}{\mathcal{K}_1}\left(2\mathcal{K}_1+\mathcal{K}_1\right)$}. 

\vspace{3mm}
\paragraph{Screening in Vector Charge Theory:
\label{appendix:A4}}

Consider an ``electrolyte'' of vector charges responding to a test charge ${\bf Q}$. The potential energy arising from this test charge is: 
\begin{align}
    -{\bf Q}\cdot\boldsymbol{\varphi} ~
\end{align}
where $\boldsymbol{\varphi}$ is the electrostatic potential.
The charge density, $\boldsymbol{\rho}$, includes the induced charge, and therefore depends on $\boldsymbol{\varphi}$ via the generalization of the Debye-H{\"u}ckel equation  \cite{chaikin1995principles} to the vector charge theory of a charge-neutral system with the induced charge density screening the test charge. The resultant  Debye-H{\"u}ckel Equation for the vector charge theory  is given by:

\begin{align}
    \partial_i\left[\partial_i\varphi_j+\partial_j\varphi_i\right]= -\rho_j = - \langle \rho_j \rangle  \left[- e^{\beta{\bf Q}\cdot\boldsymbol{\varphi}} + 1\right] ~.
\end{align}
Here, $\langle \boldsymbol{\rho} \rangle $ is the average background charge density and $\beta$ is the inverse temperature. For small values of the potential, the above equation simplifies to
\begin{align}
    \partial_i\left[\partial_i\varphi_j+\partial_j\varphi_i\right]\approx \beta \langle \rho_j \rangle Q_{k} \varphi_k
\end{align}
Fourier transforming this gives
\begin{align}
    \left[|q|^2\delta_{kj}+q_k q_j+\kappa_{jk}\right]\varphi_k=0
\end{align}
where $\kappa_{jk}= - \beta Q_j \langle \rho_{k} \rangle$ denotes the inverse of an  anisotropic screening length. The screened Green's function is given by
\begin{align}
  \mathcal{G}_{jk}^{\,\rm scr}(\boldsymbol{q})=\left[|q|^2 \delta_{jk}+q_j q_k + \kappa_{jk}\right]^{-1}
\end{align}

In 3D,  the real space screened Green's function is
{\small
\begin{align}
    \mathcal{G}^{3D\,\rm scr}_{i j} (\boldsymbol{r}) = \frac{\delta_{i j}}{4 \pi r} - (\partial_i \partial_j - \kappa_{i j})\left[\frac{1 - e^{- r \sqrt{\frac{|\kappa|}{2}} }}{4 \pi r \kappa}\right] ~,
\end{align}
}%
for $Tr(\kappa_{i j}) = \kappa > 0$,  and 
{\small
\begin{align}
    \mathcal{G}^{3D \, \rm scr}_{i j} (\boldsymbol{r}) = \frac{\delta_{i j}}{4 \pi r} - (\partial_i \partial_j - \kappa_{i j})\left[\frac{1 - \cos \left( r \sqrt{\frac{|\kappa|}{2}} \right)}{4 \pi r \kappa}\right] ~,
\end{align}
}%
for $\kappa < 0$. In 2 dimensions, 
{\small

\begin{align}
 \mathcal{G}^{2D \, \rm scr}_{i j}(\boldsymbol{r}) = &\left( \frac{\delta_{i j}}{2 \pi} \right) \ln\left( \frac{r}{C_0}\right)\nonumber\\
 &- \left(\frac{\partial_i \partial_j - \kappa_{i j}}{2 \pi \kappa} \right) \left[\ln\left(\frac{r}{C_1}\right) + 2 \pi K_0 \left( r \sqrt{\frac{\kappa}{2}}\right) \right]   
\end{align}
}%

Where $K_0$ is the modified Bessel function of the second kind of order zero.

For $\kappa > 0$ the $K_0$ term decays, but for $\kappa < 0$ the term can be written as 
{\small
\begin{align}
K_0 \left( i r \sqrt{\frac{|\kappa|}{2}} \right) = \frac{-\pi}{2} \left[ i J_0 \left( r \sqrt{\frac{|\kappa|}{2}} \right) + Y_0 \left( r \sqrt{\frac{|\kappa|}{2}} \right) \right]
\end{align}
}
which has oscillating behavior. $J_0$ is the Bessel function of the first kind of order zero and $Y_0$ is the Bessel function of the second kind of order zero.  

It is clear from these calculations that the screening in VCT is not characterized by a single length scale but by a screening tensor, $\kappa_{ij}$ that represents the strong anisotropies dictated by the charge-angular momentum conservation: a vector charge can only move along its own direction.
\vspace{-4mm}

\section{Additional numerical results for the stress-stress correlations in three dimensions}

Here, we have presented the comparisons between the VCTG predictions and numerical results for the remaining $15$ stress-stress correlations in three dimensions. The comparisons are done on a system of $27000$ grains at packing fraction $\phi = 0.69$. For clarity, the results are given in the Hammer projection~\cite{Hammer} coordinate system. 

\begin{figure*}%[!htbp]
\vspace{-4mm}
\includegraphics[width=0.99\textwidth]{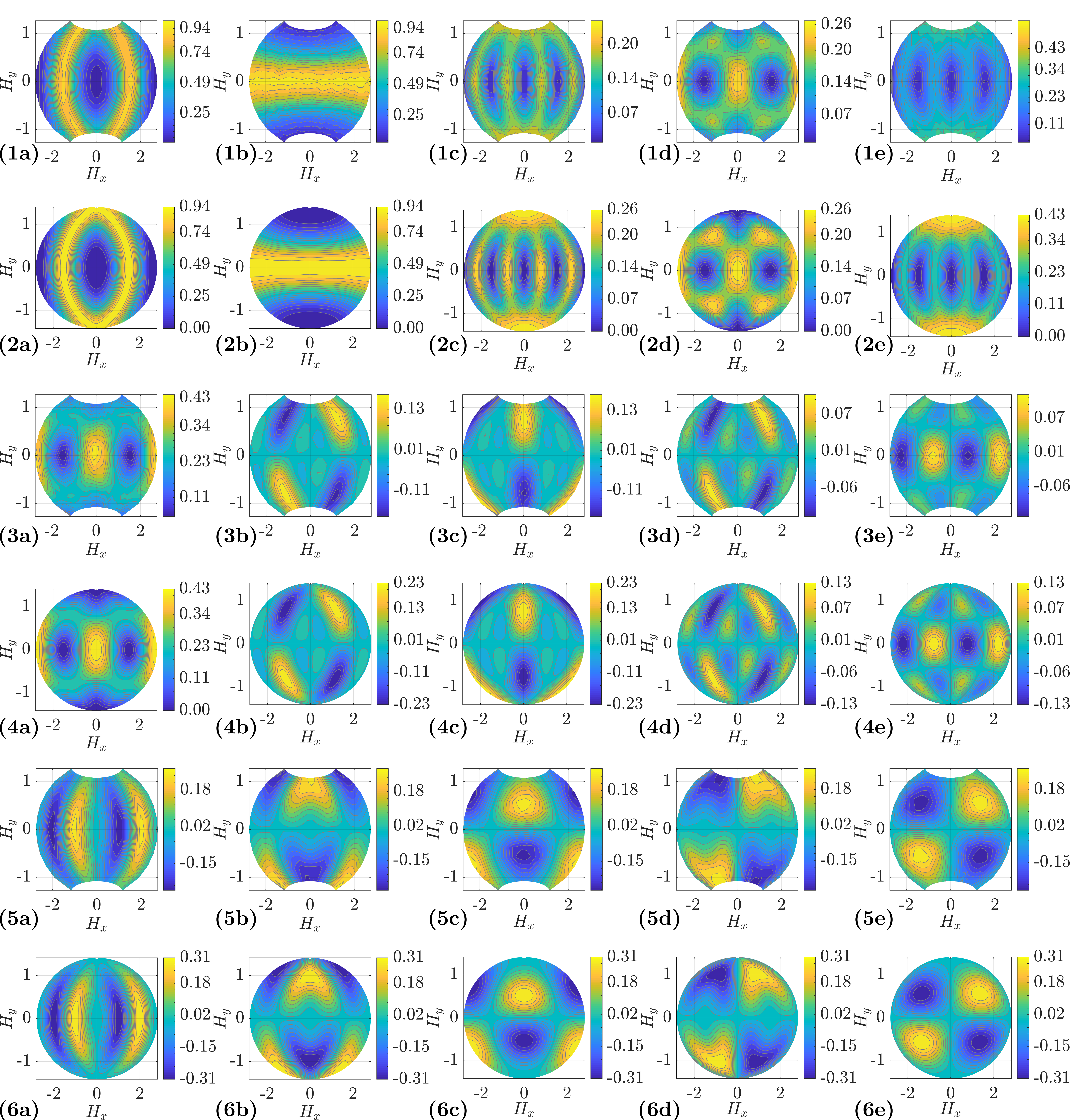}
\caption{Comparison of the correlation functions obtained from numerical simulations with the theoretical predictions in 3D, for the $15$ remaining correlations (Figs.~\ref{fig:3d_correlations_fig2},~\ref{fig:3d_correlations_fig3}). The explicit forms for the correlations are given in the Appendix (see Eq.~\eqref{eq:3d_correlations_explicit}). The panels show, respectively: numerical (1a) and theoretical (2a) results for $C_{xxxx}$, numerical (1b) and theoretical (2b) results for $C_{zzzz}$, numerical (1c) and theoretical (2c) results for $C_{xyxy}$, numerical (1d) and theoretical (2d) results for $C_{yzyz}$, numerical (1e) and theoretical (2e) results for $C_{xxyy}$, numerical (3a) and theoretical (4a) results for $C_{yyzz}$, numerical (3b) and theoretical (4b) results for $C_{xxyz}$, numerical (3c) and theoretical (4c) results for $C_{xzyy}$, numerical (3d) and theoretical (4d) results for $C_{xyxz}$, numerical (3e) and theoretical (4e) results for $C_{xzyz}$. Rows $5$ and $6$ show respectively, the numerical and theoretical predictions for correlations of the form $C_{iiij}$: (5a) and (6a) for $C_{xxxy}$, (5b) and (6b) for $C_{xxxz}$, (5c) and (6c) for $C_{xzzz}$, (5d) and (6d) for $C_{yyyz}$, (5e) and (6e) for $C_{yzzz}$. ~\label{fig:3d_full}}
\vspace{-5mm}
\end{figure*}

%%%%%%%%%%%%%%%%%%%%%

\FloatBarrier

%\bibliography{granular_stress}

\bibliographystyle{unsrt}

\end{document}

\title{Supplemental Material for\\ ``Tensor Electromagnetism and Emergent Elasticity in Jammed Solids''}

\author{Jishnu N. Nampoothiri}
\affiliation{Martin Fisher School of Physics, Brandeis University, Waltham, MA 02454 USA}
\affiliation{Centre for Interdisciplinary Sciences, Tata Institute of Fundamental Research, Hyderabad 500107, India}
\email{jishnu@brandeis.edu}
\author{Michael D'Eon}
\affiliation{Martin Fisher School of Physics, Brandeis University, Waltham, MA 02454 USA}
\email{mdeon@brandeis.edu}
\author{Kabir Ramola}
\affiliation{Centre for Interdisciplinary Sciences, Tata Institute of Fundamental Research, Hyderabad 500107, India}
\email{kabirramola@gmail.com}
\author{Bulbul Chakraborty}
\affiliation{Martin Fisher School of Physics, Brandeis University, Waltham, MA 02454 USA}
\email{bulbul@brandeis.edu}
\author{Subhro Bhattacharjee}
\affiliation{International Centre for Theoretical Sciences, Tata Institute of Fundamental Research, Bengaluru 560089, India}
\email{subhro@icts.res.in}

\date{\today}

\begin{abstract}
In this document we provide supplemental figures and details related to the results presented in the main text.
\end{abstract}

\maketitle

%{\color{red} Move next two sections (4 and 5) to SI}
\section{The multipole expansion and the dielectric~\label{app:pol}}

{Similar to ordinary electrodynamics~\cite{jackson}, the multipole expansion of the electrostatic potential, $\varphi_i(\boldsymbol{r})$, can be obtained from Eq.~\eqref{main-eq:potgreen} in the appendix, by expanding $\mathcal{G}^{3D}_{ij}(\boldsymbol{r}-\boldsymbol{r}')$ given by Eq.~\eqref{main-eq:potgreenexp} and $\mathcal{G}^{2D}_{ij}(\boldsymbol{r}-\boldsymbol{r}')$, given by Eq.~\eqref{main-eq:potgreenexp2D} for a localized charge distribution in the limit of $|\boldsymbol{r}'|/\boldsymbol{r}\ll 1$, by expanding $1/{|\boldsymbol{r}-\boldsymbol{r}'|}$.  Here we present the results for 3D.  Starting with the expansion: }
% whence we expand $1/{|\boldsymbol{r}-\boldsymbol{r}'|}$.to get
\begin{align}
    \varphi_i(\boldsymbol{r})=\sum_{p}\varphi_i^{(p)}(\boldsymbol{r})
\end{align}
$\varphi^{(1)}_i(\boldsymbol{r})$ is the monopole contribution given by Eq.~\eqref{main-eq:point_potential} in the appendix and 
{\small
\begin{align}
     \varphi^{(2)}_i(\boldsymbol{r})=\frac{1}{8\pi r^3}\left[3\left(\delta_{ij}+\frac{r_ir_j}{r^2}\right)r_kP_{jk} -\left(r_j P_{ji} + r_i P_{jj}\right)\right]
\end{align}
}%
is the dipole contribution with the net dipole moment of the charge distribution being given by Eq.~\eqref{main-eq:dipole_exp} in the main text.
% \begin{align}
%     P_{jk}=\int d^3\boldsymbol{r'}~r'_k~\rho_j(\boldsymbol{r'})
% \end{align} 

For two equal and opposite charges, $\pm \boldsymbol{Q}$, at $\pm {\boldsymbol{d}}/{2}$ such that {\small$\rho_i(\boldsymbol{r})=Q_i\delta^3(\boldsymbol{r}-\frac{\boldsymbol{d}}{2})-Q_i\delta^3(\boldsymbol{r}+\frac{\boldsymbol{d}}{2})$}, 
% \begin{align}
%   \rho_i(\boldsymbol{r})=Q_i\delta^3(\boldsymbol{r}-\frac{\boldsymbol{d}}{2})-Q_i\delta^3(\boldsymbol{r}+\frac{\boldsymbol{d}}{2})  
%   \label{eq:dipole_pot}
% \end{align}
we get the dipole moment of this zero net charge configuration as
\begin{align}
    P_{jk}=Q_jd_k
\end{align}
Note that the dipole moment is not symmetric in its indices unlike the electric (and magnetic) field. We can separate it into symmetric and antisymmetric parts by taking symmetric and ansisymmetric combinations of $P_{jk}$ and $P_{kj}$. %Separating it into the symmetric and the antisymmetric component we get
% \begin{align}
%     P_{jk}=P_{jk}^s+P_{jk}^A=\frac{Q_jd_k+Q_kd_j}{2}+\frac{Q_jd_k-Q_kd_j}{2}
% \end{align}
 In a system with many dipole moments the electrostatic potential due to them is given by
\begin{align}
    \varphi_i^{(2)}(\boldsymbol{r})=\int d^3\boldsymbol{r}'~\partial_k \mathcal{G}^{3D}_{ij}(\boldsymbol{r}-\boldsymbol{r}')\mathcal{P}_{jk}(\boldsymbol{r}')
\end{align}
where $\mathcal{G}^{3D}_{ij}(\boldsymbol{r}-\boldsymbol{r}')$ is given by Eq. \eqref{main-eq:potgreenexp} of the appendix and $\mathcal{P}_{ij}(\boldsymbol{r})$ is the dipole density which is related to the induced volume charge density via Eq.~\eqref{main-eq:bound_dipole} in the main text. In addition, similar to ordinary electromagnetism~\cite{jackson}, there is an induced surface charge-density given by $\rho^{\rm surf}_i=\hat{n}_j\mathcal{P}_{ij}$, where $\hat{\bf n}$ is normal to the surface.
\vspace{-3mm}

\vspace{3mm}

\section{Response and Correlations~\label{app:VCTG_response_correlation}}

%%%%%%%%%%%%%%%%%%%%

%\subsection{Solving Gauss's Law in 2D}

{Here we present an alternative (to Section~\ref{main-sec:correlations} of the main text) way of calculating the stress-stress correlations within VCTG using the potential $\varphi_a$ formulation that is readily extended to calculation of the stress response to an external point-force that is studied numerically in Section~\ref{main-sec:numerical}. Unlike in Section~\ref{main-sec:correlations}, however, we use use the language of VCT and hence calculate the electric-displacement correlators and responses which can then be translated to stress via the mapping $D_{ij}\leftrightarrow \sigma_{ij}$ in Eq.~\eqref{main-eq_correspondance}.}

%The response functions are most readily calculated by using the action formulation. In the electromagnetic picture, $D_{ij} = \Lambda^{-1}_{ijkl} E_{kl}$ and follows Gauss's Law for free charges. We will calculate the response and correlations for $D$ to compare them to the stress in the numerics.
%{\color{red} BC: I am trying to stick to Lagrangian instead of action because that is what we used in the main text.}

The partition function {for the electrostatics of VCT for an external charge distribution, $\rho_j^{\rm ext}$,}  is given by

\begin{equation}
    \mathcal{Z}[\rho_j^{\rm ext}] = \int [\mathcal{D}D] ~ \delta(\partial_i D_{ij} - \rho_j^{\rm ext}) ~ e^{-\mathcal{S}}
    \label{eq_freeeng}
\end{equation}
{where the Landau-Ginzburg action is given by Eq.~\eqref{main-eq:lagrangian_density}, which in the language of VCT is given by}
\begin{equation}
    \mathcal{S} = \frac{1}{2g}\int d^d \boldsymbol{r} ~ D_{ij} \Lambda_{ijkl} D_{kl} ~.
\end{equation}
{where we have used Eq.~\eqref{main-eq:E_D_relation}.}

%Here, we have assumed a diagonal form for $\Lambda$, $\Lambda^{-1}=K\mathcal{I}$, and introduced $g$ as a coupling constant to fix the dimensions of $\mathcal{S}$.
The field {for the electrostatics} equations \eqref{main-eq:VCTG_field_eqns} can be recovered by {extremising the above action and} setting $g=1$. {Implementing the Delta function constraint for the Gauss's law in Eq.~\eqref{eq_freeeng}} in terms of a Lagrange Multiplier, {$\varphi_a$,} we get
\begin{equation}
    \mathcal{Z}[\rho_j^{\rm ext}] = \int [\mathcal{D}D] [\mathcal{D}\varphi] ~ e^{-\tilde{S}}
\end{equation}
where, 
{\small
\begin{align}
    \tilde{S} %=& ~ i\int d^d \boldsymbol{r} ~ \varphi_i \rho_i + \int d^d \boldsymbol{r} \left[ \frac{1}{2g} D_{ij} \Lambda_{ijkl} D_{kl} + i D_{ij} \partial_i \varphi_j \right] \\
    =& ~ i\int d^d \boldsymbol{r} ~ \varphi_i \rho_i^{\rm ext} + \int d^d \boldsymbol{r} \left[\frac{1}{2g} D_{ij} \Lambda_{ijkl} D_{kl} + i D_{ij} \mathcal{J}_{ij} \right]
    \label{eq_action2}
\end{align}
}%
{where we have used $D_{ij} = D_{ji}$ and} 
\begin{equation}
    \mathcal{J}_{ij} = \frac{1}{2} ( \partial_i \varphi_j + \partial_j \varphi_i)
\end{equation}
It is simpler to compute {the average displacement field} in Fourier space given that
\begin{equation}
    \langle D_{ij}(\boldsymbol{r}) \rangle_{\rho^{\rm ext}} = \int \frac{d^d \boldsymbol{q}}{(2 \pi)^d} ~ e^{i \bf q\cdot r} \langle D_{ij} (\boldsymbol{q}) \rangle_{\rho^{\rm ext}}
\end{equation}
where
\begin{equation}
     \langle D_{ij} (\boldsymbol{q}) \rangle_{\rho^{\rm ext}} = \frac{1}{\mathcal{Z}[\rho^{\rm ext}]} \int [\mathcal{D}D] [\mathcal{D}\varphi] D_{ij} (\boldsymbol{q}) e^{-\tilde{S'}}
\end{equation}
{where}
\begin{multline}
    \tilde{S'} =  \int \frac{d^d \boldsymbol{q}}{(2 \pi)^d} ~ \left[ i \varphi_i (-q) \rho_i (q) +  \frac{1}{2g} D_{ij} (q) \Lambda_{ijkl} D_{kl} (-q) \right. \\ 
    \left. + i D_{ij} (-q) \mathcal{J}_{ij} (q) \right]
\end{multline}
{is the Fourier transform of Eq.~\eqref{eq_action2}. Using the standard method of sources, this can be re-written as}
{\small
\begin{equation}
    \langle D_{ij} (\boldsymbol{q}) \rangle_{\rho^{\rm ext}} = \frac{1}{\mathcal{Z}[\rho^{\rm ext}]} \int [\mathcal{D}D] [\mathcal{D}\varphi] ~ \frac{1}{-i} \frac{\delta}{\delta \mathcal{J}_{ij} (-q)} e^{-\tilde{S'}}
    \label{eq_dresponse}
\end{equation}
}%
{Completing the integral over $D_{ij}$, we get
\begin{equation}
    \langle D_{ij} (\boldsymbol{q}) \rangle_{\rho^{\rm ext}} = \frac{C_1}{\mathcal{Z}[\rho^{\rm ext}]} \int [\mathcal{D}\varphi] ~ \frac{1}{-i} \frac{\delta}{\delta \mathcal{J}_{ij} (-q)} e^{-\tilde{S''}}
\end{equation}
where $C_1$ is an overall constant and 
{\small
\begin{equation}
    \tilde{S''} =   \int \frac{d^d \boldsymbol{q}}{(2 \pi)^d} \left( \frac{g}{2} \mathcal{J}_{ij} (q) \Lambda^{-1}_{ijkl} \mathcal{J}_{kl} (-q) + i \varphi_i (-q) \rho_i (q) \right)
\end{equation}
}%
such that
{\small
\begin{align}
     \langle D_{ij} (\boldsymbol{q}) \rangle_{\rho^{\rm ext}} %=& \frac{C_1}{\mathcal{Z}[\rho]} \int [D\varphi] ~ \frac{1}{-i} \frac{\delta}{\delta \mathcal{J}_{ij} (-q)} e^{-\mathcal{S}} \\ 
     =& \frac{C_1}{\mathcal{Z}[\rho^{\rm ext}]} \int [\mathcal{D}\varphi] ~ \left[\frac{g (\Lambda^{-1}_{ijab} + \Lambda^{-1}_{abij}) \mathcal{J}_{ab} (q) }{2 i} \right] e^{-\tilde{S''}}
\end{align}
}%
with 
}
\begin{equation}
    \mathcal{J}_{ij} (\boldsymbol{q}) = \left(\frac{-i}{2} \right) ( q_i \varphi_j (\boldsymbol{q}) + q_j \varphi_i (\boldsymbol{q}) ).
\end{equation}
{Finally completing the $\varphi$ integral, we get
}
% \begin{align}
%      \langle D_{ij} (\boldsymbol{q}) \rangle_{\rho^{\rm ext}} =& \frac{C_1}{\mathcal{Z}[\rho]} \int [D\varphi] ~ \left( \frac{- g (\Lambda^{-1}_{abij}+\Lambda^{-1}_{ijab})}{4} \right) (q_a \varphi_b + q_b \varphi_a) e^{-\mathcal{S}} \nonumber \\ =& \frac{1}{\mathcal{Z} [\rho]} \left(\frac{g(\Lambda^{-1}_{abij} +\Lambda^{-1}_{ijab})}{4 i}\right) \left( q_a \frac{\delta}{\delta \rho_b} + q_b \frac{\delta}{\delta \rho_a} \right) \mathcal{Z} [\rho]
% \end{align}
\begin{equation}
     \langle D_{ij} (\boldsymbol{q}) \rangle_{\rho^{\rm ext}} = G_{ijk}({\bf q})~\rho^{\rm ext}_k ({\bf q}) ~,
     \label{eq:genresponse}
\end{equation}
{where $G_{ijk}({\bf q})$ is given by Eq.~\eqref{main-eq_stress_resp}.}
% \begin{equation}
%     \mathcal{C}_{ij} = -g q_a q_b \Lambda^{-1}_{iajb}
% \end{equation}

Turning to correlations, similar to Eq.~\eqref{eq_dresponse}, they are given by
\begin{widetext}
\begin{align}
\langle D_{ij} (q) D_{kl} (-q) \rangle_{\rho^{\rm ext}} &= \frac{-1}{\mathcal{Z}[\rho^{\rm ext}]} \int [\mathcal{D}D][\mathcal{D}\varphi]\frac{\delta}{\delta\mathcal{J}_{ij}(-q)}\frac{\delta}{\delta \mathcal{J}_{kl}(q)} e^{-\tilde{S}'} \\
\intertext{This expression is evaluated similarly to Eq.~\eqref{eq:genresponse} and also for $\rho^{\rm ext}=0$}
\langle D_{ij} (q) D_{kl} (-q) \rangle_{\rho^{\rm ext}=0} &= \left(\frac{g(\Lambda^{-1}_{ijkl}+\Lambda_{kl ij}^{-1})}{2}\right)- g^2 (\Lambda^{-1}_{klab}+\Lambda^{-1}_{abkl}) (\Lambda^{-1}_{mnij}+\Lambda^{-1}_{ijmn})(q_a q_m \mathcal{C}^{-1}_{nb}).
\end{align}

which is same as Eqs.~\eqref{main-eq:2d_stress_correlations} and \eqref{main-eq:3d_stress_correlations} in 2D and 3D respectively for $g=1$.

{For the response in 2D, with $\Lambda^{-1}$ given by Eq.~\eqref{main-eq:lambda_2d_iso}, the explicit form of the Green's functions that control the response to an external force is given by
\begin{align}
    {G}_{xxx}(\boldsymbol{q})&={\mathcal{G}}_{xxx}^{\,2D}(\boldsymbol{q})-\left(\frac{\nu}{1-\nu}\right){\mathcal{G}}_{yyx}^{\,2D}(\boldsymbol{q}), \nonumber \\
    {G}_{xxy}(\boldsymbol{q})&={\mathcal{G}}_{xxy}^{\,2D}(\boldsymbol{q})+\left(\frac{\nu}{1-\nu}\right){\mathcal{G}}_{xyx}^{\,2D}(\boldsymbol{q}), \nonumber \\
    {G}_{yyx}(\boldsymbol{q})&={\mathcal{G}}_{yyx}^{\,2D}(\boldsymbol{q})+\left(\frac{\nu}{1-\nu}\right){\mathcal{G}}_{xyy}^{\,2D}(\boldsymbol{q}), \nonumber \\
    {G}_{yyy}(\boldsymbol{q})&={\mathcal{G}}_{yyy}^{\,2D}(\boldsymbol{q})-\left(\frac{\nu}{1-\nu}\right){\mathcal{G}}_{xxy}^{\,2D}(\boldsymbol{q}), \label{eq:Greens_fn_dielectric_vac} \\
    {G}_{xyx}(\boldsymbol{q})&={\mathcal{G}}_{xyx}^{\,2D}(\boldsymbol{q})+\left(\frac{\nu}{1-\nu}\right){\mathcal{G}}_{xxy}^{\,2D}(\boldsymbol{q}), \nonumber \\
    {G}_{xyy}(\boldsymbol{q})&={\mathcal{G}}_{xyy}^{\,2D}(\boldsymbol{q})+\left(\frac{\nu}{1-\nu}\right){\mathcal{G}}_{yyx}^{\,2D}(\boldsymbol{q}). \nonumber 
\end{align}
with $\mathcal{G}^{2D}_{ijk}({\bf q})$ being given by Eq.~\eqref{main-eq:2d_response_vacuum_general}.
}

\end{widetext}

\section{Additional details of the numerical simulations~\label{app:numerical}}

%{\color{red} Methiods, Characterization of Packings can move to SI}
\subsection{Methods~\label{app:numerical_methods}}

We generate granular packings in 2D and 3D following a variation of the standard O'Hern protocol used in references~\cite{corey_packing_protocol,kabir_network_paper}. In this study, we focus on packings of soft frictionless bidisperse grains. To generate a packing at mechanical equilibrium, we begin with a $50:50$ mixture of grains with diameter ratio $1:1.4$, interacting via the harmonic soft sphere potential given by
\begin{equation}
    V\left(\boldsymbol{r}_{ij}\right)=    \begin{cases}
    \frac{\epsilon}{2}\left(1-\frac{|\boldsymbol{r}_{ij}|}{a_{ij}}\right)^2~~~~ \textrm{for}~~~|\boldsymbol{r}_{ij}|< a_{ij}\\
    0~~~~~~~~~~~~~~~~~~~~~\textrm{for}~~~|\boldsymbol{r}_{ij}|\geq a_{ij}.
    \end{cases}
    \label{eq:soft_sphere_potential}
\end{equation}
Here, $\boldsymbol{r}_{i},~\boldsymbol{r}_{j}$ denote the positions of the grains ($\boldsymbol{r}_{ij}\equiv\boldsymbol{r}_i-\boldsymbol{r}_j$) and $a_{ij} = a_i+a_j$ the sum of the radii of the grains. We choose to work in units where the energy scale $\epsilon=1$. 

\begin{figure*}[!htbp!]
\includegraphics[width=0.9\textwidth]{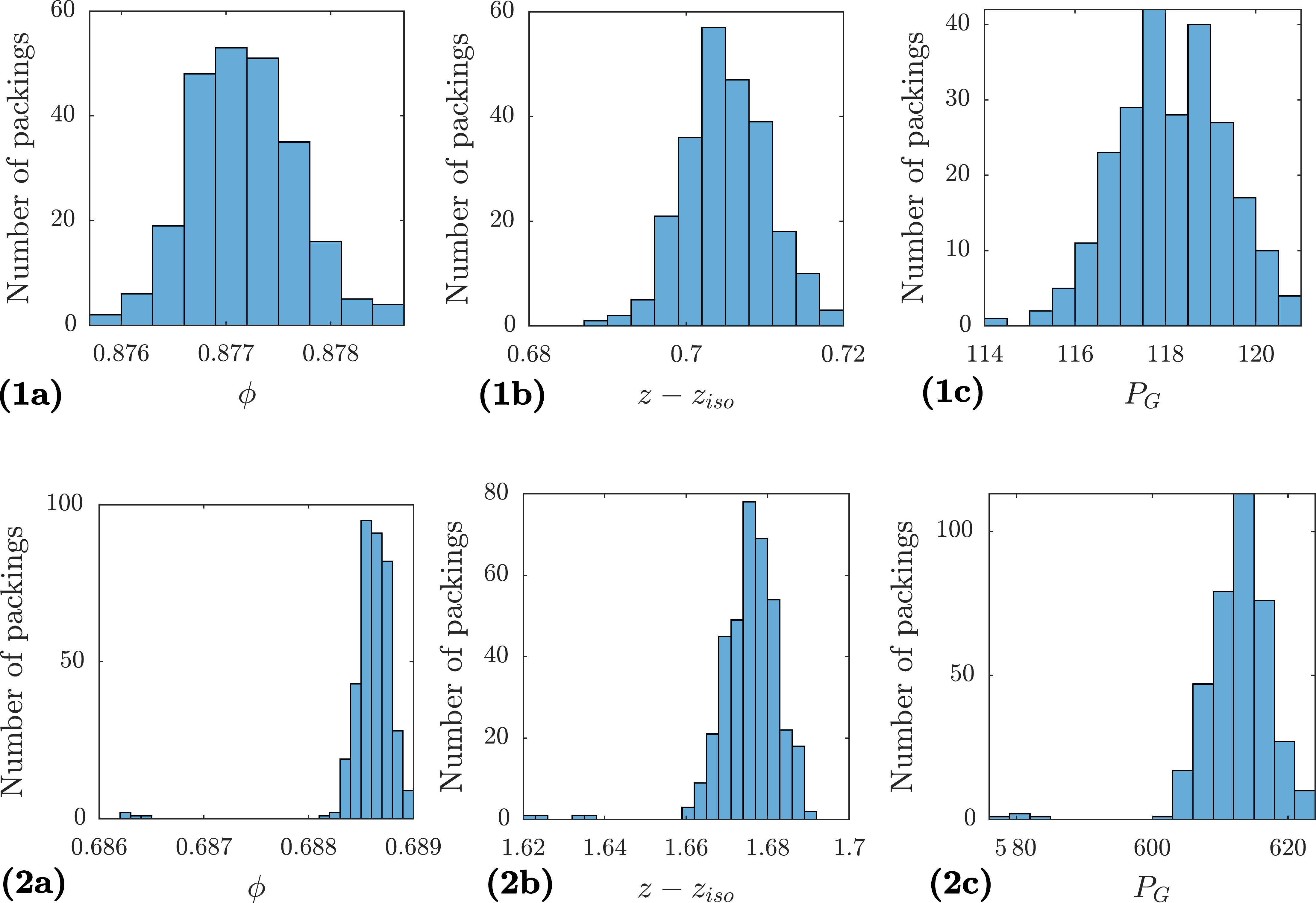}
\caption{Measured characteristics of the ensemble. The first row shows the results for $N_G=8192$ disks in 2D with a mean area $A_G=1.07\times10^{-4}$ ($239$ configurations) and the second row shows the characteristics for packings of $N_G=27000$ grains in 3D with a mean volume $A_G=2.56\times10^{-5}$ ($374$ configurations). The columns shows respectively, the histograms of (a) the packing fraction (b) the average contact number $z-z_{iso}$ and (c) the average pressure per grain $P_G$ respectively. The Gaussian nature of these distributions indicate that the ensemble can be well described by its average quantities ($\llangle P_G \rrangle$, $\llangle z-z_{iso} \rrangle$) indicating that the fixed area ensemble we use, is well behaved. For more details see Table~\ref{main-tbl:stiffness_consts} of the main text. Note that we are working in units where the energy scale $\epsilon=1$ (see Eq.~\eqref{eq:soft_sphere_potential}) and box size $L=1$.}
~\label{fig:packing_char}
\vspace{-3mm}
\end{figure*}

The generation of the packing ensemble begins with a target packing fraction $\phi_0$. 
At the initial time step, the grains are randomly distributed in a simulation box of linear dimension $L=1$, with periodic boundary conditions. Given the system size (number of particles $N_G$), the starting packing fraction is used to determine the radii of the grains. The actual packing fraction for a realization is typically lower than this due to the presence of rattlers. This procedure results in an {\it ensemble} of packings defined by $A_G$, which is the mean of the area (volume in 3D) of the two species of grains used in the packing ensemble.

A configuration at mechanical equilibrium is obtained from the initial random placement of grains by relaxing the positions of the grains to an energy minimum using FIRE~\cite{FIRE,FIRE2.0} or conjugate gradient methods. A feature of the FIRE algorithm is that the minimization procedure allows for climbing over shallow energy minima due to the presence of the inertia in the relaxation dynamics. This feature while not being an issue for generating an ensemble of packings at mechanical equilibrium, is an impediment to computing the linear response of a specific packing to an external force. Therefore, we employ the conjugate gradient algorithm for computing the response of a given packing to an external force, despite its slower performance in comparison to the FIRE algorithm.

The predictions of VCTG for stress-stress correlations,  
% and response and the observations of the granular system can be made by comparing the quantities of interest such as the stress-stress correlation functions, stress distribution, 
and the stress response to an external force can be carried out either in real space or in Fourier space. As the VCTG is a continuum theory of disorder-averaged physical quantities, the comparisons are performed at a coarse grained level, and averaged over multiple packings.
% Additionally, the presence of inherent spatial disorder in these packings necessitates some form of coarse graining.
We implement the coarse graining through: 1) a standard  procedure in real space or 2) in Fourier space by imposing a large $q$ cutoff. Since we measure lengths in terms of the simulation box size, which is set to $1$, the distance between any two valid points in the Fourier space must be greater than $2\pi$. 
Additionally, since the smallest microscopic unit in our systems is a grain, there cannot be any stress fluctuations occurring at length scales shorter than $2a_{min}$ where `$a_{min}$' is the radius of the smallest disk/sphere in the packing. This results in a natural large $|\boldsymbol{q}|$ cutoff for the theory.

The main quantity of interest in a given packing is the set of grain level force-moment tensors given by
\begin{equation}
\hat{\sigma}_{g}^{p}=\sum_{c=1}^{n_c^g} \boldsymbol{r}_c^{\, g} \otimes \boldsymbol{f}_c^g.
\label{eq:grain_level_force_moment}
\end{equation}

Here, `$p$' denotes a particular configuration of $N_G^p$ disks and $g$ denotes a particular grain in the packing. As we are interested only in grains that belong to the contact network, the number of grains $N_G^p$ fluctuates from packing to packing due to the presence of rattlers. $\hat{\sigma}_{g}^{p}$ denotes the force moment tensor for grain `$g$' in the packing, $\boldsymbol{r}_c^{\, g}$ its position and $\boldsymbol{f}_c^g$ the force at the contact `$c$' belonging to the grain `$g$', respectively.

Stress correlations are most readily computed in Fourier space as the key features of the theoretical predictions are prominently displayed in $C_{ijkl}(\boldsymbol{q})$ as discussed in the main text. We begin by computing the stress distribution in Fourier space:
\begin{equation}
\hat{\sigma}^{p}\left(\boldsymbol{q}\right)=\sum_{g=1}^{N_G^p} \hat{\sigma}_{g}^{p} \exp{\left(i\boldsymbol{q}\cdot\boldsymbol{r}_{g}^{\, p}\right)}.
\end{equation}
The stress correlation functions can be obtained from the stress fields as
%\langle\sigma_{ij}\left(\boldsymbol{q}\right)\rangle=&\frac{1}{n_p}\sum_{p=1}^{n_p}  \sigma_{ij}^{p}\left(\boldsymbol{q}\right) \\
\begin{eqnarray}
\nonumber
C_{ijkl}\left(\boldsymbol{q}\right) &=& \llangle\Delta\sigma_{ij}\left(\boldsymbol{q}\right) \Delta\sigma_{kl}\left(-\boldsymbol{q}\right)\rrangle \\
\nonumber
&=& \frac{1}{n_p}\sum_{p=1}^{n_p} \Delta\sigma_{ij}^p\left(\boldsymbol{q}\right)\Delta\sigma_{kl}^p\left(-\boldsymbol{q}\right);\\ 
\Delta\sigma_{ij}^p\left(\boldsymbol{q}\right) &=& \sum_{g=1}^{N_G^p} \exp{\left(i\boldsymbol{q}\cdot\boldsymbol{r}_{g}^{\, p}\right)} \left(\sigma_{ij,g}^{p}-\bar{\sigma}_{ij}^{p}\right)
\end{eqnarray}
where $\bar{\sigma}_{ij}^{p}$ is the average grain force moment tensor for the packing `$p$'. The double angular braces $\llangle ~...~ \rrangle$, denote the successive spatial averaging (for each packing) and disorder averaging (over multiple packings in the ensemble). In the case of stress correlations, the spatial averaging is implemented through coarse graining. Here, we subtract the average force moment tensor of the particular packing instead of subtracting the overall average stress tensor due to sample to sample fluctuations in the average force moment tensor. The set of packings that we have used for this work have a very narrow distribution of both average force moment tensor and packing fractions, making the results from either procedure sufficiently close to each other~(see Fig.~\ref{fig:packing_char}).

%A qualitative difference between the vacuum correlations and the dielectric predictions arises only if there are non zero ``off-diagonal" elements in $\hat{\Lambda}^{-1}$.

%From the forms of the correlation functions, as discussed in the main text (see discussion below Eq.~\eqref{eq:2d_stress_correlations}) we see that the theoretical forms for the correlations of the stress tensor are independent of $|\boldsymbol{q}|$. Therefore, it is better to analyze the goodness-of-fit of the angular component of the correlation functions, which is obtained by integrating the numerical results over all $|\boldsymbol{q}|$. This procedure is advantageous since the noise is reduced by the radial integration. Furthermore, since the underlying granular system is discrete, we expect deviations from the continuum predictions for $|\boldsymbol{q}| \approx \mathcal{O}(\pi/2a)$ where $a$ is the radius of the smaller grain in our bidisperse system. To reduce this effect, we perform the comparison by integrating the numerical results up to $|\boldsymbol{q}|=q_{max}$. This is equivalent to performing a coarse graining in real space with a coarse graining length $2\pi/q_{max}$.

\subsection{Characterization of packings~\label{app:packing_chars}}

In this section, we provide some additional details of the relevant physical properties characterizing the ensemble of packings used in our study. As mentioned above, fixing $A_G$ does not fix the packing fraction $\phi$ exactly because of the presence of rattlers. However, it is very narrowly distributed. 
% the set of packings used for generating the stress-stress correlations in two and three dimensions. 
The characteristic quantities of interest in a packing are the average number of contacts per grain, $z$, and the average grain level pressure $P_G$. These quantities fluctuate in the ensemble, and their distributions are shown in Fig. \ref{fig:packing_char}. 
The grain-level pressure is obtained from:
\begin{equation*}
    P_G=\frac{1}{A_G} \sum_{g=1}^{N_G^p} \hat{\sigma}_{g}^{p}
\end{equation*}
% The actual number of grains in a packing $N_G^p$ and the actual packing fraction of the packing $\phi$, varies from packing to packing due to the presence of rattlers, making a thorough examination of the distribution of the characteristic quantities in these packings a necessity. 

% In Fig.~\ref{fig:packing_char}, we present histograms of spatially averaged contact numbers $z-z_{iso}$ and grain level pressures $P_G$, to illustrate the configuration-to-configuration variations of these quantities.
% the quantities of interest in the set of packings of soft frictionless grains used in this study. 
% The columns depict the histograms of  average contact numbers $z-z_{iso}$ and grain level pressures $P_G$, respectively. 
The first row of Fig.~\ref{fig:packing_char} depicts the data from $239$ packings of soft frictionless disks in 2D. For this set of packings, ($\phi_{J} \approx 0.84$~\cite{rcp_2d}),and the histograms show a reasonably narrow distribution about finite values.   In the second row, we present the corresponding histograms from $374$ packings of $N_G=27000$ soft frictionless grains in 3D. These set of packings are also deep inside the jammed region, ($\phi_{J} \approx 0.64$~\cite{mrj_3d}) and show a narrow distribution about finite values.

% \begin{figure*}[!htbp]
% \includegraphics[width=0.9\textwidth]{figures/3d_correlations_1.pdf}
% \caption{Comparison of the auto correlations of the diagonal components of the stress tensor, obtained from numerical simulations with the theoretical predictions, in 3D. The comparisons are done on a system of $27000$ grains at packing fraction $\phi \approx 0.688$. The numerical results are presented in the first row (1) and the the theoretical results in the second row (2). The three columns show the results for (a) $C_{xxxx}$, (b) $C_{yyyy}$ and (c) $C_{zzzz}$ respectively. \label{fig:3d_1}}
% \end{figure*}

\subsection{Numerical Details for the stress-stress correlations}

We have shown that there is excellent agreement between the numerical stress-stress correlations and the VCTG predictions using the best fit to  $\hat{\Lambda}^{-1}$ for the $C_{yyyy}(\boldsymbol{q})$ correlation function.

% In this section, we compare the the VCTG predictions obtained using the same $\hat{\Lambda}^{-1}$ and numerical results for all the $21$ distinct stress-stress correlations in Figs.~\ref{fig:3d_1}-\ref{fig:3d_6}. We have grouped these correlations by their symmetry properties for clarity. These results are presented in Hammer projection coordinates which can be easily mapped to the standard spherical polar coordinates as described in~\ref{subsub:hammer}. 

\begin{figure}[!htbp]
\includegraphics[width=0.45\textwidth]{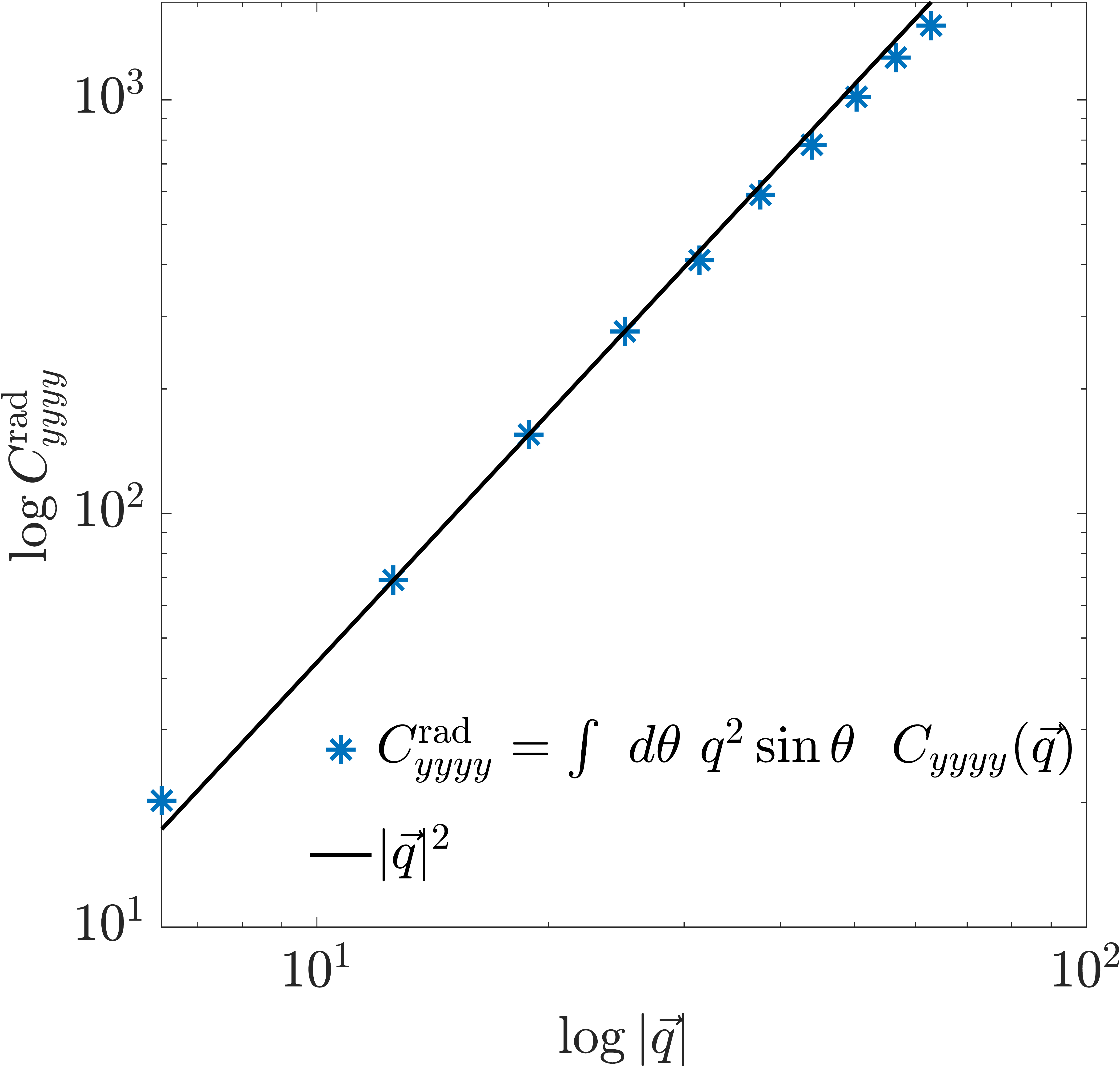}
\caption{The angular integrated numerical correlation $C_{yyyy}^{\mathrm{rad}} \equiv \int~d\Omega~q^2\sin\theta~ C_{yyyy}(\boldsymbol{q})$ (blue dots) and its comparison to the power law of $|\boldsymbol{q}|^2$ (solid black line), indicating that the correlation is independent of $|\boldsymbol{q}|$ and is purely a function of the angles i.e.  $C_{yyyy}(\boldsymbol{q}) = C_{yyyy}(\theta,\Phi)$.~\label{fig:3d_corr_radial}}
\vspace{-2mm}
\end{figure}

In this section, we present additional evidence for the presence of the pinch-point singularity at $|\boldsymbol{q}|=0$. For this purpose, we have presented the angular integral of the correlation function $C_{yyyy}(\boldsymbol{q})$ in Fig.~\ref{fig:3d_corr_radial}. From this figure, it is clear that the angular integral scales as $|\boldsymbol{q}|^2$, which is the behavior expected from a function that depends purely on angular coordinates. 

\subsubsection{Hammer Projection~\label{subsub:hammer}}

The Hammer projection is an equal area projection of the surface of a sphere. We have used these coordinates to present the three dimensional correlations to improve clarity. The relationship between the Hammer coordinates $(H_x,H_y)$ and the latitude $\alpha$ and longitude $\beta$ is given as

% \begin{figure*}[!htbp]
% \includegraphics[width=0.9\textwidth]{figures/3d_correlations_2.pdf}
% \caption{Comparison of the auto correlations of the off-diagonal components of the stress tensor, obtained from numerical simulations with the theoretical predictions, in 3D. The comparisons are done on a system of $27000$ grains at packing fraction $\phi \approx 0.688$. The numerical results are presented in the first row (1) and the the theoretical results in the second row (2). The three columns show the results for (a) $C_{xyxy}$, (b) $C_{xzxz}$ and (c) $C_{yzyz}$ respectively.
% \label{fig:3d_2}}
% \end{figure*}

\begin{align}
    H_x&=\frac{2\sqrt{2}\cos{\alpha}\sin{\frac{\beta}{2}}}{\sqrt{1+\cos{\alpha}\cos{\frac{\beta}{2}}}}\nonumber \\
    H_y&=\frac{\sqrt{2}\sin{\alpha}}{\sqrt{1+\cos{\alpha}\cos{\frac{\beta}{2}}}}.
    \vspace{3mm}
\end{align}

The latitude $\alpha$ and longitude $\beta$, are related to the spherical polar coordinate angles $(\theta,\Phi)$ as 
\begin{equation*}
    \alpha=\theta-\frac{\pi}{2},~\beta=\Phi.
    \vspace{2mm}
\end{equation*}

\subsection{Additional details on the response computation ~\label{app:numerical_response}}
\vspace{5mm}

\subsubsection{Force distribution for the response computation}

We are interested in the response of jammed packings to a point force. In our simulations, we implement a ``point force''  by distributing the total point force among all grains in a disc of radius $r_0=0.05$ centered at the origin. {A similar protocol had been used earlier in~\cite{Ellenbroek_thesis} to compute the response of jammed packings. Our protocol differs from this, by the presence of a compensating line force and the use of periodic boundary conditions.} 
Since we are interested in a  static assembly of grains,  the net force on a packing has to be zero. This condition of mechanical equilibrium necessitates a net zero external force on the system i.e.

\begin{equation*}
    \sum_g \boldsymbol{f}_g^{\,\mathrm{external}}=0.
\end{equation*}

As our simulations use periodic boundary conditions, this means that our external force distribution must include a compensating force to the point force. We have implemented this compensating point force through a line force with the linear force density chosen to enforce net zero external force. Since  we are interested in the features of the point force response, we place the compensating line charge far away from the point force at a distance of $a=0.45$ (simulation box size is $1$), such that the near field stress response resembles that of an isolated point charge.

\subsubsection{Disorder averaging protocol for response measurements}

% As the VCTG is a continuum theory, the response predictions from the VCTG are to be compared with the coarse grained stress fields in the granular system. 
% 
The response is computed as:

\begin{equation*}
    \llangle \sigma_{ij} \rrangle_{\mathrm{response}}=\llangle \sigma_{ij} \rrangle_{\mathrm{with~point~force}}-\llangle \sigma_{ij} \rrangle_{\mathrm{no~point~force}}
\end{equation*}

The disorder average in the above equation involves two, sequential averages. First, for every configuration, a spatial average is performed by moving the position of the point force to all grains within the disc of radius $r_0$ defining the ``point force'', and averaging over them.  
This step is taken, in order to minimize the effects of local contact network geometry on the result.  This defines the response of one configuration, which is then averaged over multiple jammed configurations to obtain the final result.
% The response for one configuration is then computed as the change in the coarse grained stress field between the starting and the resultant coarse grained stress field averaged over the sequence of grains in the disc. Then, the final disorder averaged response is computed as the average of this change in the coarse grained stress fields over multiple starting configurations.}

%\bibliography{../granular_stress}

\bibliographystyle{unsrt}